\documentclass[11pt]{article}
\usepackage[utf8]{inputenc}
\usepackage{geometry}
\usepackage{textcomp}
\usepackage{amstext}

\usepackage{rotating}
\usepackage{threeparttable} 
\usepackage{multirow} 
\usepackage{tablefootnote} 
\usepackage{booktabs} 
\usepackage{float} 
\usepackage{pdflscape} 

\usepackage[table]{xcolor}
\usepackage{tabularx} 
\usepackage{dcolumn} 
\newcolumntype{d}[1]{D{.}{\cdot}{#1}} 

\usepackage[toc,page]{appendix} 

\makeatletter

\providecommand{\tabularnewline}{\\}

\usepackage[singlespacing]{setspace}

\makeatother

\usepackage[bibstyle=apa,citestyle=authoryear]{biblatex}
\begin{document}
\title{Who Manipulates Data During Pandemics?\\
Evidence from Newcomb-Benford Law }
\author{Vadim S. Balashov\thanks{Corresponding author: Vadim S. Balashov, Associate Professor, Rutgers
School of Business-Camden, Rutgers University-Camden, 227 Penn Street,
Camden, NJ 08102, work (856) 225-6706, cell (225) 202-8432, e-mail:
vadim.balashov@rutgers.edu}\\
Yuxing Yan\thanks{Yuxing (Paul) Yan, Assistant Professor, SUNY at Geneseo, 1 College
Circle, Geneseo, NY 14454, (585) 245-5260, e-mail: pyan@geneseo.edu}\\
Xiaodi Zhu\thanks{Xiaodi (Coco) Zhu, Assistant Professor, New Jersey City University,
200 Hudson Street, Jersey City, NJ 07305, (201)-200-2408, e-mail:
xzhu@njcu.edu\vspace{1cm}
\protect \\
The authors report no conflicts of interest. The project received
no funding. We thank Junyi Zhang, David Pedersen, Vsevolod Strekalov,
and Arkady Kurnosov for useful comments that helped improve the paper.
We thank Aleksandr Sabirov for swine flu (H1N1) 2009-2010 pandemic
data collection.}}

\date{July 25, 2020}
\maketitle
\begin{abstract}
We use the Newcomb-Benford law to test if countries have manipulated
reported data during the COVID-19 pandemic. We find that democratic
countries, countries with the higher gross domestic product (GDP)
per capita, higher healthcare expenditures, and better universal healthcare
coverage are less likely to deviate from the Newcomb-Benford law.
The relationship holds for the cumulative number of reported deaths
and total cases but is more pronounced for the death toll. The findings
are robust for second-digit tests, for a sub-sample of countries with
regional data, and in relation to the previous swine flu (H1N1) 2009–2010
pandemic. The paper further highlights the importance of independent
surveillance data verification projects.

\ \ 



\newpage{}
\end{abstract}

\section{Introduction}

On March 11, 2020, the World Health Organization (WHO)
declared the novel coronavirus disease 2019 (COVID-19) a pandemic.
With tens of millions of confirmed cases and over two million of deaths,
this pandemic has spurred a great number of controversies,
including many related to the accuracy of the data countries report.
Mass media organizations around the globe argue that many countries
have continued to manipulate the data for political or other gains.\footnote{See \textcite{meyer2020experts}, \textcite{romaniuk2020can}, \textcite{alwine2020manipulation},
\textcite{economist2020tracking}, \textcite{sassoon2020florida},
\cite{speak2020what}, \textcite{wood2020}, and \textcite{cambell2020china},
among others.}

In this paper, we study the association between the accuracy of COVID-19
data reported by countries and their macroeconomic and political indicators.
Our results show that countries that are more functional democracies,
have higher income, and stronger healthcare systems report more accurate
data. The relationship exists for the cumulative number of confirmed
cases and for the cumulative number of reported deaths; however, the
results are more pronounced for the number of deaths, indicating that
less developed countries are more likely to manipulate mortality data.

To gauge data accuracy, we use compliance with the Newcomb-Benford
law (NBL), which is an observation that in many naturally occurring
collections of numbers the first digit is not uniformly distributed.
The numeral “1” will be the leading digit around one-third of the
time; the numeral “2” will be the leading digit 18\% of the time;
and each subsequent numeral, “3” through “9,” will be the leading
digit with decreasing frequency. One property of the NBL is that manipulated
or fraudulent data deviate significantly from the theoretical NBL
distribution. Due to the ease of its application and straightforward
approach, the law has been extensively used to detect fraud and data
manipulations. It has been applied to accounting, finance, macroeconomic,
and forensic data to test for data manipulation and fraud. We apply
the NBL to COVID-19 data for 185 countries affected by the pandemic.
For each country, we first identify the period of exponential growth
when the data are expected to obey the NBL. After the country’s data
reach a plateau, the number of cases stabilizes, and the data are
not expected to obey the NBL. During the growth period for each country,
we calculate four goodness-of-fit measures to estimate compliance
with the NBL and use these measures as proxies for data manipulation.

We then study the relationship between our proxies for accuracy of
data and indicators of the strength of the economy, democratic institutions,
and healthcare systems. Specifically, we use the regression analysis
to find the association between goodness-of-fit measures and the Economist
Intelligence Unit Democracy Index, the gross domestic product (GDP)
per capita, healthcare expenditures as percentage of GDP, and the
Universal Health Coverage Index (UHC). Previous studies have shown
in other settings that countries with weaker democracies and less
economic development are more likely to manipulate data and have lower
transparency.\footnote{See \textcite{adsera2003you}, \textcite{egorov2009resource}, \textcite{magee2015reconsidering},
\textcite{gehlbach2014government}, and \textcite{guriev2019informational}
for studies on data manipulation. See \textcite{mitchell1998sources},
\textcite{broz2002political}, \textcite{djankov2003owns}, \textcite{islam2006does},
\textcite{lebovic2006democracies}, and \textcite{fearon2011self}
for transparency. \textcite{judge2009detecting} analyze the quality
of survey data using the NBL and find that survey data in developing
countries is of poor quality while data from developed countries is
of better quality.} The governments of such countries fabricate data for political gains
and to consolidate power. Autocratic governments control most mass
media outlets and often censor inconvenient facts that undermine the
ruling regime. COVID-19 presents such a case because the wide spread
of the pandemic and high death tolls would send a negative signal
to citizens and indicate that the government is incompetent. Autocratic
governments would try to downplay the scale of the pandemic for the
sake of appearances.

Our main hypothesis is that countries with weaker democracies, and
weaker economic and healthcare systems will have lower data accuracy
as measured by the NBL goodness-of-fit statistic. Our results in many
tests support the hypothesis. We find that our four goodness-of-fit
measures (which measure deviations from the theoretical distribution
as given by the NBL) are negatively correlated with macroeconomic
indicators (Democracy Index, GDP, healthcare expenditures, and UHC).
The results are true for the cumulative number of cases and the cumulative
number of reported deaths. We also find the result is more pronounced
for the reported number of deaths than for the number of confirmed
cases. This indicates that, on average, autocratic regimes and poorer
countries are more prone to manipulate death tolls than the total
number of citizens infected.

We conduct a series of robustness tests and find that our results
are not driven by the specific period in which we calculate the goodness-of-fit
measures, by small countries, by countries with a small number of
cases or deaths, or by countries with extreme deviations from the
NBL. We also show that the same relationship between proxies for accuracy
of data and economic indicators is observed when we apply the NBL
to second digits.

One concern of our study is that the proxies for data accuracy are
calculated based on limited sample sizes for individual countries.
To resolve this potential problem, we confirm our findings for the
sub-sample of 50 countries that provide regional data (at a state
or province level). Regional data increase the sample size from which
we calculate our statistics substantially and heighten the precision
of our accuracy measures.

We find a similar relationship for the previous swine flu (H1N1) pandemic
of 2009–2010. We repeat our analysis for 35 countries that reported
weekly data of the number of confirmed cases and the number of deaths
to Pan American Health Organization (PAHO). We discover support for
the negative relationship between deviations from the NBL and the
selected developmental indicators.

There is substantial body of literature assessing the tendency of
misreporting COVID-19 surveillance pandemic data by countries using
different statistical techniques, such as case fatality rates, excess
mortality rates, the variance of reported data, the clustering of
data, and even trends in search engines.\footnote{See \textcite{polson2020manipulated}, \textcite{aron2020pandemic},
\textcite{roukema2020anti}, \textcite{goutte2020macroeconomic},
\textcite{economist2020tracking}, and \textcite{dragan2020}.} The inherent problem with these COVID-19 studies is that they rely
on uniform approaches to measure confirmed cases and COVID-19-related
deaths across countries and periods within one country. Even though
countries are expected to follow the same guidelines provided by the
WHO when reporting cases, many variations exist (and sometimes appear
in states and regions within a country) in how they collect and report
data. Any comparisons of raw numbers—like the total number of confirmed
cases, the number of deaths, and mortality rates—among countries may
be also problematic because they are most likely driven by the difference
in other variables, like the number of tests conducted, the strength
of the healthcare system, demographic composition, and reporting standards.
Correct testing would require controlling for all those hard-to-observe
variables. This makes comparisons between countries difficult.

One helpful property of the NBL is its tolerance to different data
generating processes between observations (in our case, countries).
As long as data in each country are expected to obey the NBL, cross-observation
comparisons are possible. This means that we can apply the test even
if countries differ in how they measure COVID-19 cases and related
deaths. The test is also free of country-specific differences, including
public policies used to stop the pandemic, like quarantines, social
distancing, testing, and availability of treatment. The NBL test is
only sensitive to human intervention and manipulation of data in otherwise
naturally occurring processes.

Some studies apply the NBL to individual countries to test if a given
country has been falsifying data during the pandemic (\cite{idrovo2020data,jaskson2020national,peng2020statistical}).
We indicate that such tests are problematic because they largely depend
on the sample size. We stress that our paper is not aimed to answer
questions whether a particular government manipulates data. We use
the same approach to calculate goodness-of-fit measures for all countries
and then compare countries cross-sectionally based on developmental
indicators. Our results indicate that data from autocracies and poorer
countries should be trusted less, in line with the previous literature
(\cite{hollyer2011democracy}). It should also be noted that the NBL
test is not directional. However, it is unreasonable to believe that
if a country does not comply with the NBL, the government would willingly
manipulate data to inflate the number of cases or deaths. Neither
does the divergence from the NBL provide us specifics on how the data
are being manipulated. We cannot ex ante predict which first digits
will be over- or underinflated. For example, if the country’s true
number of cases is in the 2,000s and the government tries to falsify
data to look smaller and reports high 1,900s, the first digit “one”
will be over-represented in this country’s statistics. However, if
the country’s numbers are in the 1,000s and the governments falsifies
data into the 900s, then the first digit “one” will be under-represented.

Our paper contributes to the literature in several ways. First, our
paper helps to resolve the controversy about different countries’
data manipulation during pandemics and provides estimates of how widespread
it is around the globe. Using the NBL and COVID-19 data, we document
that about one-third of the 185 countries affected by the pandemic
indeed seem to misreport their data. Second, our study shows which
data, if any, countries are more likely to misreport. We document
that governments tend to downplay the news with the highest negative
impact, i.e., the death toll, to the highest degree. To a slightly
lower degree, countries tend to manipulate the total number of confirmed
cases. We find no indication, on average, of systematic data manipulation
for the number of conducted COVID-19 tests and the number of cured
cases. Third, we are the first study to use the NBL to show that the
strength of healthcare systems, as measured by healthcare expenditures
and the Universal Healthcare Index, are linked to the government’s
ability to provide reliable data during pandemics.

Finally, our paper contributes to the political economy and comparative
economics literature. We are the first to document the cross-sectional
link between macroeconomic and political regime indicators and the
tendency to misreport data during pandemics. We show that authoritarian
regimes and countries with low GDP per capita are more likely to falsify
data. Thus, this study provides additional evidence of the link between
democracy and transparency that is often taken for granted. Combined,
the results are consistent with previous findings in the literature
that authoritarian regimes and poorer countries’ governments manipulate
information to avoid negative news that may undermine their power.

Our study has broad implications. First, we provide evidence that
the data supplied during pandemics may be of low quality, especially
from autocracies and poorer countries, and we suggest that caution
should be used when interpreting and using the data. Second, the study
highlights the importance of initiatives to externally verify data
provided by governments, including independent surveillance data verification
projects.\footnote{An example of such a project for economic data would be the Billion
Prices Project (BPP) by Alberto Cavallo and Roberto Rigobon at MIT
Sloan and Harvard Business School.} Finally, we provide new evidence on the applicability of the NBL
to detect data fabrication.

The paper is organized as follows: Section 2 presents the review of
related literature and develops our main hypothesis, Section 3 discusses
the NBL of anomalous numbers, Section 4 describes our sample and variables,
Section 5 provides major results for the study, including robustness
checks, and Section 6 concludes .

\section{Literature Review and Hypothesis Development}

Studies have long posed questions about whether democratic regimes
provide more reliable data to the public than autocracies in both
theoretical and empirical settings. For many, the intuitive answer
is “yes.” However, this depends on the definition of “democracy.”
If democracy is defined only through electoral competition (e.g.,
\cite{schumpeter1942capitalism,Przeworski1999Democracy}), then the
link between data reliability and a particular political regime is
not obvious. Some authors argue that the expected relationship is
actually reverse: greater vulnerability to public disapproval within
democracies may lead to their higher tendency to falsify data (\cite{kono2006optimal,mani2007democracy}).
Most studies, however, show that democracies indeed are more transparent.
They argue that it is authoritarian regimes that are more vulnerable
to negative information and have more incentives to distort and manipulate
information that undermines their image. In addition, such regimes
usually have control over mass media organizations and therefore have
more capabilities to exercise control. \textcite{guriev2019informational}
show that modern authoritarian regimes do not use ideological propaganda
and political repression to the same extent as dictators in the twentieth
century. Instead, in the twenty-first century, information is the
key factor to obtain and retain power. Authoritarian regimes ground
their legitimacy and support from citizens in strong economic performance
and successful domestic and foreign affairs. When news that undermines
this image is released, it threatens the survival of the autocrat.
It hurts democratic leaders as well, but democratic regimes depend
more on voter welfare, which, in turn, is contingent on available
information (\cite{hollyer2011democracy}). Therefore, democracies
are more inclined to disclose truthful information. In addition, authoritarian
regimes exert much tighter control over information supplied to the
public, and as such they have easier ways to distort data. Autocrats
use data manipulation to improve their public image and prolong their
stay in office.

Indeed, \textcite{gehlbach2014government} demonstrate that government
media ownership increases media bias.\footnote{That bias, however, comes at a cost because citizens tend to trust
the news less, and because of the reduced advertising revenue. } In line with this argument, \textcite{rosenas2019how} propose that
autocrats are more likely to manipulate data for which it is more
difficult for citizens to obtain hard external information benchmarks.
Autocrats manipulate such data with censorship or falsifications.
Easily verifiable data are only framed to improve the image of the
government. The authors provide examples of how citizens can easily
benchmark news about income and market prices and, therefore, the
government resorts to a tender narrative when reporting such news;
whereas domestic politics and international affairs are hard to verify,
and the government more easily falsifies these data. COVID-19 provides
a unique setting to test a related hypothesis. Pandemic surveillance
data are hard to acquire independently by citizens because they lack
access to the necessary large-scale data collection and medical facilities.
At the same time, the news that the disease is raging and is widespread
under authoritarian rule would be an indicator of the inefficiency
or failure of the government. The death toll is even more damaging
to the image of the autocrat, who sees such news as a threat and tries
to downplay the scale of the problem. We therefore formulate the following
two hypotheses:

\bigskip{}

\emph{Hypothesis 1: Democratic regimes are less likely to manipulate
pandemic surveillance data.}\medskip{}

\emph{Hypothesis 2: The link between democratic regimes and data manipulation
is more pronounced for the reported death toll.}

\bigskip{}

Extant empirical studies find support for the hypothesis that democracies
provide more reliable data in different settings. For example, \textcite{bueno2003logic}
argue that countries with larger “winning coalitions”—i.e., democracies—are
more transparent than countries with small winning coalitions. By
analyzing the reported by governments tax revenues and national income
data, they find support for their hypothesis. \textcite{hollyer2011democracy}
use a model to back their similar hypothesis that in democratic regimes,
governments are more willing to disclose policy information. Their
empirical test is based on the willingness of governments to report
data to the World Bank’s World Development Indicators. \textcite{rosenas2019how}
use a corpus of daily news reports from Russia’s largest state-owned
television network, Channel 1, and find that the state-owned media
systematically frames facts to make the government look better. 

Some authors embrace extreme positions and claim that the reliability
of data supplied by the government should be a measure of the country’s
democracy level, and that elections and pluralism alone are not enough
(\cite{dahl1971polyarchy}). This is because, for elections to be
fair, voters should make informed decisions, and informed decisions
are only possible in regimes that provide reliable data to voters.
Studies, thus, differentiate between two measures of democracy: a
“thin” measure, or minimalist, i.e., covering only the election process
and freedoms; and a “thick” measure that includes more general concepts
like transparency and culture. Indeed, if democracy is defined, at
least partially, through transparency, then any findings regarding
the link between the two will be trivial. Therefore, the measure of
democracy, preferably, should not include the degree of transparency.

We address this issue by constructing several measures of democracy.
We start with the widely used \emph{Economist Intelligence Unit} Democracy
Index. We also study its five components: electoral pluralism, functioning
of the government, political participation, political culture, and
civil liberties. “Stripping” the Democracy Index helps to evaluate
the political component of democracy that is not directly related
to transparency: while some components are more likely to be related
to transparency (like political culture), others are not (like electoral
pluralism or the functioning of the government). We also use alternative
democracy measures, e.g., the \emph{Freedom House} Electoral Democracy
Index and their broader democracy measure (which includes political
freedom and civil liberties). The \emph{Freedom House} Electoral Democracy
Index is the “thin” definition of democracy, and should be unrelated
to transparency to avoid spurious correlation.

We also adopt other measures that may explain countries’ tendencies
to manipulate data. \textcite{hollyer2011democracy} maintain that
GDP per capita is a measure of the “\emph{ability} of the governments
to collect and disseminate high-quality statistical data.” We therefore
include the GDP per capita in our tests. Because our setup has been
created during the pandemic and the testing is done on surveillance
data, we use two other proxies for each country’s ability to collect
and report reliable health-related data: health expenditures as a
percentage of GDP, and the Universal Health Coverage Index. We thus
formulate our third hypothesis as follows:

\bigskip{}

\emph{Hypothesis 3A: Countries with higher GDP per capita levels are
less likely to manipulate pandemic surveillance data.}

\medskip{}

\emph{Hypothesis 3B: Countries with higher levels of health expenditures
as percentage of GDP are less likely to manipulate pandemic surveillance
data.}

\medskip{}

\emph{Hypothesis 3C: Countries with higher levels of the Universal
Health Coverage Index are less likely to manipulate pandemic surveillance
data.}

\bigskip{}

To gauge data manipulation, we use compliance with the NBL. We compare
countries cross-sectionally to find if there is a relationship between
developmental indicators and goodness-of-fit to the NBL measures.
We are not the only paper to use the NBL to test the validity of reported
data during COVID-19. Several other concurrent studies employ a similar
approach.\footnote{See \textcite{jaskson2020national,idrovo2020data,koch2020benford,peng2020statistical,zhang2020testing}.}
However, these papers usually select one or a few countries and apply
the NBL to test if there is any evidence of manipulation in a given
country’s data. The authors use the cutoff values from the chi-squared
distribution (or similar distributions) and give a “yes-or-no” type
of answer to their binary research question. In many cases, the goodness-of-fit
measures are calculated with substantial errors, and many studies
do not provide estimates for the statistical significance of the test
or its power. In addition, these test statistics and inference results
greatly depend on the sample size. With large enough sample sizes,
the null hypothesis of compliance with the NBL will be rejected in
almost every case. Some studies estimate their test statistic at the
country level, some studies estimate it at a regional or state level,
and some studies use county-level data. This leads to contradictory
findings among these studies even when looking at the same country.
Any inferences from such tests are also problematic.

Our approach is different. We use all countries affected by COVID-19.
For each test, we employ the same approach for all countries to calculate
the test statistic. We make any inferences from the NBL test only
in comparison. We study the link between compliance with the NBL and
economic indicators. The unit of observation in our analysis is the
country because it is this relationship between the proxies for data
manipulation and the democratic and economic indicators that we find
significant in most tests. To our best knowledge, this is the first
paper to examine the cross-section of all countries and compare them
based on developmental indicators when analyzing data manipulation
during pandemics.

Not many studies apply the NBL in an international setting, though
there are several notable exceptions. \textcite{nye2007political}
indicate that international macroeconomic data generally conforms
the NBL. They find, however, that for non-OECD (African) countries,
the data do not conform with the law, which raises questions about
data quality and manipulation in these countries. \textcite{gonzalez2009benford}
uses a similar approach to test the annual IMF data, but finds no
connection between independent assessments of data quality and adherence
to the first-digit NBL in different country groups. The limitation
of these studies, however, is that they group countries based on geographical
proximity, instead of some logical choice of economic indicators.
\textcite{michalski2013countries} provide a theoretical model and
empirical findings that some countries strategically misreport their
economic data for short-term government gains. The authors reveal
that some groups of countries (i.e. countries with fixed exchange
rate regimes, high negative net foreign asset positions, negative
current account balances, or greater vulnerability to capital flow
reversal) are more likely to falsify macroeconomic data than others.
Our paper is different in that it applies the NBL to the pandemic
data in the international setting. The countries in our study are
grouped based on developmental, economic, and political indicators.
We contribute to this body of literature by providing additional evidence
that some types of countries are more likely to falsify not only macroeconomic
data but also surveillance data during pandemics. 

\section{Newcomb-Benford Law of Anomalous Numbers}

In many naturally occurring processes, the resulting data have the
leading significant digit that is not uniformly distributed. The distribution
is monotonically decreasing, with ``1'' being the most common first
digit, and ``9'' being the least common. The law was formally stated
by \textcite{newcomb1881note} and \textcite{benford1938law}. A set
of numbers is said to follow the NBL if the first digit $d$ occurs
with probability $P(d)=log_{10}(1+\frac{1}{d}).$\footnote{The law can be extended to digits beyond the first. In general, for
the $n\text{th}$ digit, $n\geq2$, the probability is given by $P(d)=\mathop{\sum_{k=10^{n-2}}^{10^{n-1}-1}log_{10}(1+\frac{1}{10k+d}).}$} This gives the following probabilities for observing the first and
second digits:

\vspace{1cm}

\begin{tabular}{|c|c|c|c|c|c|c|c|c|c|c|}
\hline 
Digit & 0 & 1 & 2 & 3 & 4 & 5 & 6 & 7 & 8 & 9\tabularnewline
\hline 
\hline 
First & - & 30.1\% & 17.6\% & 12.5\% & 9.7\% & 7.9\% & 6.7\% & 5.8\% & 5.1\% & 4.6\%\tabularnewline
\hline 
Second & 12.0\% & 11.4\% & 10.9\% & 10.4\% & 10.0\% & 9.7\% & 9.3\% & 9.0\% & 8.8\% & 8.5\%\tabularnewline
\hline 
\end{tabular}

\vspace{1cm}

The data are expected to follow the NBL when the logarithms of values
are uniformly and randomly distributed. The NBL accurately describes
many real-life sets of numerical data, including lengths of rivers,
stock prices, street addresses, accounting data, populations, physical
constants, and regression coefficients (\cite{diekmann2007not}).
Data generated from many distributions and integer sequences have
been shown to closely obey the NBL, including Fibonacci numbers, powers
of numbers, exponential growth, many ratio distributions, and the
$F$-distribution with low degrees of freedom.\footnote{For more examples, see \textcite{hill1995statistical,hill1998first,leemis2000survival,formann2010newcomb,morrow2010}.} 

Not all distributions generate data that follow the law. For example,
uniform distribution, normal distribution, and square roots of numbers
do not obey it. For the data to obey the NBL, several criteria should
be satisfied (\cite{durtschi2004effective,diekmann2007not,tam2007breaking}):
\begin{itemize}
\item Data span several orders of magnitude and are relatively uniform over
such orders
\item The mean is greater than the median, with a positive skewness
\item Naturally occurring processes, the data which are the result of multiplicative
fluctuations, and data that is not influenced by human intervention
\end{itemize}
The last requirement, i.e., the fact that human intervention usually
generates data that violates the NBL, has led to its usefulness in
detecting fraud and data manipulation. Studies have shown that when
humans intervene with the data generating process that is expected
to comply with the NBL, compliance stops. For example, \textcite{diekmann2007not}
and \textcite{horton2018detecting} show that when scientific data
are fabricated, they do not conform with the NBL. The researchers
find that retracted accounting papers significantly deviate from the
NBL relative to a control group of papers. \textcite{cantu2010supervised}
and \textcite{breunig2011searching} reveal the same effect for electoral
data. In a similar spirit, \textcite{kaiser2019benford} uncovers
how discrepancies from the target NBL distribution can be used to
test reliability among survey data sets.

The NBL has been extensively used to detect fraud in accounting, finance,
and macroeconmic data. \textcite{nigrini2012benford} and \textcite{stambaugh2012using}
show that fraudulent trading records and fabricated returns do not
comply with the the NBL, whereas naturally occurring data do. \textcite{rauch2013libor}
apply the NBL to the London Inter-bank Offered Rate (LIBOR) rates
and successfully detect manipulated data. \textcite{o2017offsite}
study the determinants of fraudulent behavior among failed banks between
1989 and 2015. They use the second-digit NBL to identify those banks
whose financial statements suggest tampering and purposeful misstatements.
Their results suggest that insider abuse and fraud at banks are detectable
through an NBL analysis of bank financial data. \textcite{hussain2010application}
detects any possible data errors, irregularities, or fraud by applying
the NBL to the credit bureau data of commercial banks. By analyzing
five European equity market indices, \textcite{kalaichelvan2012critical}
find evidence that substantiates the criticism for the use of the
uniformity assumption for tests at the 1,000 level in favor of a distribution
consistent with the NBL.

Overall, the NBL has been used to detect fraud in: a) scientific studies
(\cite{geyer2004detecting,diekmann2007not,judge2009detecting,horton2018detecting}),
b) accounting (\cite{varian1972benfords,durtschi2004effective,suh2011effective,nigrini2012benford,stambaugh2012using,horton2018detecting}),
c) election data (\cite{mebane2006election,tam2007breaking,cantu2010supervised,breunig2011searching,deckert2011benford}),
d) macroeconomic data (\cite{nye2007political,gonzalez2009benford,hussain2010application,rauch2011fact,kalaichelvan2012critical,rauch2013libor,michalski2013countries,o2017offsite}),
e) forensic analysis (\cite{pinilla2018pinocchio}), f) tax evasion
(\cite{nigrini1996taxpayer,demir2018forensics}), g) toxic release
inventory (\cite{marchi2006assessing}), h) reported data during pandemics
(\cite{idrovo2011performance,gomez2016monitoring,idrovo2020data}).

Another useful property of the data obeying the NBL is that it is
scale invariant, i.e., it is independent of the measurement units.
This makes it a powerful tool when testing data from different sources
(i.e., countries, companies). The NBL is also not the same as the
imprecision (or variance) of the data. The data may well be very noisy
but still is expected to conform with the law, as long as there is
no deliberate falsification of data. For example, if a country’s data
are collected with error or irregularly but there are no data manipulations,
the first digit should still adhere to the NBL. In our application,
it means that countries may differ in the way they count COVID-19
cases or deaths, but as long as the data for each country is expected
to obey the NBL, we can test the data for the goodness-of-fit to the
NBL. 

\section{Data and Variables}

\subsection{Goodness-of-Fit Measures}

To measure how well the data comply with the NBL, we use several goodness-of-fit
measures. The most intuitive and commonly used is the chi-squared
statistic:

\begin{equation}
Chi\text{-\ensuremath{sq.=\mathop{\sum_{d=1}^{9}\frac{(O_{d}-E_{d})^{2}}{E_{d}}}}},
\end{equation}
where $O_{d}$ and $E_{d}$ are observed and expected by the NBL frequencies
for digit $d$, respectively. Chi-squared, however, has several problems:
it has low statistical power when used with small sample sizes and
enormous power with large sample sizes. Therefore, we use alternative
measures of goodness-of-fit proposed in extant studies. We use a modified
version of the \textcite{kuiper1960tests} test proposed by \textcite{stephens1970use}
and \textcite{giles2007benford} that is less dependent on the sample
size $N$:

\begin{equation}
Kuiper=(D_{N}^{+}+D_{N}^{-})\left[\sqrt{N}+0.155+\frac{0.24}{\sqrt{N}}\right],\text{\ensuremath{\text{ and}}}
\end{equation}

\begin{equation}
D_{N}^{+}=\sup_{-\infty<x<+\infty}|F_{N}(x)-F_{0}(x)|\text{ and \ensuremath{D_{N}^{-}=\sup_{-\infty<x<+\infty}|F_{0}(x)-F_{N}(x)|}},
\end{equation}
where $F_{N}(x)$ and $F_{0}(x)$ are the observed cumulative distribution
functions ($cdf$) of leading digits and the $cdf$ of the data that
comply with the NBL. In addition, we calculate the $\text{\ensuremath{M}}$-statistic
proposed by \textcite{leemis2000survival}:

\begin{equation}
M=\max_{d=1}^{9}|o_{d}-e_{d}|\sqrt{N}
\end{equation}
and the $D$-statistic proposed by \textcite{tam2007breaking}:

\begin{equation}
D=\sqrt{N\sum_{d=1}^{9}(o_{d}-e_{d})^{2}}
\end{equation}
where $o_{d}$ and $e_{d}$ are the proportions of observations with
$d$ as the first digit and proportions expected by the NBL, respectively.
The latter two measures are also insensitive to sample size. We calculate
each goodness-of-fit measure for two variables, the cumulative number
of confirmed cases and the cumulative number of reported deaths. In
unreported tests we also analyze two other variables: the number of
cured cases and the number of conducted COVID-19 tests, and find insignificant
results.

\subsection{Sample Description and Developmental Indicators}

We first collect daily data from John Hopkins University for the cumulative
number of confirmed cases, the cumulative number of cured cases, and
the cumulative number of deaths \footnote{https://coronavirus.jhu.edu/map.html. Downloadable database is available
at https://github.com/CSSEGISandData/COVID-19.} between January 22, 2020 and June 10, 2020. We also obtain the number
of conducted tests from Our World in Data.\footnote{https://ourworldindata.org/coronavirus-testing.}
Studies have shown that naturally occurring processes comply well
with the NBL when the data grow exponentially or close to it (\cite{leemis2000survival,formann2010newcomb}).
Once the data reach the plateau, they are no longer expected to obey
the NBL. Hence, for the data to comply with the NBL, we select the
growth part using the following approach. Because data show weekly
seasonality, we first compute seven-day moving averages (MA) for the
new daily number of confirmed cases. Then, for each country, we identify
the date with the highest MA number of new daily confirmed cases.
If there are several dates with the same maximum, we use the earliest
as the cutoff. For our main analyses, we use data before the obtained
cutoff for each country.\footnote{In unreported tests, we also use modified approaches. We find the
maximum ratio MA(number of new daily cases)/(Days since the first
case for the country) and MA(number of new daily cases)/(Days since
the latest nonzero case for the country). The results are robust to
alternative definitions of the cutoff date.}

For developmental indicators, we select the following four proxies
for democratic and economic development widely used in the literature:
the \emph{Economist Intelligence Unit} Democracy Index, GDP per capita,
healthcare expenditures as a percentage of GDP, and the Universal
Health Coverage Index. The Democracy Index ($EIU$) is a weighted-average
of answers to 60 questions from expert assessments grouped into five
categories: electoral pluralism, functioning of the government, political
participation, political culture, and civil liberties. The index is
aimed to measure the degree of democracy of a country. In addition
to the Democracy Index, we use GDP per capita as a proxy for the country’s
ability to provide precise data. We also take the country’s healthcare
spending as a percentage of GDP and its Universal Health Coverage
Index as proxies for the strength of each country’s healthcare system.
We download countries’ democracy indices from the \emph{Economist
Intelligence Unit} for 2019.\footnote{https://www.eiu.com/topic/democracy-index. }
We collect the Gross Domestic Product ($GDP$) per capita, healthcare
expenditures as percentage of GDP ($HE\text{\_}GDP$), and Universal
Health Coverage Index ($UHC$) for 2017 from the World Bank.\footnote{https://data.worldbank.org/. Downloadable database is available from
https://databank.worldbank.org/indicator/NY.GDP.MKTP.KD.ZG/1ff4a498/Popular-Indicators.
At the time we collected the data, many countries still did not have
the World Bank data available for 2018 or 2019. 2017 is the latest
year for which the data are available for all countries. } We also acquire 2019 population data for each country from Worldometer.\footnote{https://www.worldometers.info/.}
A total of 185 countries with available data were affected by COVID-19.
The summary for each country can be found in Appendix A1. Variable
definitions can be found in Appendix A2.

We find that we cannot reject the NBL distribution for the entire
world population for the cumulative number of confirmed cases using
the 1\% significance level (Appendix A1).\footnote{The 1\% threshold for all four measures are: 20.09 for Chi-squared,
2.00 for Kuiper, 1.21 for M, and 1.57 for D.} For the cumulative number of reported deaths, however, we reject
the hypothesis of compliance with the NBL for the total world numbers.
This indicates that, on average, countries are more likely to falsify
death tolls, and are less likely to falsify the confirmed number of
cases. Using country-level data, we also find that between 37 and
62 countries (depending on the goodness-of-fit measure used) out of
185 deviate from the NBL when reporting the confirmed cases. Between
50 and 71 countries deviate from the NBL when reporting the number
of deaths.\footnote{The values go up to between 62 and 103 and between 72 and 99 countries
for the number of cases and deaths, respectively, when the 10\% level
of significance is used. Note also that switching from country-level
data to state data or county-level data increases the statistics significantly
(see Table A1.2 for the case of the U.S. county-level data.}

Table 1 provides descriptive statistics for the major variables in
our analyses. For the confirmed number of cases, the goodness-of-fit measures show that the average country’s
data are about borderline in terms of compliance with the NBL: they
are consistent with the NBL if 1\% level of significance is used, but
are not consistent if 10\% is used. Observe further that the corresponding mean goodness-of-fit
measures for the number of deaths are higher than for the number of
confirmed cases and are lagrely not consistent with the NBL at 1\%. This indicates that, ceteris paribus, countries are
more prone to manipulate data on death rates. We must stress, however, that any inferences about
data manipulation simply based on individual goodness-of-fit statistics
is questionable because they largely depend on the selected sample
size (country versus state, state versus county level data). In our
analyses, therefore, we aim to compare countries cross-sectionally.

The average country in our sample has over 42 million people, slightly
less than \$6,000 in terms of GDP per capita,\footnote{Note that we use the natural logarithm of the population and GDP values
when calculating averages for Table 1.} with roughly 6\% of the GDP spent on healthcare expenditures, a democracy
index of around 55 on the scale between zero and 100, and around 65\%
of the population are covered by universal health care (Table 1).
The average sample size used to estimate the goodness-of-fit measures
per country is slightly over 61 days for the number of confirmed cases
and is around 40 for the number of deaths (until the end of the growth
period).

{\centering
	\begin{table}[!htb]
		\begin{threeparttable}
			\footnotesize
			\renewcommand{\arraystretch}{1.25}
			\label{tab0:SummaryStat}
			\caption{Descriptive Statistics}
			\begin{tabular*}{\textwidth}{l@{\extracolsep{\fill}}llllll}
				\toprule
				\textbf{Variables}& \textbf{Obs} & \textbf{Mean} & \textbf{Min} & \textbf{Med} & \textbf{Max} & \textbf{Std.} \\
				
				\midrule
				
				\textbf{Chi-sq. Conf.} & 185 & 19.55\tnote{**} & \phantom{9}1.40 & 13.60\tnote{*} & 129.38\tnote{***} & 20.22 \\
				\textbf{Kuiper Conf.} & 185 & \phantom{9}1.48 & \phantom{9}0.39 & \phantom{9}1.37 & \phantom{99}4.70\tnote{***} & \phantom{9}0.73 \\
				\textbf{M Conf.} & 185 & \phantom{9}1.08\tnote{**} & \phantom{9}0.29 & \phantom{9}0.90\tnote{*} & \phantom{99}4.54\tnote{***} & \phantom{9}0.69 \\
				\textbf{D Conf.} & 185 & \phantom{9}1.44\tnote{**} & \phantom{9}0.42 & \phantom{9}1.31\tnote{*} & \phantom{99}5.04\tnote{***} & \phantom{9}0.74 \\
				
				\textbf{Chi-sq. Death} & 160 & 29.35\tnote{***,3} & \phantom{9}1.71 & 17.22\tnote{**} & 261.49\tnote{***} & 37.35 \\
				\textbf{Kuiper Death} & 160 & \phantom{9}1.75\tnote{**,3} & \phantom{9}0.32 & \phantom{9}1.54 & \phantom{99}5.81\tnote{***} & \phantom{9}1.07 \\
				\textbf{M Death} & 160 & \phantom{9}1.32\tnote{***,3} & \phantom{9}0.19 & \phantom{9}1.00\tnote{**} & \phantom{99}5.68\tnote{***} & \phantom{9}0.99 \\
				\textbf{D Death} & 160 & \phantom{9}1.72\tnote{***,3} & \phantom{9}0.39 & \phantom{9}1.46\tnote{**} & \phantom{99}6.10\tnote{***} & \phantom{9}1.07 \\

				\textbf{EIU} & 163 & 54.84 & 13.20 & 56.50 & \phantom{9}98.70 & 21.98 \\	
				\textbf{ln(GDP)} & 178 & \phantom{9}8.68 & \phantom{9}5.68 & \phantom{9}8.65 & \phantom{9}12.03 & \phantom{9}1.45 \\														
				\textbf{HE\_GDP} & 174 & \phantom{9}6.44 & \phantom{9}1.18 & \phantom{9}6.23 & \phantom{9}17.06 & \phantom{9}2.57 \\
				\textbf{UHC} & 175 & 64.44 & 25.00 & 69.00 & \phantom{9}89.00 & 15.68 \\
				
				\textbf{No. of Days Conf.} & 185 & 61.24 & \phantom{9}1.00 & 61.00 & 136.00 & 30.61 \\
				\textbf{No. of Days Death} & 185 & 39.59 & \phantom{9}0.00 & 36.00 & 124.00 & 29.74 \\
				\textbf{ln(Population)} & 182 & 15.84 & 10.43 & 16.08 & \phantom{9}21.09 & \phantom{9}2.02 \\

				\bottomrule
			\end{tabular*}
			\begin{tablenotes}
				\singlespacing
				\item[] The table presents the mean value of goodness-of-fit measures (Chi-square, Kuiper, M and D) for the cumulative number of confirmed and death cases, developmental indicators (EIU, ln(GDP), HE\_GDP, UHC), and other variables. The number of observations vary due to missing values. The original dataset is included in Appendix A1. ***, ** and * denote goodness-of-fit measures that correspond to significant differences from the theoretical NBL distribution at 1\%, 5\% and 10\% level, respectively. We also analyze the difference between the Confirmed an the Death mean values for each goodness-of-fit measures using the \textit{t}-test. 3, 2, and 1 indicate significant difference between the Confirmed and the Death cases at the 1\%, 5\% and 10\% level, respectively.
				All variable definitions are in Appendix A2. 
			\end{tablenotes}
	
		\end{threeparttable}
	\end{table}
\par
}


%
Table 2 provides mean values for our goodness-of-fit measures for
the four quartiles of each of our independent variables: $EIU$, ln($GDP$),
$HE\text{\_}GDP$, and $UHC$. The quartiles for the $EIU$ Democracy
Index roughly correlate with the definitions of the four regime types:
full democracy, flawed democracy, hybrid regime, and authoritarian
regime. The table shows a general monotonic trend for the data to
deviate more from the Newcomb-Benford distribution as we move from
the highest quartile to the lowest. For the top quartiles, we cannot
reject the hypothesis that countries manipulate confirmed cases or
death data at the 1\% level. For the bottom quartile, however, we
reject that hypothesis about half the time for the number of confirmed
cases, and almost every time for the number of deaths. In a univariate
setting, this is consistent with our three hypotheses. We also find
that for the cumulative number of deaths, the difference between goodness-of-fit
measures for the top and bottom quartiles is always significant.
\begin{sidewaystable}
	\centering

	\begin{threeparttable}[h]
		\footnotesize
		\renewcommand{\arraystretch}{1.2}
		\label{tab0:SummaryQuartile}
		\caption{Mean Values of Goodness-of-Fit Measures}
		\begin{tabular*}{\linewidth}{l|@{\extracolsep{\fill}}llll | llll}
			\toprule
			\textbf{Indicator}& \textbf{Chi-sq. Conf.} & \textbf{Kuiper Conf.} & \textbf{M Conf.} & \textbf{D Conf.} & \textbf{Chi-sq. Death} & \textbf{Kuiper Death} & \textbf{M Death} & \textbf{D Death} \\
			
			\midrule
			
			\textbf{EIU Largest} & 15.76\tnote{**} & 1.32 & 1.02\tnote{**} & 1.32\tnote{*} & 16.13\tnote{**} & 1.25 & 0.95\tnote{*} & 1.20 \\	
			\textbf{EIU Q3} & 19.42\tnote{**} & 1.53 & 1.12\tnote{**} & 1.48\tnote{**} & 29.54\tnote{***} & 1.73\tnote{*} & 1.03\tnote{**} & 1.42\tnote{**} \\
			\textbf{EIU Q2} & 20.06\tnote{**} & 1.49 & 1.00\tnote{**} & 1.43\tnote{**} & 35.58\tnote{***} & 2.03\tnote{***} & 1.52\tnote{***} & 1.90\tnote{***} \\
			\textbf{EIU Smallest} & 20.99\tnote{***} & 1.48 & 1.09\tnote{**} & 1.45\tnote{**} & 36.95\tnote{***,3} & 1.96\tnote{**,3} & 1.36\tnote{***,1} & 1.78\tnote{***,2} \\	
			
			\midrule

			\textbf{GDP Largest} & 13.91\tnote{*} & 1.24 & 0.97\tnote{**} & 1.27\tnote{*} & 14.71\tnote{*} & 1.21  & 0.86\tnote{*} & 1.12 \\
			\textbf{GDP Q3} & 17.28\tnote{**} & 1.41 & 0.98\tnote{**} & 1.30\tnote{*} & 22.57\tnote{***} & 1.56  & 0.95\tnote{*} & 1.26\tnote{*} \\	
			\textbf{GDP Q2} & 19.67\tnote{**} & 1.47 & 1.09\tnote{**} & 1.46\tnote{**} & 36.99\tnote{***} & 2.02\tnote{***}  & 1.32\tnote{***} & 1.70\tnote{***} \\	
			\textbf{GDP Smallest} & 26.20\tnote{***,3} & 1.74\tnote{*,3} & 1.24\tnote{***,3} & 1.71\tnote{***,3} & 39.41\tnote{***,2} & 2.09\tnote{***,3}  & 1.44\tnote{***,3} & 1.87\tnote{***,3} \\	
			
			\midrule

			\textbf{HE\_GDP Largest} & 15.41\tnote{*} & 1.38 & 1.04\tnote{**} & 1.35\tnote{**} & 16.32\tnote{**} & 1.29  & 0.91\tnote{*} & 1.20 \\
			\textbf{HE\_GDP Q3} & 17.02\tnote{**} & 1.34 & 0.91\tnote{*} & 1.27\tnote{*} & 24.83\tnote{***} & 1.67\tnote{*}  & 1.20\tnote{**} & 1.56\tnote{**} \\
			\textbf{HE\_GDP Q2} & 20.71\tnote{***} & 1.58 & 1.22\tnote{***} & 1.58\tnote{***} & 29.74\tnote{***} & 1.66\tnote{*}  & 1.01\tnote{**} & 1.33\tnote{**} \\
			\textbf{HE\_GDP Smallest} & 23.14\tnote{***,2} & 1.56 & 1.11\tnote{**} & 1.54\tnote{**} & 46.18\tnote{***,3} & 2.29\tnote{***,3} & 1.37\tnote{***,2} & 1.78\tnote{***,2} \\
			
			\midrule

			\textbf{UHC Largest} & 13.90\tnote{*} & 1.25 & 0.92\tnote{*} & 1.23\tnote{*} & 15.22\tnote{**} & 1.24  & 0.91\tnote{*} & 1.18 \\
			\textbf{UHC Q3} & 15.55\tnote{**} & 1.39 & 1.01\tnote{**} & 1.34\tnote{**} & 22.04\tnote{***} & 1.52  & 0.96\tnote{*} & 1.32\tnote{*} \\
			\textbf{UHC Q2} & 28.36\tnote{***} & 1.69\tnote{*} & 1.32\tnote{***} & 1.70\tnote{***} & 44.07\tnote{***} & 2.17\tnote{***} & 1.28\tnote{***} & 1.63\tnote{***} \\
			\textbf{UHC Smallest} & 22.26\tnote{***,2} & 1.61\tnote{2} & 1.09\tnote{**} & 1.54\tnote{**,2} & 39.81\tnote{***,3} & 2.13\tnote{***,3} & 1.53\tnote{***,3} & 1.98\tnote{***,3} \\

			\bottomrule
		\end{tabular*}
		\begin{tablenotes}
			\singlespacing
			\item The table presents the mean values of goodness-of-fit measures (Chi-square, Kuiper, M and D) for the cumulative number of confirmed and death cases by four quartiles of developmental indicators (EIU, ln(GDP), HE\_GDP, UHC). Smallest, Q2, Q3 and Largest represent the values partitioned by quartile 25\%, 50\%, 75\%. ***, ** and * denote goodness-of-fit measures that correspond to significant differences from the theoretical NBL distribution at 1\%, 5\% and 10\% level, respectively. We also analyze the difference between the Smallest an the Largest quartiles for each indicator using the \textit{t}-test. 3, 2, and 1 indicate significant difference between the Smallest and the Largest quartiles at the 1\%, 5\% and 10\% level, respectively. All variable definitions are in Appendix A2.
			
			
		\end{tablenotes}
		
	\end{threeparttable}

\end{sidewaystable}


Table 3 provides Pearson correlation coefficients between major variables.
The four major economic indicators, especially $EIU$, ln($GDP$),
and $UHC$, are highly correlated, with correlation coefficients ranging
between 0.59 and 0.85 (all values are statistically significant).
$HE\text{\_}GDP$ is also correlated with the other indicators, with
correlation coefficients ranging between 0.37 and 0.46 (also significant).
These variables are most likely proxies for the same indicator, the
development level of a country, and therefore—to avoid multicollinearity—we
include only one indicator at a time in our analysis.\footnote{In unreported tests we put all economic indicators together in one
equation on the right-hand-side and test for collinearity. The result
shows the Condition Number is over 46, indicative of serious collinearity.} The four goodness-of-fit measures are also highly correlated with
each other. The total number of confirmed cases and the country's
population are also significantly correlated. Univariate results in
Table 3 also show that all goodness-of-fit measures are negatively
correlated with the four economic indicators, with 22 out of 32 correlation
coefficients being significant (all correlation coefficients for the
cumulative number of deaths are significant).

{\centering
\begin{sidewaystable}
	\begin{center}
		\begin{threeparttable}[h]
			\caption{Correlation Matrix}
			\label{tab0:Correlation}
			\centering
			\footnotesize
			\renewcommand\arraystretch{1.2}
			\setlength{\tabcolsep}{3pt}
			\begin{tabular*}{0.87\linewidth}{@{\extracolsep{\fill}}ll}
				\toprule
				\begin{tabular}[t]{@{\extracolsep{\fill}}ll}
					\textbf{Variables}\\
					\cmidrule(r){1-2}
					\textbf{Chi-sq. Conf.} & (1)\\
					\textbf{Kuiper Conf.} & (2)\\
					\textbf{M Conf.} & (3)\\
					\textbf{D Conf.} & (4) \\
					
					\textbf{Chi-sq. Death} & (5)\\
					\textbf{Kuiper Death} & (6)\\
					\textbf{M Death} & (7)\\
					\textbf{D Death} & (8)\\
					
					\textbf{EIU} & (9)\\
					\textbf{ln(GDP)} & (10)\\
					\textbf{HE\_GDP}  & (11)\\
					\textbf{UHC}  & (12)\\
					
					\textbf{ln(Population)}  & (13)\\
					
				\end{tabular} &
				
				\begin{tabular}[t]{@{\extracolsep{\fill}}lllllllllllll}
					\multicolumn{1}{c}{(1)} & \multicolumn{1}{c}{(2)} & \multicolumn{1}{c}{(3)} & \multicolumn{1}{c}{(4)} & \multicolumn{1}{c}{(5)} & \multicolumn{1}{c}{(6)} & \multicolumn{1}{c}{(7)} & \multicolumn{1}{c}{(8)} & \multicolumn{1}{c}{(9)} & \multicolumn{1}{c}{(10)} & \multicolumn{1}{c}{(11)} & \multicolumn{1}{c}{(12)} & \multicolumn{1}{c}{(13)} \\
					\cmidrule(r){1-13}
					& \phantom{-}0.89\tnote{***} & \phantom{-}0.81\tnote{***} & \phantom{-}0.91\tnote{***} & \phantom{-}0.48\tnote{***} & \phantom{-}0.53\tnote{***} & \phantom{-}0.56\tnote{***} & \phantom{-}0.54\tnote{***} & -0.03 & -0.21\tnote{**} & -0.13 & -0.19\tnote{**} & -0.12   \\
					
					& & \phantom{-}0.88\tnote{***} & \phantom{-}0.95\tnote{***} & \phantom{-}0.46\tnote{***} & \phantom{-}0.47\tnote{***} & \phantom{-}0.47\tnote{*} & \phantom{-}0.48\tnote{***} & -0.01 & -0.21\tnote{**} & -0.09 & -0.16\tnote{*} & -0.10 \\
					
					& &  & \phantom{-}0.96\tnote{***} & \phantom{-}0.46\tnote{***} & \phantom{-}0.45\tnote{***} & \phantom{-}0.44\tnote{***} & \phantom{-}0.44\tnote{***} & \phantom{-}0.05 & -0.08 & -0.05 & -0.06 & -0.05 \\
					
					& &  &  & \phantom{-}0.49\tnote{***} & \phantom{-}0.50\tnote{***} & \phantom{-}0.51\tnote{***} & \phantom{-}0.50\tnote{***} & \phantom{-}0.00 & -0.18\tnote{**} & -0.08 & -0.15\tnote{*} & -0.08 \\
					
					& 	&  &  &  & \phantom{-}0.86\tnote{***} & \phantom{-}0.81\tnote{***} & \phantom{-}0.86\tnote{***} & -0.18\tnote{**} & -0.27\tnote{***} & -0.27\tnote{***} & -0.28\tnote{***} & -0.16\tnote{*} \\
					
					& &  &  &  &  & \phantom{-}0.94\tnote{***} & \phantom{-}0.97\tnote{***} & -0.23\tnote{***} & -0.35\tnote{***} & -0.30\tnote{***} & -0.34\tnote{***} & -0.23\tnote{***} \\
					
					& &  &  &  &  &  & \phantom{-}0.98\tnote{***} & -0.20\tnote{**} & -0.30\tnote{***} & -0.25\tnote{***} & -0.32\tnote{***} & -0.28\tnote{***} \\
					
					& &  &  &  &  &  &  & -0.25\tnote{***} & -0.36\tnote{***} & -0.31\tnote{***} & -0.37\tnote{***} & -0.26\tnote{***} \\
					
					& &  &  &  &  &  &  &  & \phantom{-}0.64\tnote{***} & \phantom{-}0.46\tnote{***} & \phantom{-}0.59\tnote{***} & -0.11 \\
					
					& &  &  &  &  &  &  &  &  & \phantom{-}0.37\tnote{***} & \phantom{-}0.85\tnote{***} & -0.15\tnote{*} \\
					
					& &  &  &  &  &  &  &  &  &  & \phantom{-}0.45\tnote{***} & -0.04 \\
					
					& &  &  &  &  &  &  &  &  &  &  & -0.07 \\

				\end{tabular} \\
				\bottomrule	
			\end{tabular*}
			\begin{tablenotes}
				\singlespacing
				\item[] The table shows the correlation matrix with Pearson correlation coefficients of major variables. ***, **, * denote significance at the 1\%, 5\%, 10\% level, respectively. All variable definitions are in Appendix A2.
			\end{tablenotes}
		\end{threeparttable}

	\end{center}
\end{sidewaystable}

\par
}

\section{Results}

\subsection{Goodness-of-fit and Economic Indicators}

We start with the simple ordinary least squares (OLS) regression model
where our goodness-of-fit measures appear on the left-hand-side and
economic indicators are on the right-hand-side:

\begin{eqnarray}
Goodness\text{-\ensuremath{of\text{-\ensuremath{fit}}}}_{i} & = & \beta_{0}+\beta_{1}\mathbf{Indicator}{}_{i}+\beta_{2}\ln(Population){}_{i}+\nonumber \\
 &  & +\beta_{3}Number\text{\_\ensuremath{of\text{\_\ensuremath{Days_{i}}}}}+\varepsilon_{i},\label{eq:1}
\end{eqnarray}
where $Indicator_{i}$ denotes one of the four economic indicators:
$EIU$, ln($GDP)$, $HE\text{\_}GDP$, or $UHC$. Higher values of
the goodness-of-fit measures indicate greater deviation from the NBL.
If more developed countries are less likely to manipulate data, we
expect the coefficient $\beta_{1}$ to be negative.

How well the data for each country are \emph{expected} to obey the
NBL depends on the span. For example, countries with higher populations
and more confirmed cases or deaths are expected to follow the NBL
more closely. To control for that, we include the natural logarithm
of the country’s total population.\footnote{Alternatively, we include the very correlated number of confirmed
cases (or deaths). We find that the results are qualitatively and
quantitatively the same when we used alternative control variables
(untabulated).} Even though the Kupier, M, and D-statistics are more independent
of the sample size, goodness-of-fit measures may still be affected
by the sizes of the samples used to estimate them. To control for
the sample size effect, we include $Number\text{\_\ensuremath{of\text{\_\ensuremath{Days_{i}}}}}$,
which is the number of days with nonzero confirmed cases (or the number
of days with nonzero deaths) between January 22, 2020 and the cutoff
date for the growth part for each country.

The results of estimating Equation \ref{eq:1} are presented in Table
4. Panel A provides estimates for the cumulative number of confirmed
cases. All but one coefficient in front of economic indicators are
negative. Coefficients for ln($GDP$) and $UHC$ are always significant.
The coefficient for $EIU$ is significant only when the chi-squared
goodness-of-fit measure is used, and the coefficient for $HE\text{\_}GDP$
lacks significance in all tests. Panel B provides estimates for the
cumulative number of deaths. All coefficients are negative, and all
are significant. The magnitude of coefficients for the number of deaths
is also much higher than that for the number of confirmed cases. We
find that the coefficients for corresponding economic indicators are
statistically different from each other between Panels A and B in
each case in Table 4.

{\centering

\begin{sidewaystable}
	\centering
	
	\begin{threeparttable}[h]
		
		\def\sym#1{\ifmmode^{#1}\else\(^{#1}\)\fi}
		
		\caption{Main Results: OLS Regressions}\label{tab_res:goodness_ols} 
		\scriptsize
		\renewcommand{\arraystretch}{0.4}
		\begin{tabular*}{0.9\linewidth}{@{\extracolsep{\fill}}l|l|l}
			\toprule[1.5pt]\midrule[0.3pt]
			
			\begin{tabular}[t]{ll}
				
				& \textcolor{white}{1}\\
				\arrayrulecolor{white}\midrule
				\multicolumn{2}{c}{\textbf{Variable}} \\
				
				\arrayrulecolor{black}\midrule

				\textbf{EIU}\tnote{{\textcolor{white}{*}}} \\
				\textcolor{white}{()} \\
				\addlinespace[3pt]
				
				\textbf{ln(Population)}\tnote{{\textcolor{white}{*}}} \\
				\textcolor{white}{()} \\
				\addlinespace[2pt]
				
				\textbf{No. of Days}\tnote{{\textcolor{white}{*}}}  \\
				\textcolor{white}{()} \\
				\addlinespace[1.8pt]
				
				\textbf{Sample Size} \\
				\addlinespace[2.8pt]
				
				\textbf{Adj. R\textsuperscript{2}} \\
				\midrule

				\textbf{ln(GDP)}\tnote{{\textcolor{white}{*}}} \\
				\textcolor{white}{()} \\
				\addlinespace[2pt]
				
				\textbf{ln(Population)}\tnote{{\textcolor{white}{*}}} \\
				\textcolor{white}{()} \\
				\addlinespace[1.8pt]
				
				\textbf{No. of Days}\tnote{{\textcolor{white}{*}}}  \\
				\textcolor{white}{()} \\
				\addlinespace[1.8pt]
				
				\textbf{Sample Size} \\
				\addlinespace[2.8pt]
				
				\textbf{Adj. R\textsuperscript{2}} \\
				\midrule

				\textbf{HE\_GDP}\tnote{{\textcolor{white}{*}}} \\
				\textcolor{white}{()} \\
				\addlinespace[2pt]
				
				\textbf{ln(Population)}\tnote{{\textcolor{white}{*}}} \\
				\textcolor{white}{()} \\
				\addlinespace[1.8pt]
				
				\textbf{No. of Days}\tnote{{\textcolor{white}{*}}}  \\
				\textcolor{white}{()} \\
				\addlinespace[1.8pt]
				
				\textbf{Sample Size} \\
				\addlinespace[2.8pt]
				
				\textbf{Adj. R\textsuperscript{2}} \\
				\midrule

				\textbf{UHC}\tnote{{\textcolor{white}{*}}} \\
				\textcolor{white}{()} \\
				\addlinespace[3pt]
				
				\textbf{ln(Population)}\tnote{{\textcolor{white}{*}}} \\
				\textcolor{white}{()} \\
				\addlinespace[1.8pt]
				
				\textbf{No. of Days}\tnote{{\textcolor{white}{*}}}  \\
				\textcolor{white}{()} \\
				\addlinespace[1.8pt]
				
				\textbf{Sample Size} \\
				\addlinespace[2.8pt]
				
				\textbf{Adj. R\textsuperscript{2}} \\

			\end{tabular} &

			\begin{tabular}[t]
				{
					D{.}{.}{-1}
					D{.}{.}{-1}
					D{.}{.}{-1}
					D{.}{.}{-1}
				}

				\multicolumn{4}{c}{\textbf{Panel A. Confirmed Cases}}  \\
				
				\midrule
				\multicolumn{1}{c}{\textbf{Chi-squared}} & \multicolumn{1}{c}{\textbf{Kuiper}} & \multicolumn{1}{c}{\textbf{M}} & \multicolumn{1}{c}{\textbf{D}} \\

				\midrule
				
				-9.54\tnote{*} & -0.19 & -0.02 & -0.15 \\
				( 0.08) & ( 0.22) & ( 0.46) & ( 0.27) \\
				\addlinespace[3pt]
				
				-3.41\tnote{***} & -0.12\tnote{***} & -0.08\tnote{**} & -0.11\tnote{***} \\
				( 0.00) & ( 0.00) & ( 0.02) & ( 0.00) \\
				\addlinespace[3pt]
				
				20.78\tnote{***} &  0.93\tnote{***} &  0.75\tnote{***} &  0.87\tnote{***} \\
				( 0.00) & ( 0.00) & ( 0.00) & ( 0.00) \\
				\addlinespace[3pt]
				
				\multicolumn{1}{c}{162} & \multicolumn{1}{c}{162} & \multicolumn{1}{c}{162} & \multicolumn{1}{c}{162} \\
				\addlinespace[3pt]
				10.05\% & 11.57\% & 7.12\% & 9.39\% \\

				\midrule
				-3.41\tnote{***} & -0.13\tnote{***} & -0.07\tnote{**} & -0.12\tnote{***} \\
				( 0.00) & ( 0.00) & ( 0.02) & ( 0.00) \\
				\addlinespace[3pt]
				
				-3.93\tnote{***} & -0.15\tnote{***} & -0.11\tnote{***} & -0.14\tnote{***} \\
				( 0.00) & ( 0.00) & ( 0.00) & ( 0.00) \\
				\addlinespace[3pt]
				
				23.13\tnote{***} &  0.99\tnote{***} &  0.83\tnote{***} &  0.95\tnote{***} \\
				( 0.00) & ( 0.00) & ( 0.00) & ( 0.00) \\
				\addlinespace[3pt]
				
				\multicolumn{1}{c}{176} & \multicolumn{1}{c}{176} & \multicolumn{1}{c}{176} & \multicolumn{1}{c}{176} \\
				\addlinespace[3pt]
				16.06\% & 20.39\% & 11.98\% & 17.05\% \\

				\midrule
				-0.53 & -0.01 &  0.00 & -0.01\tnote{{\textcolor{white}{*}}} \\
				( 0.17) & ( 0.38) & ( 0.45) & ( 0.34) \\
				\addlinespace[3pt]
				
				-3.16\tnote{***} & -0.13\tnote{***} & -0.10\tnote{***} & -0.12\tnote{***} \\
				( 0.00) & ( 0.00) & ( 0.00) & ( 0.00) \\
				\addlinespace[3pt]
				
				21.40\tnote{***} &  0.99\tnote{***} &  0.81\tnote{***} &  0.92\tnote{***} \\
				( 0.00) & ( 0.00) & ( 0.00) & ( 0.00) \\
				\addlinespace[3pt]
				
				\multicolumn{1}{c}{173} & \multicolumn{1}{c}{173} & \multicolumn{1}{c}{173} & \multicolumn{1}{c}{173} \\
				\addlinespace[3pt]
				10.94\% & 13.67\% & 9.17\% & 11.15\% \\

				\midrule
				-24.33\tnote{***} & -0.81\tnote{***} & -0.46\tnote{*} & -0.81\tnote{***} \\
				(  0.01) & ( 0.01) & ( 0.08) & ( 0.01) \\
				\addlinespace[3pt]
				
				-3.97\tnote{***} & -0.14\tnote{***} & -0.10\tnote{***} & -0.13\tnote{***} \\
				(  0.00) & ( 0.00) & ( 0.00) & ( 0.00) \\
				\addlinespace[3pt]
				
				22.75\tnote{***} &  0.98\tnote{***} &  0.81\tnote{***} &  0.93\tnote{***} \\
				(  0.00) & ( 0.00) & ( 0.00) & ( 0.00) \\
				\addlinespace[3pt]
				
				\multicolumn{1}{c}{174} & \multicolumn{1}{c}{174} & \multicolumn{1}{c}{174} & \multicolumn{1}{c}{174} \\
				\addlinespace[3pt]
				14.01\% & 16.57\% & 10.27\% & 14.29\% \\

			\end{tabular} &
			
			\begin{tabular}[t]
				{
					D{.}{.}{-1}
					D{.}{.}{-1}
					D{.}{.}{-1}
					D{.}{.}{-1}
				}

				\multicolumn{4}{c}{\textbf{Panel B. Death Cases}}  \\
				
				\midrule
				\multicolumn{1}{c}{\textbf{Chi-squared}} & \multicolumn{1}{c}{\textbf{Kuiper}} & \multicolumn{1}{c}{\textbf{M}} & \multicolumn{1}{c}{\textbf{D}} \\

				\midrule

				-33.61\tnote{***} & -1.25\tnote{***} & -1.09\tnote{***} & -1.36\tnote{***} \\
				(  0.01) & ( 0.00) & ( 0.00) & ( 0.00) \\
				\addlinespace[3pt]
				
				-6.39\tnote{***} & -0.24\tnote{***} & -0.24\tnote{***} & -0.26\tnote{***} \\
				(  0.00) & ( 0.00) & ( 0.00) & ( 0.00) \\
				\addlinespace[3pt]
				
				37.45\tnote{***} &  1.11\tnote{***} &  0.69\tnote{**} &  0.96\tnote{***} \\
				(  0.00) & ( 0.00) & ( 0.01) & ( 0.00) \\
				\addlinespace[3pt]
				
				\multicolumn{1}{c}{148} & \multicolumn{1}{c}{148} & \multicolumn{1}{c}{148} & \multicolumn{1}{c}{148} \\
				\addlinespace[3pt]
				11.63\% & 18.08\% & 15.43\% & 19.00\% \\

				\midrule
				-6.53\tnote{***} & -0.25\tnote{***} & -0.22\tnote{***} & -0.27\tnote{***} \\
				( 0.00) & ( 0.00) & ( 0.00) & ( 0.00) \\
				\addlinespace[3pt]
				
				-5.71\tnote{***} & -0.22\tnote{***} & -0.19\tnote{***} & -0.21\tnote{***} \\
				( 0.00) & ( 0.00) & ( 0.00) & ( 0.00) \\
				\addlinespace[3pt]
				
				36.00\tnote{***} &  1.07\tnote{***} &  0.66\tnote{**} &  0.92\tnote{***} \\
				( 0.00) & ( 0.00) & ( 0.01) & ( 0.00) \\
				\addlinespace[3pt]
				
				\multicolumn{1}{c}{155} & \multicolumn{1}{c}{155} & \multicolumn{1}{c}{155} & \multicolumn{1}{c}{155} \\
				\addlinespace[3pt]
				13.89\% & 22.99\% & 16.97\% & 22.21\% \\

				\midrule
				-3.21\tnote{***} & -0.11\tnote{***} & -0.09\tnote{***} & -0.11\tnote{***} \\
				( 0.00) & ( 0.00) & ( 0.00) & ( 0.00) \\
				\addlinespace[3pt]
				
				-4.47\tnote{***} & -0.17\tnote{***} & -0.15\tnote{***} & -0.17\tnote{***} \\
				( 0.01) & ( 0.00) & ( 0.00) & ( 0.00) \\
				\addlinespace[3pt]
				
				31.12\tnote{***} &  1.00\tnote{***} &  0.61\tnote{**} &  0.83\tnote{***} \\
				( 0.00) & ( 0.00) & ( 0.03) & ( 0.01) \\
				\addlinespace[3pt]
				
				\multicolumn{1}{c}{151} & \multicolumn{1}{c}{151} & \multicolumn{1}{c}{151} & \multicolumn{1}{c}{151} \\
				\addlinespace[3pt]
				11.24\% & 16.50\% & 11.56\% & 15.33\% \\

				\midrule
				-59.61\tnote{***} & -2.09\tnote{***} & -1.87\tnote{***} & -2.33\tnote{***} \\
				(  0.00) & ( 0.00) & ( 0.00) & ( 0.00) \\
				\addlinespace[3pt]
				
				-5.49\tnote{***} & -0.21\tnote{***} & -0.19\tnote{***} & -0.21\tnote{***} \\
				(  0.00) & ( 0.00) & ( 0.00) & ( 0.00) \\
				\addlinespace[3pt]
				
				38.18\tnote{***} &  1.19\tnote{***} &  0.77\tnote{***} &  1.04\tnote{***} \\
				(  0.00) & ( 0.00) & ( 0.01) & ( 0.00) \\
				\addlinespace[3pt]
				
				\multicolumn{1}{c}{155} & \multicolumn{1}{c}{155} & \multicolumn{1}{c}{155} & \multicolumn{1}{c}{155} \\
				\addlinespace[3pt]
				13.73\% & 20.66\% & 16.63\% & 21.52\% \\

			\end{tabular} \\
			
			\midrule[0.3pt]\bottomrule[1.5pt]
		\end{tabular*}

		\begin{tablenotes}
			\singlespacing
			\item[] The table presents the main results using COVID-19 pandemic data. We estimate equation 6 using OLS for first-digit goodness-of-fit measures. Panel A shows the results for the cumulative number of confirmed cases, while panel B shows the results for the cumulative number of deaths. To avoid small coefficients, we divide EIU, UHC, and No. of Days values by 100 for all models. Sample sizes vary due to missing values. \textit{P}-values for a one-tailed \textit{t}-test are in parentheses. ***, ** and * denote significance at the 1\%, 5\% and 10\% levels, respectively. All variable definitions are in Appendix A2. 
		\end{tablenotes}
		
	\end{threeparttable}
\end{sidewaystable}

\par
}

To disentangle the political and other components of the \emph{Economist
Intelligence Unit} Democracy Index, we then analyze the five components
of the index separately: electoral pluralism ($ELECT$), functioning
of the government ($GVMT$), political participation ($PART$), political
culture ($CULT$), and civil liberties ($LIBERT$). The results are
presented in Table 5. We find that the results for the overall $EIU$
index are driven by its three components: electoral pluralism, functioning
of the government, and civil liberties (the significance of these
coefficients coincides in each case with the significance of the overall
index), but not political participation or political culture (these
coefficients are never significant). We also substitute the $EIU$
measure with the “thin” definition of the democracy, i.e., the \emph{Freedom
House} Electoral Democracy Index. We use the reported by \emph{Freedom
House} dummy variable for electoral democracy ($FH\text{\_}DEM$),
as well as the sum of their measures of political freedom and civil
liberties ($FH\text{\_}AV$).\footnote{Available at https://freedomhouse.org/report/freedom-world.}
Again, for the cumulative number of cases, the coefficients are negative
but lack significance. For the cumulative number of deaths, the coefficients
are negative and significant. The results show that the findings are
not driven by the choice of the democracy measure or by the spurious
transparency component in the index.

{\centering
\begin{sidewaystable}
	\centering
	
	\begin{threeparttable}[h]
		
		\def\sym#1{\ifmmode^{#1}\else\(^{#1}\)\fi}
		
		\caption{Main Results: Using Alternative Democracy Measures}\label{tab_res:goodness_eiu1} 
		\tiny
		\renewcommand{\arraystretch}{0.25}
		\begin{tabular*}{0.9\linewidth}{@{\extracolsep{\fill}}l|l|l}
			\toprule[1.5pt]\midrule[0.3pt]
			
			\begin{tabular}[t]{ll}
				
				& \textcolor{white}{1}\\
				\arrayrulecolor{white}\midrule
				\multicolumn{2}{c}{\textbf{Variable}} \\
				
				\arrayrulecolor{black}\midrule

				\textbf{ELECT}\tnote{{\textcolor{white}{*}}} \\
				\textcolor{white}{()} \\
				\addlinespace[4pt]
				
				\textbf{ln(Population)}\tnote{{\textcolor{white}{*}}} \\
				\textcolor{white}{()} \\
				\addlinespace[3pt]
				
				\textbf{No. of Days}\tnote{{\textcolor{white}{*}}}  \\
				\textcolor{white}{()} \\
				\addlinespace[2.5pt]
				
				\textbf{Sample Size} \\
				\addlinespace[3.5pt]
				
				\textbf{Adj. R\textsuperscript{2}} \\
				\midrule

				\textbf{GVMT}\tnote{{\textcolor{white}{*}}} \\
				\textcolor{white}{()} \\
				\addlinespace[4pt]
				
				\textbf{ln(Population)}\tnote{{\textcolor{white}{*}}} \\
				\textcolor{white}{()} \\
				\addlinespace[3pt]
				
				\textbf{No. of Days}\tnote{{\textcolor{white}{*}}}  \\
				\textcolor{white}{()} \\
				\addlinespace[2.5pt]
				
				\textbf{Sample Size} \\
				\addlinespace[3.5pt]
				
				\textbf{Adj. R\textsuperscript{2}} \\
				\midrule

				\textbf{PART}\tnote{{\textcolor{white}{*}}} \\
				\textcolor{white}{()} \\
				\addlinespace[4pt]
				
				\textbf{ln(Population)}\tnote{{\textcolor{white}{*}}} \\
				\textcolor{white}{()} \\
				\addlinespace[3pt]
				
				\textbf{No. of Days}\tnote{{\textcolor{white}{*}}}  \\
				\textcolor{white}{()} \\
				\addlinespace[2.5pt]
				
				\textbf{Sample Size} \\
				\addlinespace[3.5pt]
				
				\textbf{Adj. R\textsuperscript{2}} \\
				\midrule

				\textbf{CULT}\tnote{{\textcolor{white}{*}}} \\
				\textcolor{white}{()} \\
				\addlinespace[4pt]
				
				\textbf{ln(Population)}\tnote{{\textcolor{white}{*}}} \\
				\textcolor{white}{()} \\
				\addlinespace[3pt]
				
				\textbf{No. of Days}\tnote{{\textcolor{white}{*}}}  \\
				\textcolor{white}{()} \\
				\addlinespace[2.5pt]
				
				\textbf{Sample Size} \\
				\addlinespace[3.5pt]
				
				\textbf{Adj. R\textsuperscript{2}} \\

				\midrule
				
				\textbf{LIBERT}\tnote{{\textcolor{white}{*}}} \\
				\textcolor{white}{()} \\
				\addlinespace[4pt]
				
				\textbf{ln(Population)}\tnote{{\textcolor{white}{*}}} \\
				\textcolor{white}{()} \\
				\addlinespace[3pt]
				
				\textbf{No. of Days}\tnote{{\textcolor{white}{*}}}  \\
				\textcolor{white}{()} \\
				\addlinespace[2.5pt]
				
				\textbf{Sample Size} \\
				\addlinespace[3.5pt]
				
				\textbf{Adj. R\textsuperscript{2}} \\

			\end{tabular} &

			\begin{tabular}[t]
				{
					D{.}{.}{-1}
					D{.}{.}{-1}
					D{.}{.}{-1}
					D{.}{.}{-1}
				}

				\multicolumn{4}{c}{\textbf{Panel A. Confirmed Cases}}  \\
				
				\midrule
				\multicolumn{1}{c}{\textbf{Chi-squared}} & \multicolumn{1}{c}{\textbf{Kuiper}} & \multicolumn{1}{c}{\textbf{M}} & \multicolumn{1}{c}{\textbf{D}} \\

				\midrule
				
				-5.50\tnote{*} & -0.11 & -0.05 & -0.10\tnote{{\textcolor{white}{*}}} \\
				( 0.09) & ( 0.23) & ( 0.37) & ( 0.26) \\
				\addlinespace[4pt]
				
				-3.37\tnote{***} & -0.12\tnote{***} & -0.08\tnote{**} & -0.11\tnote{***} \\
				( 0.00) & ( 0.00) & ( 0.02) & ( 0.00) \\
				\addlinespace[4pt]
				
				21.11\tnote{***} &  0.94\tnote{***} &  0.75\tnote{***} &  0.87\tnote{***} \\
				( 0.00) & ( 0.00) & ( 0.00) & ( 0.00) \\
				\addlinespace[4pt]
				
				\multicolumn{1}{c}{161} & \multicolumn{1}{c}{161} & \multicolumn{1}{c}{161} & \multicolumn{1}{c}{161} \\
				\addlinespace[4pt]
				9.86\% & 11.52\% & 7.19\% & 9.39\% \\

				\midrule
				-7.29 & -0.19 & -0.02 & -0.12\tnote{{\textcolor{white}{*}}} \\
				( 0.10) & ( 0.19) & ( 0.46) & ( 0.29) \\
				\addlinespace[4pt]
				
				-3.29\tnote{***} & -0.12\tnote{***} & -0.07\tnote{**} & -0.11\tnote{***} \\
				( 0.00) & ( 0.00) & ( 0.02) & ( 0.00) \\
				\addlinespace[4pt]
				
				20.95\tnote{***} &  0.93\tnote{***} &  0.75\tnote{***} &  0.87\tnote{***} \\
				( 0.00) & ( 0.00) & ( 0.00) & ( 0.00) \\
				\addlinespace[4pt]
				
				\multicolumn{1}{c}{161} & \multicolumn{1}{c}{161} & \multicolumn{1}{c}{161} & \multicolumn{1}{c}{161} \\
				\addlinespace[4pt]
				9.73\% & 11.64\% & 7.13\% & 9.33\% \\

				\midrule
				-10.86\tnote{*} & -0.13 &  0.03 & -0.16 \\
				(  0.09) & ( 0.33) & ( 0.46) & ( 0.30) \\
				\addlinespace[4pt]
				
				-3.20\tnote{***} & -0.11\tnote{***} & -0.07\tnote{**} & -0.10\tnote{***} \\
				(  0.00) & ( 0.00) & ( 0.02) & ( 0.00) \\
				\addlinespace[4pt]
				
				20.94\tnote{***} &  0.94\tnote{***} &  0.75\tnote{***} &  0.87\tnote{***} \\
				(  0.00) & ( 0.00) & ( 0.00) & ( 0.00) \\
				\addlinespace[4pt]
				
				\multicolumn{1}{c}{161} & \multicolumn{1}{c}{161} & \multicolumn{1}{c}{161} & \multicolumn{1}{c}{161} \\
				\addlinespace[4pt]
				9.91\% & 11.32\% & 7.13\% & 9.31\% \\

				\midrule
				-8.52 & -0.17 &  0.09 & -0.10\tnote{{\textcolor{white}{*}}} \\
				( 0.18) & ( 0.30) & ( 0.39) & ( 0.39) \\
				\addlinespace[4pt]
				
				-3.31\tnote{***} & -0.12\tnote{***} & -0.07\tnote{**} & -0.11\tnote{***} \\
				( 0.00) & ( 0.00) & ( 0.02) & ( 0.00) \\
				\addlinespace[4pt]
				
				20.32\tnote{***} &  0.92\tnote{***} &  0.76\tnote{***} &  0.86\tnote{***} \\
				( 0.00) & ( 0.00) & ( 0.00) & ( 0.00) \\
				\addlinespace[4pt]
				
				\multicolumn{1}{c}{161} & \multicolumn{1}{c}{161} & \multicolumn{1}{c}{161} & \multicolumn{1}{c}{161} \\
				\addlinespace[4pt]
				9.32\% & 11.36\% & 7.17\% & 9.20\% \\

				\midrule
				-8.60\tnote{*} & -0.18 & -0.06 & -0.17 \\
				( 0.06) & ( 0.19) & ( 0.39) & ( 0.22) \\
				\addlinespace[4pt]
				
				-3.49\tnote{***} & -0.12\tnote{***} & -0.08\tnote{**} & -0.11\tnote{***} \\
				( 0.00) & ( 0.00) & ( 0.02) & ( 0.00) \\
				\addlinespace[4pt]
				
				21.21\tnote{***} &  0.94\tnote{***} &  0.75\tnote{***} &  0.87\tnote{***} \\
				( 0.00) & ( 0.00) & ( 0.00) & ( 0.00) \\
				\addlinespace[4pt]
				
				\multicolumn{1}{c}{161} & \multicolumn{1}{c}{161} & \multicolumn{1}{c}{161} & \multicolumn{1}{c}{161} \\
				\addlinespace[4pt]
				10.15\% & 11.64\% & 7.17\% & 9.49\% \\

			\end{tabular} &
			
			\begin{tabular}[t]
				{
					D{.}{.}{-1}
					D{.}{.}{-1}
					D{.}{.}{-1}
					D{.}{.}{-1}
				}

				\multicolumn{4}{c}{\textbf{Panel B. Death Cases}}  \\
				
				\midrule
				\multicolumn{1}{c}{\textbf{Chi-squared}} & \multicolumn{1}{c}{\textbf{Kuiper}} & \multicolumn{1}{c}{\textbf{M}} & \multicolumn{1}{c}{\textbf{D}} \\

				\midrule

				-24.76\tnote{***} & -0.85\tnote{***} & -0.76\tnote{***} & -0.91\tnote{***} \\
				(  0.00) & ( 0.00) & ( 0.00) & ( 0.00) \\
				\addlinespace[4pt]
				
				-6.41\tnote{***} & -0.24\tnote{***} & -0.24\tnote{***} & -0.26\tnote{***} \\
				(  0.00) & ( 0.00) & ( 0.00) & ( 0.00) \\
				\addlinespace[4pt]
				
				37.95\tnote{***} &  1.15\tnote{***} &  0.72\tnote{**} &  0.99\tnote{***} \\
				(  0.00) & ( 0.00) & ( 0.01) & ( 0.00) \\
				\addlinespace[4pt]
				
				\multicolumn{1}{c}{147} & \multicolumn{1}{c}{147} & \multicolumn{1}{c}{147} & \multicolumn{1}{c}{147} \\
				\addlinespace[4pt]
				13.15\% & 19.28\% & 16.93\% & 20.30\% \\

				\midrule
				-29.06\tnote{***} & -1.14\tnote{***} & -0.97\tnote{***} & -1.22\tnote{***} \\
				(  0.01) & ( 0.00) & ( 0.00) & ( 0.00) \\
				\addlinespace[4pt]
				
				-6.15\tnote{***} & -0.23\tnote{***} & -0.23\tnote{***} & -0.25\tnote{***} \\
				(  0.00) & ( 0.00) & ( 0.00) & ( 0.00) \\
				\addlinespace[4pt]
				
				37.17\tnote{***} &  1.10\tnote{***} &  0.68\tnote{**} &  0.94\tnote{***} \\
				(  0.00) & ( 0.00) & ( 0.01) & ( 0.00) \\
				\addlinespace[4pt]
				
				\multicolumn{1}{c}{147} & \multicolumn{1}{c}{147} & \multicolumn{1}{c}{147} & \multicolumn{1}{c}{147} \\
				\addlinespace[4pt]
				11.76\% & 19.07\% & 15.92\% & 19.92\% \\

				\midrule
				-23.67\tnote{*} & -0.83\tnote{**} & -0.77\tnote{**} & -0.96\tnote{**} \\
				(  0.08) & ( 0.04) & ( 0.04) & ( 0.02) \\
				\addlinespace[4pt]
				
				-6.06\tnote{***} & -0.23\tnote{***} & -0.22\tnote{***} & -0.24\tnote{***} \\
				(  0.00) & ( 0.00) & ( 0.00) & ( 0.00) \\
				\addlinespace[4pt]
				
				39.47\tnote{***} &  1.20\tnote{***} &  0.76\tnote{***} &  1.04\tnote{***} \\
				(  0.00) & ( 0.00) & ( 0.01) & ( 0.00) \\
				\addlinespace[4pt]
				
				\multicolumn{1}{c}{147} & \multicolumn{1}{c}{147} & \multicolumn{1}{c}{147} & \multicolumn{1}{c}{147} \\
				\addlinespace[4pt]
				9.15\% & 13.62\% & 11.76\% & 14.14\% \\

				\midrule
				-3.31 & -0.49 & -0.41 & -0.59\tnote{{\textcolor{white}{*}}} \\
				( 0.43) & ( 0.17) & ( 0.21) & ( 0.13) \\
				\addlinespace[4pt]
				
				-6.23\tnote{***} & -0.24\tnote{***} & -0.23\tnote{***} & -0.25\tnote{***} \\
				( 0.00) & ( 0.00) & ( 0.00) & ( 0.00) \\
				\addlinespace[4pt]
				
				40.82\tnote{***} &  1.19\tnote{***} &  0.76\tnote{**} &  1.02\tnote{***} \\
				( 0.00) & ( 0.00) & ( 0.01) & ( 0.00) \\
				\addlinespace[4pt]
				
				\multicolumn{1}{c}{147} & \multicolumn{1}{c}{147} & \multicolumn{1}{c}{147} & \multicolumn{1}{c}{147} \\
				\addlinespace[4pt]
				7.89\% & 12.25\% & 10.28\% & 12.37\% \\

				\midrule
				-32.27\tnote{***} & -1.19\tnote{***} & -1.01\tnote{***} & -1.26\tnote{***} \\
				(  0.00) & ( 0.00) & ( 0.00) & ( 0.00) \\
				\addlinespace[4pt]
				
				-7.03\tnote{***} & -0.26\tnote{***} & -0.25\tnote{***} & -0.28\tnote{***} \\
				(  0.00) & ( 0.00) & ( 0.00) & ( 0.00) \\
				\addlinespace[4pt]
				
				38.70\tnote{***} &  1.17\tnote{***} &  0.74\tnote{***} &  1.01\tnote{***} \\
				(  0.00) & ( 0.00) & ( 0.01) & ( 0.00) \\
				\addlinespace[4pt]
				
				\multicolumn{1}{c}{147} & \multicolumn{1}{c}{147} & \multicolumn{1}{c}{147} & \multicolumn{1}{c}{147} \\
				\addlinespace[4pt]
				12.58\% & 19.55\% & 16.34\% & 20.28\% \\

			\end{tabular} \\
			
			\midrule[0.3pt]\bottomrule[1.5pt]
		\end{tabular*}

		\begin{tablenotes}
			\singlespacing
			\item[] 
		\end{tablenotes}
		
	\end{threeparttable}
\end{sidewaystable}

\begin{sidewaystable}
	\centering
	
	\begin{threeparttable}[h]
		
		\def\sym#1{\ifmmode^{#1}\else\(^{#1}\)\fi}
		
		\tiny
		\renewcommand{\arraystretch}{0.25}
		\begin{tabular*}{0.9\linewidth}{@{\extracolsep{\fill}}l|l|l}
			\toprule[1.5pt]\midrule[0.3pt]
			
			\begin{tabular}[t]{ll}
				
				& \textcolor{white}{1}\\
				\arrayrulecolor{white}\midrule
				\multicolumn{2}{c}{\textbf{Variable}} \\
				
				\arrayrulecolor{black}\midrule

				\textbf{FH\_DEM}\tnote{{\textcolor{white}{*}}} \\
				\textcolor{white}{()} \\
				\addlinespace[4pt]
				
				\textbf{ln(Population)}\tnote{{\textcolor{white}{*}}} \\
				\textcolor{white}{()} \\
				\addlinespace[3pt]
				
				\textbf{No. of Days}\tnote{{\textcolor{white}{*}}}  \\
				\textcolor{white}{()} \\
				\addlinespace[2.5pt]
				
				\textbf{Sample Size} \\
				\addlinespace[3.5pt]
				
				\textbf{Adj. R\textsuperscript{2}} \\
				\midrule

				\textbf{FH\_AV}\tnote{{\textcolor{white}{*}}} \\
				\textcolor{white}{()} \\
				\addlinespace[4pt]
				
				\textbf{ln(Population)}\tnote{{\textcolor{white}{*}}} \\
				\textcolor{white}{()} \\
				\addlinespace[3pt]
				
				\textbf{No. of Days}\tnote{{\textcolor{white}{*}}}  \\
				\textcolor{white}{()} \\
				\addlinespace[2.5pt]
				
				\textbf{Sample Size} \\
				\addlinespace[3.5pt]
				
				\textbf{Adj. R\textsuperscript{2}} \\

			\end{tabular} &

			\begin{tabular}[t]
				{
					D{.}{.}{-1}
					D{.}{.}{-1}
					D{.}{.}{-1}
					D{.}{.}{-1}
				}

				\multicolumn{4}{c}{\textbf{Panel A. Confirmed Cases}}  \\
				
				\midrule
				\multicolumn{1}{c}{\textbf{Chi-squared}} & \multicolumn{1}{c}{\textbf{Kuiper}} & \multicolumn{1}{c}{\textbf{M}} & \multicolumn{1}{c}{\textbf{D}} \\
				\midrule

				-2.51 & -0.04 &  0.01 & -0.04\tnote{{\textcolor{white}{*}}} \\
				(   0.20) & ( 0.34) & ( 0.46) & ( 0.37) \\
				\addlinespace[4pt]
				
				-3.28\tnote{***} & -0.12\tnote{***} & -0.09\tnote{***} & -0.11\tnote{***} \\
				(   0.00) & ( 0.00) & ( 0.00) & ( 0.00) \\
				\addlinespace[4pt]
				
				23.39\tnote{***} &  1.00\tnote{***} &  0.82\tnote{***} &  0.96\tnote{***} \\
				(   0.00) & ( 0.00) & ( 0.00) & ( 0.00) \\
				\addlinespace[4pt]
				
				\multicolumn{1}{c}{181} & \multicolumn{1}{c}{181} & \multicolumn{1}{c}{181} & \multicolumn{1}{c}{181} \\
				\addlinespace[4pt]
				9.54\% & 12.95\% & 8.96\% & 10.92\% \\

				\midrule
				-7.75\tnote{*} & -0.21 & -0.13 & -0.18 \\
				( 0.06) & ( 0.12) & ( 0.23) & ( 0.17) \\
				\addlinespace[4pt]
				
				-3.42\tnote{***} & -0.13\tnote{***} & -0.10\tnote{***} & -0.12\tnote{***} \\
				( 0.00) & ( 0.00) & ( 0.00) & ( 0.00) \\
				\addlinespace[4pt]
				
				22.84\tnote{***} &  0.99\tnote{***} &  0.81\tnote{***} &  0.94\tnote{***} \\
				( 0.00) & ( 0.00) & ( 0.00) & ( 0.00) \\
				\addlinespace[4pt]
				
				\multicolumn{1}{c}{181} & \multicolumn{1}{c}{181} & \multicolumn{1}{c}{181} & \multicolumn{1}{c}{181} \\
				\addlinespace[4pt]
				10.39\% & 13.52\% & 9.24\% & 11.34\% \\

			\end{tabular} &
			
			\begin{tabular}[t]
				{
					D{.}{.}{-1}
					D{.}{.}{-1}
					D{.}{.}{-1}
					D{.}{.}{-1}
				}

				\multicolumn{4}{c}{\textbf{Panel B. Death Cases}}  \\
				
				\midrule
				\multicolumn{1}{c}{\textbf{Chi-squared}} & \multicolumn{1}{c}{\textbf{Kuiper}} & \multicolumn{1}{c}{\textbf{M}} & \multicolumn{1}{c}{\textbf{D}} \\
				\midrule

				-12.55\tnote{**} & -0.54\tnote{***} & -0.49\tnote{***} & -0.58\tnote{***} \\
				(    0.02) & (  0.00) & (  0.00) & (  0.00) \\
				\addlinespace[4pt]
				
				-5.35\tnote{***} &  -0.21\tnote{***} &  -0.18\tnote{***} &  -0.21\tnote{***} \\
				(    0.00) & (  0.00) & (  0.00) & (  0.00) \\
				\addlinespace[4pt]
				
				40.22\tnote{***} &   1.26\tnote{***} &   0.83\tnote{***} &   1.11\tnote{***} \\
				(    0.00) & (  0.00) & (  0.00) & (  0.00) \\
				\addlinespace[4pt]
				
				\multicolumn{1}{c}{157} & \multicolumn{1}{c}{157} & \multicolumn{1}{c}{157} & \multicolumn{1}{c}{157} \\
				\addlinespace[4pt]
				9.84\% & 16.84\% & 13.09\% & 16.39\% \\

				\midrule
				-30.42\tnote{***} & -1.09\tnote{***} & -0.94\tnote{***} & -1.17\tnote{***} \\
				(  0.00) & ( 0.00) & ( 0.00) & ( 0.00) \\
				\addlinespace[4pt]
				
				-5.69\tnote{***} & -0.21\tnote{***} & -0.19\tnote{***} & -0.21\tnote{***} \\
				(  0.00) & ( 0.00) & ( 0.00) & ( 0.00) \\
				\addlinespace[4pt]
				
				35.75\tnote{***} &  1.11\tnote{***} &  0.70\tnote{**} &  0.95\tnote{***} \\
				(  0.00) & ( 0.00) & ( 0.01) & ( 0.00) \\
				\addlinespace[4pt]
				
				\multicolumn{1}{c}{157} & \multicolumn{1}{c}{157} & \multicolumn{1}{c}{157} & \multicolumn{1}{c}{157} \\
				\addlinespace[4pt]
				12.54\% & 19.30\% & 14.49\% & 18.97\% \\

			\end{tabular} \\
			
			\midrule[0.3pt]\bottomrule[1.5pt]
		\end{tabular*}

		\begin{tablenotes}
			\singlespacing
			\item[] The table presents results using alternative democracy measures. We estimate equation 6 using OLS for first-digit goodness-of-fit measures. Panel A shows the results for the cumulative number of confirmed cases, while panel B shows the results for the cumulative number of deaths. To avoid small coefficients, we divide ELEC, GVMT, PART, EULT, LIBERT, FH\_AV, and No. of Days values by 100 for all models. Sample sizes vary due to missing values. \textit{P}-values for a one-tailed \textit{t}-test are in parentheses. ***, ** and * denote significance at the 1\%, 5\% and 10\% levels, respectively. All variable definitions are in Appendix A2. 
		\end{tablenotes}
		
	\end{threeparttable}
\end{sidewaystable}

\par
}

We interpret the data as being consistent with the argument that more
democratic and more highly developed countries are less likely to
deviate from the NBL when reporting pandemic data. Specifically, countries
with higher GDP per capita and universal health coverage are less
likely to manipulate their data during COVID-19. For the democracy
index ($EIU$) and health expenditures as percentage of GDP ($HE\text{\_}GDP$),
we find convincing evidence only for the number of deaths. We conclude
that the relationship is more pronounced for the total number of deaths
than for the number of confirmed cases. As predicted, the control
variable $\text{ln}(Population)_{i}$ is negative and significant
in all regressions: countries with higher populations (and total number
of cases) deviate less from the NBL. The results are also economically
significant: an increase of one standard deviation in the economic
indicators, on average, results in a 0.25 increase of the standard
deviation in the goodness-of-fit measures. This value is roughly the
same for the number of confirmed cases and for the number of deaths,
across all economic indicators.

Instead of looking at the linear relationship between the goodness-of-fit
measures, one could look at the probability of a country’s data to
deviate from the NBL. To do that, we first identify the critical values
(at the 1\% significance level) for each goodness-of-fit measure and
create a set of four dummy variables, $Chi-sq.$, $Kuiper$, $M$,
and $D$, where each dummy variable equals one if the corresponding
goodness-of-fit measure is above the critical value (i.e., we reject
the null hypothesis that the data obeys the NBL), and zero otherwise.
We then estimate Equation \ref{eq:1} using the logit model. Again,
most coefficients for the economic indicators are negative, and all
coefficients for ln($GDP$) and $UHC$ are significant. For the death
toll, all coefficients for all economic indicators are negative and
significant. For brevity, we omit the table with the results. We again
conclude that more developed countries and countries with better health
systems follow the NBL more closely, and the relationship is more
pronounced for the number of reported deaths than for the number of
confirmed cases. We also assert that the findings are not driven by
the choice of the model.

Finally, we conduct the same tests for the cumulative number of cured
cases and the number of COVID-19 tests conducted. On average, we cannot
reject the null hypothesis that countries manipulated data on cured
cases or the number of conducted tests. The regression results are
also not significant, indicating systematic cross-sectional difference
between countries. We conclude that countries are most prone to falsify
mortality data, slightly less so the number of confirmed cases, and
that there is no evidence of systematic data falsification of cured
cases or the number of tests. The cross-sectional difference between
countries is also the strongest for the death toll, weaker for the
total number of confirmed cases, and is insignificant for the number
of cured cases and tests.

\subsection{Robustness Analyses}

One limitation of our analysis above is that it depends on the cutoff
date for the growth part of the data. The cutoff date is estimated
using the data, the validity of which we are assessing. This creates
a possible endogeneity problem. To resolve this issue, we use several
approaches. First, instead of using a specific cutoff date for each
country, we use the same, “global,” cutoff date for all countries,
which is 80 days after January 22, 2020 (or April 11, 2020). We pick
80 days because it corresponds to the second tercile of cutoff dates
in our sample. Fewer dates will result in too small sample sizes for
many countries, especially those that were affected by the pandemic
later than others. For longer periods, too many countries will have
already reached their plateaus, and are thus no longer expected to
obey the NBL. We then calculate our four goodness-of-fit measures
using the global cutoff date and re-estimate Equation \ref{eq:1}.
The results are reported in Table 6. Panel A has the data for the
confirmed cases. All coefficients for all four economic indicators
are negative and significant in all regressions. Panel B has the data
for the death count. Again, all coefficients in all regression are
negative and significant, confirming our earlier findings.
 
{\centering
\begin{sidewaystable}
	
	\centering
	\begin{threeparttable}[h]
		\def\sym#1{\ifmmode^{#1}\else\(^{#1}\)\fi}
		
		\caption{Robustness Results: Global Cutoff}\label{tab_res:robust_glob80} 
		\scriptsize
		\renewcommand{\arraystretch}{0.4}
		\begin{tabular*}{0.9\linewidth}{@{\extracolsep{\fill}}l|l|l}
			\toprule[1.5pt]\midrule[0.3pt]
			
			\begin{tabular}[t]{ll}
				
				& \textcolor{white}{1}\\
				\arrayrulecolor{white}\midrule
				\multicolumn{2}{c}{\textbf{Variable}} \\
				
				\arrayrulecolor{black}\midrule

				\textbf{EIU}\tnote{{\textcolor{white}{*}}} \\
				\textcolor{white}{()} \\
				\addlinespace[3pt]
				
				\textbf{ln(Population)}\tnote{{\textcolor{white}{*}}} \\
				\textcolor{white}{()} \\
				\addlinespace[2pt]
				
				\textbf{No. of Days}\tnote{{\textcolor{white}{*}}}  \\
				\textcolor{white}{()} \\
				\addlinespace[1.8pt]
				
				\textbf{Sample Size} \\
				\addlinespace[2.8pt]
				
				\textbf{Adj. R\textsuperscript{2}} \\
				\midrule

				\textbf{ln(GDP)}\tnote{{\textcolor{white}{*}}} \\
				\textcolor{white}{()} \\
				\addlinespace[2pt]
				
				\textbf{ln(Population)}\tnote{{\textcolor{white}{*}}} \\
				\textcolor{white}{()} \\
				\addlinespace[1.8pt]
				
				\textbf{No. of Days}\tnote{{\textcolor{white}{*}}}  \\
				\textcolor{white}{()} \\
				\addlinespace[1.8pt]
				
				\textbf{Sample Size} \\
				\addlinespace[2.8pt]
				
				\textbf{Adj. R\textsuperscript{2}} \\
				\midrule

				\textbf{HE\_GDP}\tnote{{\textcolor{white}{*}}} \\
				\textcolor{white}{()} \\
				\addlinespace[2pt]
				
				\textbf{ln(Population)}\tnote{{\textcolor{white}{*}}} \\
				\textcolor{white}{()} \\
				\addlinespace[1.8pt]
				
				\textbf{No. of Days}\tnote{{\textcolor{white}{*}}}  \\
				\textcolor{white}{()} \\
				\addlinespace[1.8pt]
				
				\textbf{Sample Size} \\
				\addlinespace[2.8pt]
				
				\textbf{Adj. R\textsuperscript{2}} \\
				\midrule

				\textbf{UHC}\tnote{{\textcolor{white}{*}}} \\
				\textcolor{white}{()} \\
				\addlinespace[3pt]
				
				\textbf{ln(Population)}\tnote{{\textcolor{white}{*}}} \\
				\textcolor{white}{()} \\
				\addlinespace[1.8pt]
				
				\textbf{No. of Days}\tnote{{\textcolor{white}{*}}}  \\
				\textcolor{white}{()} \\
				\addlinespace[1.8pt]
				
				\textbf{Sample Size} \\
				\addlinespace[2.8pt]
				
				\textbf{Adj. R\textsuperscript{2}} \\

			\end{tabular} &

			\begin{tabular}[t]
				{
					D{.}{.}{-1}
					D{.}{.}{-1}
					D{.}{.}{-1}
					D{.}{.}{-1}
				}

				\multicolumn{4}{c}{\textbf{Panel A. Confirmed Cases}}  \\
				
				\midrule
				\multicolumn{1}{c}{\textbf{Chi-squared}} & \multicolumn{1}{c}{\textbf{Kuiper}} & \multicolumn{1}{c}{\textbf{M}} & \multicolumn{1}{c}{\textbf{D}} \\

				\midrule
				
				-42.42\tnote{***} & -0.93\tnote{***} & -0.83\tnote{***} & -0.98\tnote{***} \\
				(  0.00) & ( 0.00) & ( 0.00) & ( 0.00) \\
				\addlinespace[3pt]
				
				1.98 & -0.02 & -0.01 & -0.02\tnote{{\textcolor{white}{*}}} \\
				(  0.17) & ( 0.34) & ( 0.44) & ( 0.35) \\
				\addlinespace[3pt]
				
				67.53\tnote{***} &  1.77\tnote{***} &  1.80\tnote{***} &  1.91\tnote{***} \\
				(  0.00) & ( 0.00) & ( 0.00) & ( 0.00) \\
				\addlinespace[3pt]
				
				\multicolumn{1}{c}{159} & \multicolumn{1}{c}{159} & \multicolumn{1}{c}{159} & \multicolumn{1}{c}{159} \\
				\addlinespace[3pt]
				12.56\% & 11.85\% & 13.14\% & 13.73\% \\

				\midrule
				-8.73\tnote{***} & -0.32\tnote{***} & -0.31\tnote{***} & -0.35\tnote{***} \\
				( 0.00) & ( 0.00) & ( 0.00) & ( 0.00) \\
				\addlinespace[3pt]
				
				-2.22 & -0.17\tnote{***} & -0.16\tnote{***} & -0.18\tnote{***} \\
				( 0.10) & ( 0.00) & ( 0.00) & ( 0.00) \\
				\addlinespace[3pt]
				
				94.43\tnote{***} &  3.13\tnote{***} &  3.17\tnote{***} &  3.40\tnote{***} \\
				( 0.00) & ( 0.00) & ( 0.00) & ( 0.00) \\
				\addlinespace[3pt]
				
				\multicolumn{1}{c}{173} & \multicolumn{1}{c}{173} & \multicolumn{1}{c}{173} & \multicolumn{1}{c}{173} \\
				\addlinespace[3pt]
				11.01\% & 21.77\% & 22.67\% & 25.44\% \\

				\midrule
				-2.75\tnote{***} & -0.08\tnote{***} & -0.07\tnote{***} & -0.08\tnote{***} \\
				( 0.01) & ( 0.00) & ( 0.00) & ( 0.00) \\
				\addlinespace[3pt]
				
				1.11 & -0.05\tnote{*} & -0.04 & -0.04\tnote{*} \\
				( 0.23) & ( 0.07) & ( 0.11) & ( 0.10) \\
				\addlinespace[3pt]
				
				54.66\tnote{***} &  1.59\tnote{***} &  1.71\tnote{***} &  1.73\tnote{***} \\
				( 0.00) & ( 0.00) & ( 0.00) & ( 0.00) \\
				\addlinespace[3pt]
				
				\multicolumn{1}{c}{170} & \multicolumn{1}{c}{170} & \multicolumn{1}{c}{170} & \multicolumn{1}{c}{170} \\
				\addlinespace[3pt]
				8.17\% & 10.34\% & 11.68\% & 11.43\% \\

				\midrule
				-51.09\tnote{**} & -1.95\tnote{***} & -2.05\tnote{***} & -2.26\tnote{***} \\
				(  0.01) & ( 0.00) & ( 0.00) & ( 0.00) \\
				\addlinespace[3pt]
				
				-0.23 & -0.11\tnote{***} & -0.10\tnote{***} & -0.12\tnote{***} \\
				(  0.45) & ( 0.00) & ( 0.00) & ( 0.00) \\
				\addlinespace[3pt]
				
				75.17\tnote{***} &  2.51\tnote{***} &  2.64\tnote{***} &  2.80\tnote{***} \\
				(  0.00) & ( 0.00) & ( 0.00) & ( 0.00) \\
				\addlinespace[3pt]
				
				\multicolumn{1}{c}{171} & \multicolumn{1}{c}{171} & \multicolumn{1}{c}{171} & \multicolumn{1}{c}{171} \\
				\addlinespace[3pt]
				7.84\% & 13.74\% & 16.36\% & 17.47\% \\

			\end{tabular} &
			
			\begin{tabular}[t]
				{
					D{.}{.}{-1}
					D{.}{.}{-1}
					D{.}{.}{-1}
					D{.}{.}{-1}
				}

				\multicolumn{4}{c}{\textbf{Panel B. Death Cases}}  \\
				
				\midrule
				\multicolumn{1}{c}{\textbf{Chi-squared}} & \multicolumn{1}{c}{\textbf{Kuiper}} & \multicolumn{1}{c}{\textbf{M}} & \multicolumn{1}{c}{\textbf{D}} \\

				\midrule

				-34.03\tnote{***} & -1.63\tnote{***} & -1.40\tnote{***} & -1.54\tnote{***} \\
				(  0.00) & ( 0.00) & ( 0.00) & ( 0.00) \\
				\addlinespace[3pt]
				
				-2.94\tnote{***} & -0.15\tnote{***} & -0.14\tnote{***} & -0.15\tnote{***} \\
				(  0.00) & ( 0.00) & ( 0.00) & ( 0.00) \\
				\addlinespace[3pt]
				
				44.71\tnote{***} &  1.16\tnote{**} &  1.22\tnote{**} &  1.45\tnote{***} \\
				(  0.00) & ( 0.02) & ( 0.02) & ( 0.01) \\
				\addlinespace[3pt]
				
				\multicolumn{1}{c}{135} & \multicolumn{1}{c}{135} & \multicolumn{1}{c}{135} & \multicolumn{1}{c}{135} \\
				\addlinespace[3pt]
				15.42\% & 17.42\% & 12.54\% & 15.49\% \\

				\midrule
				-5.08\tnote{***} & -0.29\tnote{***} & -0.25\tnote{***} & -0.28\tnote{***} \\
				( 0.00) & ( 0.00) & ( 0.00) & ( 0.00) \\
				\addlinespace[3pt]
				
				-3.22\tnote{***} & -0.20\tnote{***} & -0.16\tnote{***} & -0.18\tnote{***} \\
				( 0.00) & ( 0.00) & ( 0.00) & ( 0.00) \\
				\addlinespace[3pt]
				
				56.23\tnote{***} &  2.00\tnote{***} &  1.82\tnote{***} &  2.20\tnote{***} \\
				( 0.00) & ( 0.00) & ( 0.00) & ( 0.00) \\
				\addlinespace[3pt]
				
				\multicolumn{1}{c}{142} & \multicolumn{1}{c}{142} & \multicolumn{1}{c}{142} & \multicolumn{1}{c}{142} \\
				\addlinespace[3pt]
				13.60\% & 20.94\% & 14.45\% & 18.09\% \\

				\midrule
				-2.75\tnote{***} & -0.12\tnote{***} & -0.12\tnote{***} & -0.13\tnote{***} \\
				( 0.00) & ( 0.00) & ( 0.00) & ( 0.00) \\
				\addlinespace[3pt]
				
				-1.57\tnote{**} & -0.10\tnote{***} & -0.09\tnote{***} & -0.10\tnote{***} \\
				( 0.02) & ( 0.00) & ( 0.01) & ( 0.00) \\
				\addlinespace[3pt]
				
				43.66\tnote{***} &  1.11\tnote{**} &  1.27\tnote{**} &  1.46\tnote{***} \\
				( 0.00) & ( 0.02) & ( 0.01) & ( 0.00) \\
				\addlinespace[3pt]
				
				\multicolumn{1}{c}{138} & \multicolumn{1}{c}{138} & \multicolumn{1}{c}{138} & \multicolumn{1}{c}{138} \\
				\addlinespace[3pt]
				17.74\% & 17.37\% & 16.40\% & 17.75\% \\

				\midrule
				-48.90\tnote{***} & -2.60\tnote{***} & -2.40\tnote{***} & -2.59\tnote{***} \\
				(  0.00) & ( 0.00) & ( 0.00) & ( 0.00) \\
				\addlinespace[3pt]
				
				-2.85\tnote{***} & -0.16\tnote{***} & -0.15\tnote{***} & -0.16\tnote{***} \\
				(  0.00) & ( 0.00) & ( 0.00) & ( 0.00) \\
				\addlinespace[3pt]
				
				54.81\tnote{***} &  1.78\tnote{***} &  1.87\tnote{***} &  2.15\tnote{***} \\
				(  0.00) & ( 0.00) & ( 0.00) & ( 0.00) \\
				\addlinespace[3pt]
				
				\multicolumn{1}{c}{140} & \multicolumn{1}{c}{140} & \multicolumn{1}{c}{140} & \multicolumn{1}{c}{140} \\
				\addlinespace[3pt]
				13.24\% & 18.28\% & 15.02\% & 17.76\% \\

			\end{tabular} \\
			
			\midrule[0.3pt]\bottomrule[1.5pt]
		\end{tabular*}

		\begin{tablenotes}
			\singlespacing
			\item[] The table presents robustness results using global cutoff value which is April 11, 2020. We estimate equation 6 using OLS for first-digit goodness-of-fit measures. Panel A shows the results for the cumulative number of confirmed cases, while panel B shows the results for the cumulative number of deaths. To avoid small coefficients, we divide EIU, UHC, and No. of Days values by 100 for all models. Sample sizes vary due to missing values. \textit{P}-values for a one-tailed \textit{t}-test are in parentheses. ***, ** and * denote significance at the 1\%, 5\% and 10\% levels, respectively. All variable definitions are in Appendix A2. 
		\end{tablenotes}
		
	\end{threeparttable}
\end{sidewaystable}
\par
}

Second, we use 45 days since the first case for each country. We again
select the second tercile. We re-estimate Equation \ref{eq:1} for
the new measures of the goodness-of-fit. The results are presented
in Table 7, which are largely consistent with our previous findings.\footnote{Observe that we still include the $Number\text{\_\ensuremath{of\text{\_\ensuremath{Days_{i}}}}}$
control variable. This variable in general is different from the number
of days between the first case for the country and the cutoff date
because it counts only the days with nonzero number of cases (or deaths).
For many countries, especially during the first weeks of the pandemic,
there were days with zero confirmed cases (or deaths) after the first
case.} Finally, we use the following “window” approach: instead of using
one cutoff date, we estimate a series of goodness-of-fit measures
over a range of dates (specifically, we try \textpm 1, \textpm 3,
and \textpm 5 days around the original cutoff date). We find the average
over those values, and re-run regression \ref{eq:1}. Again, the results
are unchanged (untabulated). We conclude that our results are not
driven by the specific selection of the cutoff dates for the growth
part of the data.

{\centering
\begin{sidewaystable}
	\centering
	
	\begin{threeparttable}[h]
		\def\sym#1{\ifmmode^{#1}\else\(^{#1}\)\fi}
		
		\caption{Robustness Results: 45 Days Since First Case}\label{tab_res:robust_first45} 
		\scriptsize
		\renewcommand{\arraystretch}{0.4}
		\begin{tabular*}{0.9\linewidth}{@{\extracolsep{\fill}}l|l|l}
			\toprule[1.5pt]\midrule[0.3pt]
			
			\begin{tabular}[t]{ll}
				
				& \textcolor{white}{1}\\
				\arrayrulecolor{white}\midrule
				\multicolumn{2}{c}{\textbf{Variable}} \\
				
				\arrayrulecolor{black}\midrule

				\textbf{EIU}\tnote{{\textcolor{white}{*}}} \\
				\textcolor{white}{()} \\
				\addlinespace[3pt]
				
				\textbf{ln(Population)}\tnote{{\textcolor{white}{*}}} \\
				\textcolor{white}{()} \\
				\addlinespace[2pt]
				
				\textbf{No. of Days}\tnote{{\textcolor{white}{*}}}  \\
				\textcolor{white}{()} \\
				\addlinespace[1.8pt]
				
				\textbf{Sample Size} \\
				\addlinespace[2.8pt]
				
				\textbf{Adj. R\textsuperscript{2}} \\
				\midrule

				\textbf{ln(GDP)}\tnote{{\textcolor{white}{*}}} \\
				\textcolor{white}{()} \\
				\addlinespace[2pt]
				
				\textbf{ln(Population)}\tnote{{\textcolor{white}{*}}} \\
				\textcolor{white}{()} \\
				\addlinespace[1.8pt]
				
				\textbf{No. of Days}\tnote{{\textcolor{white}{*}}}  \\
				\textcolor{white}{()} \\
				\addlinespace[1.8pt]
				
				\textbf{Sample Size} \\
				\addlinespace[2.8pt]
				
				\textbf{Adj. R\textsuperscript{2}} \\
				\midrule

				\textbf{HE\_GDP}\tnote{{\textcolor{white}{*}}} \\
				\textcolor{white}{()} \\
				\addlinespace[2pt]
				
				\textbf{ln(Population)}\tnote{{\textcolor{white}{*}}} \\
				\textcolor{white}{()} \\
				\addlinespace[1.8pt]
				
				\textbf{No. of Days}\tnote{{\textcolor{white}{*}}}  \\
				\textcolor{white}{()} \\
				\addlinespace[1.8pt]
				
				\textbf{Sample Size} \\
				\addlinespace[2.8pt]
				
				\textbf{Adj. R\textsuperscript{2}} \\
				\midrule

				\textbf{UHC}\tnote{{\textcolor{white}{*}}} \\
				\textcolor{white}{()} \\
				\addlinespace[3pt]
				
				\textbf{ln(Population)}\tnote{{\textcolor{white}{*}}} \\
				\textcolor{white}{()} \\
				\addlinespace[1.8pt]
				
				\textbf{No. of Days}\tnote{{\textcolor{white}{*}}}  \\
				\textcolor{white}{()} \\
				\addlinespace[1.8pt]
				
				\textbf{Sample Size} \\
				\addlinespace[2.8pt]
				
				\textbf{Adj. R\textsuperscript{2}} \\

			\end{tabular} &

			\begin{tabular}[t]
				{
					D{.}{.}{-1}
					D{.}{.}{-1}
					D{.}{.}{-1}
					D{.}{.}{-1}
				}

				\multicolumn{4}{c}{\textbf{Panel A. Confirmed Cases}}  \\
				
				\midrule
				\multicolumn{1}{c}{\textbf{Chi-squared}} & \multicolumn{1}{c}{\textbf{Kuiper}} & \multicolumn{1}{c}{\textbf{M}} & \multicolumn{1}{c}{\textbf{D}} \\

				\midrule
				
				-8.73 & -0.01 &  0.10 & -0.05\tnote{{\textcolor{white}{*}}} \\
				( 0.19) & ( 0.48) & ( 0.38) & ( 0.45) \\
				\addlinespace[3pt]
				
				0.92 &  0.02 &  0.03 &  0.02\tnote{{\textcolor{white}{*}}} \\
				( 0.25) & ( 0.37) & ( 0.22) & ( 0.31) \\
				\addlinespace[3pt]
				
				-5.53 &  0.27 & -0.18 & -0.43\tnote{{\textcolor{white}{*}}} \\
				( 0.48) & ( 0.47) & ( 0.48) & ( 0.46) \\
				\addlinespace[3pt]
				
				\multicolumn{1}{c}{162} & \multicolumn{1}{c}{162} & \multicolumn{1}{c}{162} & \multicolumn{1}{c}{162} \\
				\addlinespace[3pt]
				-0.99\% & -1.81\% & -1.48\% & -1.72\% \\

				\midrule
				-3.42\tnote{***} & -0.10\tnote{**} & -0.06 & -0.10\tnote{**} \\
				( 0.01) & ( 0.03) & ( 0.12) & ( 0.03) \\
				\addlinespace[3pt]
				
				-1.65\tnote{*} & -0.09\tnote{***} & -0.08\tnote{**} & -0.09\tnote{***} \\
				( 0.06) & ( 0.01) & ( 0.02) & ( 0.01) \\
				\addlinespace[3pt]
				
				81.84 &  3.55 &  2.78 &  3.01\tnote{{\textcolor{white}{*}}} \\
				( 0.26) & ( 0.21) & ( 0.27) & ( 0.25) \\
				\addlinespace[3pt]
				
				\multicolumn{1}{c}{176} & \multicolumn{1}{c}{176} & \multicolumn{1}{c}{176} & \multicolumn{1}{c}{176} \\
				\addlinespace[3pt]
				2.05\% & 2.93\% & 1.16\% & 2.72\% \\

				\midrule
				-2.12\tnote{***} & -0.06\tnote{**} & -0.05\tnote{**} & -0.05\tnote{**} \\
				( 0.00) & ( 0.02) & ( 0.05) & ( 0.03) \\
				\addlinespace[3pt]
				
				-0.86 & -0.07\tnote{**} & -0.07\tnote{**} & -0.07\tnote{**} \\
				( 0.20) & ( 0.03) & ( 0.03) & ( 0.03) \\
				\addlinespace[3pt]
				
				15.25 &  1.60 &  1.64 &  1.12\tnote{{\textcolor{white}{*}}} \\
				( 0.45) & ( 0.36) & ( 0.35) & ( 0.40) \\
				\addlinespace[3pt]
				
				\multicolumn{1}{c}{173} & \multicolumn{1}{c}{173} & \multicolumn{1}{c}{173} & \multicolumn{1}{c}{173} \\
				\addlinespace[3pt]
				2.83\% & 3.24\% & 2.14\% & 2.66\% \\

				\midrule
				-31.48\tnote{***} & -0.60\tnote{*} & -0.45 & -0.71\tnote{*} \\
				(  0.01) & ( 0.09) & ( 0.16) & ( 0.06) \\
				\addlinespace[3pt]
				
				-1.17 & -0.07\tnote{**} & -0.06\tnote{*} & -0.07\tnote{**} \\
				(  0.15) & ( 0.03) & ( 0.07) & ( 0.04) \\
				\addlinespace[3pt]
				
				65.84 &  2.63 &  2.21 &  2.16\tnote{{\textcolor{white}{*}}} \\
				(  0.30) & ( 0.27) & ( 0.31) & ( 0.31) \\
				\addlinespace[3pt]
				
				\multicolumn{1}{c}{174} & \multicolumn{1}{c}{174} & \multicolumn{1}{c}{174} & \multicolumn{1}{c}{174} \\
				\addlinespace[3pt]
				1.89\% & 1.18\% & 0.04\% & 1.21\% \\

			\end{tabular} &
			
			\begin{tabular}[t]
				{
					D{.}{.}{-1}
					D{.}{.}{-1}
					D{.}{.}{-1}
					D{.}{.}{-1}
				}
				
				\multicolumn{4}{c}{\textbf{Panel B. Death Cases}}  \\
				
				\midrule
				\multicolumn{1}{c}{\textbf{Chi-squared}} & \multicolumn{1}{c}{\textbf{Kuiper}} & \multicolumn{1}{c}{\textbf{M}} & \multicolumn{1}{c}{\textbf{D}} \\				
				
				\midrule
				
				-29.44\tnote{**} & -1.46\tnote{***} & -1.13\tnote{***} & -1.29\tnote{***} \\
				(  0.02) & ( 0.00) & ( 0.00) & ( 0.00) \\
				\addlinespace[3pt]
				
				-5.86\tnote{***} & -0.21\tnote{***} & -0.18\tnote{***} & -0.20\tnote{***} \\
				(  0.00) & ( 0.00) & ( 0.00) & ( 0.00) \\
				\addlinespace[3pt]
				
				84.18\tnote{***} &  2.18\tnote{***} &  1.42\tnote{**} &  2.12\tnote{***} \\
				(  0.00) & ( 0.00) & ( 0.03) & ( 0.00) \\
				\addlinespace[3pt]
				
				\multicolumn{1}{c}{139} & \multicolumn{1}{c}{139} & \multicolumn{1}{c}{139} & \multicolumn{1}{c}{139} \\
				\addlinespace[3pt]
				13.83\% & 21.12\% & 13.39\% & 18.89\% \\

				\midrule
				-6.59\tnote{***} & -0.28\tnote{***} & -0.24\tnote{***} & -0.27\tnote{***} \\
				( 0.00) & ( 0.00) & ( 0.00) & ( 0.00) \\
				\addlinespace[3pt]
				
				-6.18\tnote{***} & -0.25\tnote{***} & -0.19\tnote{***} & -0.23\tnote{***} \\
				( 0.00) & ( 0.00) & ( 0.00) & ( 0.00) \\
				\addlinespace[3pt]
				
				67.31\tnote{***} &  1.48\tnote{**} &  0.91 &  1.48\tnote{**} \\
				( 0.01) & ( 0.02) & ( 0.12) & ( 0.03) \\
				\addlinespace[3pt]
				
				\multicolumn{1}{c}{147} & \multicolumn{1}{c}{147} & \multicolumn{1}{c}{147} & \multicolumn{1}{c}{147} \\
				\addlinespace[3pt]
				17.49\% & 28.60\% & 17.36\% & 24.63\% \\

				\midrule
				-3.45\tnote{***} & -0.13\tnote{***} & -0.14\tnote{***} & -0.14\tnote{***} \\
				( 0.00) & ( 0.00) & ( 0.00) & ( 0.00) \\
				\addlinespace[3pt]
				
				-4.73\tnote{***} & -0.19\tnote{***} & -0.14\tnote{***} & -0.17\tnote{***} \\
				( 0.00) & ( 0.00) & ( 0.00) & ( 0.00) \\
				\addlinespace[3pt]
				
				72.71\tnote{***} &  1.75\tnote{***} &  0.96 &  1.66\tnote{**} \\
				( 0.00) & ( 0.01) & ( 0.10) & ( 0.01) \\
				\addlinespace[3pt]
				
				\multicolumn{1}{c}{143} & \multicolumn{1}{c}{143} & \multicolumn{1}{c}{143} & \multicolumn{1}{c}{143} \\
				\addlinespace[3pt]
				15.96\% & 25.29\% & 18.77\% & 22.82\% \\

				\midrule
				-61.57\tnote{***} & -2.61\tnote{***} & -2.35\tnote{***} & -2.53\tnote{***} \\
				(  0.00) & ( 0.00) & ( 0.00) & ( 0.00) \\
				\addlinespace[3pt]
				
				-6.30\tnote{***} & -0.23\tnote{***} & -0.19\tnote{***} & -0.22\tnote{***} \\
				(  0.00) & ( 0.00) & ( 0.00) & ( 0.00) \\
				\addlinespace[3pt]
				
				80.12\tnote{***} &  2.11\tnote{***} &  1.35\tnote{**} &  2.04\tnote{***} \\
				(  0.00) & ( 0.00) & ( 0.03) & ( 0.00) \\
				\addlinespace[3pt]
				
				\multicolumn{1}{c}{146} & \multicolumn{1}{c}{146} & \multicolumn{1}{c}{146} & \multicolumn{1}{c}{146} \\
				\addlinespace[3pt]
				18.65\% & 28.99\% & 20.84\% & 26.99\% \\

			\end{tabular} \\
			
			\midrule[0.3pt]\bottomrule[1.5pt]
		\end{tabular*}

		\begin{tablenotes}
			\singlespacing
			\item[] The table presents robustness results using 45 days since the first case in each country as cutoff days. We estimate equation 6 using OLS for first-digit goodness-of-fit measures. Panel A shows the results for the cumulative number of confirmed cases, while panel B shows the results for the cumulative number of deaths. To avoid small coefficients, we divide EIU, UHC, and No. of Days values by 100 for all models. Sample sizes vary due to missing values. \textit{P}-values for a one-tailed \textit{t}-test are in parentheses. ***, ** and * denote significance at the 1\%, 5\% and 10\% levels, respectively. All variable definitions are in Appendix A2. 
		\end{tablenotes}
		
	\end{threeparttable}
\end{sidewaystable}

\par
}

Another concern is that our data might be driven by countries with
few cases or few data points. To test for that, we exclude countries
with lower than 200 (500 and 1,000) total confirmed cases. We then
exclude countries with fewer than 30 (40) days of nonzero cases. We
also exclude countries with the highest 1\% (5\%) goodness-of-fit
measures. In reported tests, we find that the results are robust in
all cases. We conclude that our results are not driven by small countries,
countries with a small number of cases, or by extreme deviations from
the NBL.

\subsection{Regional Data}

Testing for compliance with the NBL requires sufficient data. For
many countries in our analysis, the goodness-of-fit measures are calculated
based on relatively small samples sizes between 40 and 140 days. Even
though we control for the sample size and conduct robustness checks,
making inferences from results based on such small sample sizes might
be problematic. The sample size may increase significantly for a country
if it reports data at a regional (state, territory, or provinces)
level. Each reported value at the regional level can then be used
to estimate goodness-of-fit measures, instead of using the country-level
data. The method has the upside that the goodness-of-fit measures
are estimated with a greater precision, though the downside is the
lack of countries that collect regional data.

Fifty out of 185 countries in our sample collect regional-level data.\footnote{For the list of countries, see Appendix A1. Regional data are from
the COVID-19 Coronavirus Map. https://covid19.health/. Downloadable
data set is at https://github.com/stevenliuyi/covid19. We check for
data consistency between the two data sources and find the high degree
of agreement.} For these countries, we re-estimate the goodness-of-fit measures
and re-run Equation \ref{eq:1}. The results are reported in Table
8.\footnote{Note the slightly smaller number of observations. For the second-digit
test to work, the data should be at least over ten.} Panel A depicts the confirmed number of cases; Panel B the number
of deaths. For the cumulative number of cases, 12 out of 16 coefficients
are negative, with eight demonstrating significance. For the cumulative
number of deaths, all coefficients are negative and significant, even
with the much smaller number of countries for these tests. The results
are consistent with our earlier finding: countries with higher democracy
indices, GPD per capita, health expenditures, and universal healthcare
coverage are less likely to manipulate pandemic data, especially the
number of deaths. We further conclude that our findings are not driven
by the errors in goodness-of-fit measures.

{\centering
\begin{sidewaystable}
	\centering
	
	\begin{threeparttable}[h]
		\def\sym#1{\ifmmode^{#1}\else\(^{#1}\)\fi}
		
		\caption{Results for 50 Countries with Regional Data}\label{tab_res:50country} 
		\scriptsize
		\renewcommand{\arraystretch}{0.4}
		\begin{tabular*}{0.9\linewidth}{@{\extracolsep{\fill}}l|l|l}
			\toprule[1.5pt]\midrule[0.3pt]
			
			\begin{tabular}[t]{ll}
				
				& \textcolor{white}{1}\\
				\arrayrulecolor{white}\midrule
				\multicolumn{2}{c}{\textbf{Variable}} \\
				
				\arrayrulecolor{black}\midrule

				\textbf{EIU}\tnote{{\textcolor{white}{*}}} \\
				\textcolor{white}{()} \\
				\addlinespace[4pt]
				
				\textbf{ln(Population)}\tnote{{\textcolor{white}{*}}} \\
				\textcolor{white}{()} \\
				\addlinespace[3pt]
				
				\textbf{No. of Days}\tnote{{\textcolor{white}{*}}}  \\
				\textcolor{white}{()} \\
				\addlinespace[2pt]
				
				\textbf{Sample Size} \\
				\addlinespace[3pt]
				
				\textbf{Adj. R\textsuperscript{2}} \\
				\midrule

				\textbf{ln(GDP)}\tnote{{\textcolor{white}{*}}} \\
				\textcolor{white}{()} \\
				\addlinespace[4pt]
				
				\textbf{ln(Population)}\tnote{{\textcolor{white}{*}}} \\
				\textcolor{white}{()} \\
				\addlinespace[3pt]
				
				\textbf{No. of Days}\tnote{{\textcolor{white}{*}}}  \\
				\textcolor{white}{()} \\
				\addlinespace[2pt]
				
				\textbf{Sample Size} \\
				\addlinespace[3pt]
				
				\textbf{Adj. R\textsuperscript{2}} \\
				\midrule

				\textbf{HE\_GDP}\tnote{{\textcolor{white}{*}}} \\
				\textcolor{white}{()} \\
				\addlinespace[4pt]
				
				\textbf{ln(Population)}\tnote{{\textcolor{white}{*}}} \\
				\textcolor{white}{()} \\
				\addlinespace[3pt]
				
				\textbf{No. of Days}\tnote{{\textcolor{white}{*}}}  \\
				\textcolor{white}{()} \\
				\addlinespace[2pt]
				
				\textbf{Sample Size} \\
				\addlinespace[3pt]
				
				\textbf{Adj. R\textsuperscript{2}} \\
				\midrule

				\textbf{UHC}\tnote{{\textcolor{white}{*}}} \\
				\textcolor{white}{()} \\
				\addlinespace[4pt]
				
				\textbf{ln(Population)}\tnote{{\textcolor{white}{*}}} \\
				\textcolor{white}{()} \\
				\addlinespace[3pt]
				
				\textbf{No. of Days}\tnote{{\textcolor{white}{*}}}  \\
				\textcolor{white}{()} \\
				\addlinespace[2pt]
				
				\textbf{Sample Size} \\
				\addlinespace[3pt]
				
				\textbf{Adj. R\textsuperscript{2}} \\

			\end{tabular} &

			\begin{tabular}[t]
				{
					D{.}{.}{-1}
					D{.}{.}{-1}
					D{.}{.}{-1}
					D{.}{.}{-1}
				}

				\multicolumn{4}{c}{\textbf{Panel A. Confirmed Cases}}  \\
				
				\midrule
				\multicolumn{1}{c}{\textbf{Chi-squared}} & \multicolumn{1}{c}{\textbf{Kuiper}} & \multicolumn{1}{c}{\textbf{M}} & \multicolumn{1}{c}{\textbf{D}} \\

				\midrule
				
				27.37 & 0.09 & 0.29 & 0.34\tnote{{\textcolor{white}{*}}} \\
				( 0.37) & (0.46) & (0.37) & (0.38) \\
				\addlinespace[4pt]
				
				1.60 & 0.01 & 0.03 & 0.02\tnote{{\textcolor{white}{*}}} \\
				( 0.44) & (0.49) & (0.41) & (0.44) \\
				\addlinespace[4pt]
				
				3.56\tnote{**} & 0.04\tnote{**} & 0.03\tnote{**} & 0.05\tnote{**} \\
				( 0.03) & (0.04) & (0.05) & (0.02) \\
				\addlinespace[4pt]
				
				\multicolumn{1}{c}{50} & \multicolumn{1}{c}{50} & \multicolumn{1}{c}{50} & \multicolumn{1}{c}{50} \\
				\addlinespace[3.6pt]
				5.09\% & 3.55\% & 2.64\% & 6.39\% \\

				\midrule
				-21.00\tnote{*} & -0.36\tnote{**} & -0.16 & -0.25 \\
				(  0.09) & ( 0.04) & ( 0.17) & ( 0.13) \\
				\addlinespace[4pt]
				
				-2.05 & -0.04 &  0.00 & -0.02\tnote{{\textcolor{white}{*}}} \\
				(  0.42) & ( 0.38) & ( 0.48) & ( 0.44) \\
				\addlinespace[4pt]
				
				2.42\tnote{*} &  0.03 &  0.02 &  0.04\tnote{*} \\
				(  0.10) & ( 0.15) & ( 0.12) & ( 0.07) \\
				\addlinespace[4pt]
				
				\multicolumn{1}{c}{50} & \multicolumn{1}{c}{50} & \multicolumn{1}{c}{50} & \multicolumn{1}{c}{50} \\
				\addlinespace[3.6pt]
				8.43\% & 9.61\% & 4.30\% & 8.85\% \\

				\midrule
				-6.08 & -0.15\tnote{**} & -0.10\tnote{**} & -0.11\tnote{*} \\
				( 0.13) & ( 0.02) & ( 0.05) & ( 0.06) \\
				\addlinespace[4pt]
				
				1.36 &  0.01 &  0.02 &  0.02\tnote{{\textcolor{white}{*}}} \\
				( 0.45) & ( 0.46) & ( 0.43) & ( 0.46) \\
				\addlinespace[4pt]
				
				2.58\tnote{*} &  0.02 &  0.02 &  0.03\tnote{*} \\
				( 0.09) & ( 0.18) & ( 0.19) & ( 0.09) \\
				\addlinespace[4pt]
				
				\multicolumn{1}{c}{49} & \multicolumn{1}{c}{49} & \multicolumn{1}{c}{49} & \multicolumn{1}{c}{49} \\
				\addlinespace[3.6pt]
				7.15\% & 12.07\% & 7.79\% & 10.38\% \\

				\midrule
				-301.29\tnote{**} & -4.29\tnote{**} & -2.21 & -3.37\tnote{*} \\
				(   0.03) & ( 0.02) & ( 0.11) & ( 0.06) \\
				\addlinespace[4pt]
				
				-0.44 & -0.01 &  0.01 &  0.00\tnote{{\textcolor{white}{*}}} \\
				(   0.48) & ( 0.46) & ( 0.47) & ( 0.49) \\
				\addlinespace[4pt]
				
				2.61\tnote{*} &  0.03\tnote{*} &  0.03\tnote{*} &  0.04\tnote{**} \\
				(   0.07) & ( 0.09) & ( 0.09) & ( 0.05) \\
				\addlinespace[4pt]
				
				\multicolumn{1}{c}{50} & \multicolumn{1}{c}{50} & \multicolumn{1}{c}{50} & \multicolumn{1}{c}{50} \\
				\addlinespace[3.6pt]
				11.70\% & 11.69\% & 5.67\% & 10.86\% \\

			\end{tabular} &
			
			\begin{tabular}[t]
				{
					D{.}{.}{-1}
					D{.}{.}{-1}
					D{.}{.}{-1}
					D{.}{.}{-1}
				}
				
				\multicolumn{4}{c}{\textbf{Panel B. Death Cases}}  \\
				
				\midrule
				\multicolumn{1}{c}{\textbf{Chi-squared}} & \multicolumn{1}{c}{\textbf{Kuiper}} & \multicolumn{1}{c}{\textbf{M}} & \multicolumn{1}{c}{\textbf{D}} \\				
				
				\midrule
				
				-220.31\tnote{***} & -4.74\tnote{***} & -3.64\tnote{***} & -4.51\tnote{***} \\
				(   0.01) & ( 0.00) & ( 0.00) & ( 0.00) \\
				\addlinespace[4pt]
				
				13.16 &  0.23 &  0.02 &  0.13\tnote{{\textcolor{white}{*}}} \\
				(   0.14) & ( 0.13) & ( 0.44) & ( 0.25) \\
				\addlinespace[4pt]
				
				-1.77 & -0.01 & -0.02 & -0.02\tnote{{\textcolor{white}{*}}} \\
				(   0.26) & ( 0.38) & ( 0.32) & ( 0.35) \\
				\addlinespace[4pt]
				
				\multicolumn{1}{c}{30} & \multicolumn{1}{c}{30} & \multicolumn{1}{c}{30} & \multicolumn{1}{c}{30} \\
				\addlinespace[3.6pt]
				19.78\% & 32.99\% & 23.23\% & 27.44\% \\

				\midrule
				-52.00\tnote{***} & -0.78\tnote{***} & -0.36\tnote{*} & -0.70\tnote{**} \\
				(  0.00) & ( 0.01) & ( 0.08) & ( 0.01) \\
				\addlinespace[4pt]
				
				15.52\tnote{*} &  0.32\tnote{*} &  0.13 &  0.23 \\
				(  0.09) & ( 0.06) & ( 0.23) & ( 0.13) \\
				\addlinespace[4pt]
				
				-3.35 & -0.03 & -0.02 & -0.03\tnote{{\textcolor{white}{*}}} \\
				(  0.10) & ( 0.26) & ( 0.33) & ( 0.25) \\
				\addlinespace[4pt]
				
				\multicolumn{1}{c}{30} & \multicolumn{1}{c}{30} & \multicolumn{1}{c}{30} & \multicolumn{1}{c}{30} \\
				\addlinespace[3.6pt]
				28.71\% & 25.12\% & 0.90\% & 16.75\% \\

				\midrule
				-19.65\tnote{***} & -0.34\tnote{***} & -0.20\tnote{**} & -0.31\tnote{***} \\
				(  0.00) & ( 0.00) & ( 0.02) & ( 0.00) \\
				\addlinespace[4pt]
				
				19.45\tnote{*} &  0.38\tnote{**} &  0.15 &  0.28\tnote{*} \\
				(  0.05) & ( 0.03) & ( 0.19) & ( 0.08) \\
				\addlinespace[4pt]
				
				-4.09\tnote{*} & -0.05 & -0.03 & -0.05 \\
				(  0.07) & ( 0.15) & ( 0.21) & ( 0.15) \\
				\addlinespace[4pt]
				
				\multicolumn{1}{c}{29} & \multicolumn{1}{c}{29} & \multicolumn{1}{c}{29} & \multicolumn{1}{c}{29} \\
				\addlinespace[3.6pt]
				31.47\% & 34.29\% & 11.57\% & 25.88\% \\

				\midrule
				-639.03\tnote{***} & -9.79\tnote{***} & -4.42\tnote{**} & -8.42\tnote{***} \\
				(   0.00) & ( 0.00) & ( 0.04) & ( 0.00) \\
				\addlinespace[4pt]
				
				17.61\tnote{**} &  0.35\tnote{**} &  0.14 &  0.26\tnote{*} \\
				(   0.04) & ( 0.03) & ( 0.20) & ( 0.09) \\
				\addlinespace[4pt]
				
				-2.15 & -0.01 & -0.01 & -0.02\tnote{{\textcolor{white}{*}}} \\
				(   0.17) & ( 0.37) & ( 0.40) & ( 0.36) \\
				\addlinespace[4pt]
				
				\multicolumn{1}{c}{30} & \multicolumn{1}{c}{30} & \multicolumn{1}{c}{30} & \multicolumn{1}{c}{30} \\
				\addlinespace[3.6pt]
				45.25\% & 38.17\% & 5.52\% & 26.45\% \\

			\end{tabular} \\
			
			\midrule[0.3pt]\bottomrule[1.5pt]
		\end{tabular*}

		\begin{tablenotes}
			\singlespacing
			\item[] The table presents the results using regional data from 50 selected countries. We estimate equation 6 using OLS for first-digit goodness-of-fit measures. Panel A shows the results for the cumulative number of confirmed cases, while panel B shows the results for the cumulative number of deaths. To avoid small coefficients, we divide EIU, UHC, and No. of Days values by 100 for all models. Sample sizes vary due to missing values. \textit{P}-values for a one-tailed \textit{t}-test are in parentheses. ***, ** and * denote significance at the 1\%, 5\% and 10\% levels, respectively. All variable definitions are in Appendix A2. 
		\end{tablenotes}
		
	\end{threeparttable}
\end{sidewaystable}
\par
}

\subsection{The Case of the United States}

The United States of America collects data not only at a state level,
but also at the county level. This enables a deeper analysis of individual
states. The Unites States is classified as a “flawed democracy” by
the Economist Intelligence Unit Democracy Index, with around \$60,000
GDP per capita, an unprecedented 17\% of GDP spent on healthcare expenditures,
and a high value of the Universal Health Coverage Index: 84. Appendix
A1 shows that there is no systematic indication of COVID-19 data manipulation,
either for the cumulative number of confirmed cases, or for the cumulative
number of reported deaths: all goodness-of-fit measures are below
critical values at the 1\% significance level. Yet, much controversy
exists as to whether individual states manipulate COVID-19 data \parencite{Smith2020states,King2020Florida}.
Mass media outlets argue that state governments downplay the spread
of the virus for political gains. It is not clear, however, if there
is a cross-sectional difference in the political parties of the state’s
governments (i.e., the legislative branch, or senate, and the executive
branch, or governor) in a particular state’s misreporting of data.
To see if political systems drive the differences between states,
we conduct a modified analysis, when we treat each state as a separate
“country.” We measure the GDP per capita and $HE\text{\_}GDP$ variables
at the state level. We then substitute the $EIU$ regime indicator
with three alternative political indicators: $Won$, a dummy variable
indicator that the incumbent U.S. president (Republican) won the state
during the previous election; $Senate$, a dummy variable indicator
that the state legislature derives its majority from the same political
party as the current U.S. president; and $Governor$, a dummy variable
indicator that the governor of the state is from the same party as
the incumbent U.S. president.

We re-estimate the end of the pandemic growth period and, using county
level data, re-calculate the eight goodness-of-fit measures (four
for the cumulative number of confirmed cases, and four for the cumulative
number of deaths). We then estimate Equation \ref{eq:1}. The results
are shown in Table 9. The ln($GDP$) is mostly insignificant. The
negative sign of $HE\text{\_}GDP$ in all tests is consistent with
our previous findings, albeit the coefficients lack significance in
most regressions. $Won$ and $Senate$ are positive and significant
in all tests, indicating that states that voted for the incumbent
U.S. president and have governments led by the same political party
are more likely to manipulate the pandemic data. We also show that
the results for the two variables are stronger for the death toll
than the number of confirmed cases, consistent with our earlier findings.
We highlight here that the sample size for the U.S. tests is much
smaller than the original sample, consisting only of 50 states. Finally,
the $Governor$ variable is insignificant throughout.
 
{\centering
\begin{sidewaystable}
	\centering
	
	\begin{threeparttable}[h]
		\def\sym#1{\ifmmode^{#1}\else\(^{#1}\)\fi}
		
		\caption{Results for US State Data}\label{tab_res:usState} 
		\tiny
		\renewcommand{\arraystretch}{0.4}
		\begin{tabular*}{0.85\linewidth}{@{\extracolsep{\fill}}l|l|l}
			\toprule[1.5pt]\midrule[0.3pt]
			
			\begin{tabular}[t]{l}
				
				\textcolor{white}{1}\\
				\arrayrulecolor{white}\midrule
				\multicolumn{1}{c}{\textbf{Variable}} \\
				
				\arrayrulecolor{black}\midrule

				\textbf{ln(GDP)}\tnote{{\textcolor{white}{*}}} \\
				\textcolor{white}{()} \\
				\addlinespace[4pt]
				
				\textbf{ln(Population)}\tnote{{\textcolor{white}{*}}} \\
				\textcolor{white}{()} \\
				\addlinespace[2.5pt]
				
				\textbf{No. of Days}\tnote{{\textcolor{white}{*}}}  \\
				\textcolor{white}{()} \\
				\addlinespace[2.5pt]
				
				\textbf{Sample Size} \\
				\addlinespace[3.6pt]
				
				\textbf{Adj. R\textsuperscript{2}} \\
				\midrule

				\textbf{HE\_GDP}\tnote{{\textcolor{white}{*}}} \\
				\textcolor{white}{()} \\
				\addlinespace[4pt]
				
				\textbf{ln(Population)}\tnote{{\textcolor{white}{*}}} \\
				\textcolor{white}{()} \\
				\addlinespace[2.5pt]
				
				\textbf{No. of Days}\tnote{{\textcolor{white}{*}}}  \\
				\textcolor{white}{()} \\
				\addlinespace[2.5pt]
				
				\textbf{Sample Size} \\
				\addlinespace[3.6pt]
				
				\textbf{Adj. R\textsuperscript{2}} \\
				\midrule

				\textbf{Won\_Rep}\tnote{{\textcolor{white}{*}}} \\
				\textcolor{white}{()} \\
				\addlinespace[4pt]
				
				\textbf{ln(Population)}\tnote{{\textcolor{white}{*}}} \\
				\textcolor{white}{()} \\
				\addlinespace[2.5pt]
				
				\textbf{No. of Days}\tnote{{\textcolor{white}{*}}}  \\
				\textcolor{white}{()} \\
				\addlinespace[2.5pt]
				
				\textbf{Sample Size} \\
				\addlinespace[3.6pt]
				
				\textbf{Adj. R\textsuperscript{2}} \\
				\midrule

				\textbf{Senate\_Rep}\tnote{{\textcolor{white}{*}}} \\
				\textcolor{white}{()} \\
				\addlinespace[4pt]
				
				\textbf{ln(Population)}\tnote{{\textcolor{white}{*}}} \\
				\textcolor{white}{()} \\
				\addlinespace[2.5pt]
				
				\textbf{No. of Days}\tnote{{\textcolor{white}{*}}}  \\
				\textcolor{white}{()} \\
				\addlinespace[2.5pt]
				
				\textbf{Sample Size} \\
				\addlinespace[3.6pt]
				
				\textbf{Adj. R\textsuperscript{2}} \\
				\midrule

				\textbf{Governor\_Rep}\tnote{{\textcolor{white}{*}}} \\
				\textcolor{white}{()} \\
				\addlinespace[4pt]
				
				\textbf{ln(Population)}\tnote{{\textcolor{white}{*}}} \\
				\textcolor{white}{()} \\
				\addlinespace[2.5pt]
				
				\textbf{No. of Days}\tnote{{\textcolor{white}{*}}}  \\
				\textcolor{white}{()} \\
				\addlinespace[2.5pt]
				
				\textbf{Sample Size} \\
				\addlinespace[3.6pt]
				
				\textbf{Adj. R\textsuperscript{2}} \\

			\end{tabular} &

			\begin{tabular}[t]
				{
					D{.}{.}{-1}
					D{.}{.}{-1}
					D{.}{.}{-1}
					D{.}{.}{-1}
				}

				\multicolumn{4}{c}{\textbf{Panel A. Confirmed Cases}}  \\
				
				\midrule
				\multicolumn{1}{c}{\textbf{Chi-squared}} & \multicolumn{1}{c}{\textbf{Kuiper}} & \multicolumn{1}{c}{\textbf{M}} & \multicolumn{1}{c}{\textbf{D}} \\

				\midrule
				
				176.81\tnote{**} &  2.33\tnote{*} &  1.79 &  2.16 \\
				(  0.02) & ( 0.09) & ( 0.13) & ( 0.10) \\
				\addlinespace[4pt]
				
				-51.36\tnote{***} & -0.69\tnote{**} & -0.71\tnote{**} & -0.78\tnote{**} \\
				(  0.00) & ( 0.03) & ( 0.02) & ( 0.02) \\
				\addlinespace[4pt]
				
				3.41\tnote{***} &  0.06\tnote{***} &  0.04\tnote{*} &  0.05\tnote{**} \\
				(  0.00) & ( 0.01) & ( 0.06) & ( 0.02) \\
				\addlinespace[4pt]
				
				\multicolumn{1}{c}{50} & \multicolumn{1}{c}{50} & \multicolumn{1}{c}{50} & \multicolumn{1}{c}{50} \\
				\addlinespace[4pt]
				17.31\% & 7.28\% & 4.02\% & 6.62\% \\

				\midrule
				-8.57\tnote{*} & -0.12 & -0.11 & -0.13 \\
				(  0.05) & ( 0.13) & ( 0.14) & ( 0.11) \\
				\addlinespace[4pt]
				
				-54.17\tnote{***} & -0.74\tnote{**} & -0.77\tnote{**} & -0.85\tnote{**} \\
				(  0.00) & ( 0.03) & ( 0.02) & ( 0.01) \\
				\addlinespace[4pt]
				
				2.92\tnote{***} &  0.05\tnote{**} &  0.03\tnote{*} &  0.05\tnote{**} \\
				(  0.01) & ( 0.02) & ( 0.08) & ( 0.03) \\
				\addlinespace[4pt]
				
				\multicolumn{1}{c}{50} & \multicolumn{1}{c}{50} & \multicolumn{1}{c}{50} & \multicolumn{1}{c}{50} \\
				\addlinespace[4pt]
				13.82\% & 6.15\% & 3.90\% & 6.43\% \\

				\midrule
				62.66\tnote{**} &  1.51\tnote{***} &  1.29\tnote{**} &  1.40\tnote{**} \\
				(  0.02) & ( 0.01) & ( 0.01) & ( 0.01) \\
				\addlinespace[4pt]
				
				-38.53\tnote{**} & -0.46\tnote{*} & -0.52\tnote{*} & -0.57\tnote{**} \\
				(  0.01) & ( 0.09) & ( 0.06) & ( 0.05) \\
				\addlinespace[4pt]
				
				2.07\tnote{**} &  0.03\tnote{*} &  0.02 &  0.03 \\
				(  0.04) & ( 0.08) & ( 0.24) & ( 0.12) \\
				\addlinespace[4pt]
				
				\multicolumn{1}{c}{50} & \multicolumn{1}{c}{50} & \multicolumn{1}{c}{50} & \multicolumn{1}{c}{50} \\
				\addlinespace[4pt]
				16.67\% & 15.32\% & 11.34\% & 13.76\% \\

				\midrule
				63.69\tnote{**} &  1.48\tnote{***} &  1.27\tnote{**} &  1.39\tnote{**} \\
				(  0.02) & ( 0.01) & ( 0.01) & ( 0.01) \\
				\addlinespace[4pt]
				
				-39.77\tnote{**} & -0.49\tnote{*} & -0.55\tnote{**} & -0.60\tnote{**} \\
				(  0.01) & ( 0.08) & ( 0.05) & ( 0.04) \\
				\addlinespace[4pt]
				
				2.06\tnote{**} &  0.03\tnote{*} &  0.02 &  0.03 \\
				(  0.04) & ( 0.08) & ( 0.24) & ( 0.12) \\
				\addlinespace[4pt]
				
				\multicolumn{1}{c}{50} & \multicolumn{1}{c}{50} & \multicolumn{1}{c}{50} & \multicolumn{1}{c}{50} \\
				\addlinespace[4pt]
				16.92\% & 14.93\% & 11.14\% & 13.58\% \\

				\midrule
				19.16 &  0.43 &  0.30 &  0.35\tnote{{\textcolor{white}{*}}} \\
				(  0.27) & ( 0.25) & ( 0.31) & ( 0.29) \\
				\addlinespace[4pt]
				
				-41.10\tnote{**} & -0.52\tnote{*} & -0.59\tnote{**} & -0.64\tnote{**} \\
				(  0.01) & ( 0.08) & ( 0.05) & ( 0.04) \\
				\addlinespace[4pt]
				
				2.55\tnote{**} &  0.05\tnote{**} &  0.03 &  0.04\tnote{*} \\
				(  0.02) & ( 0.04) & ( 0.13) & ( 0.06) \\
				\addlinespace[4pt]
				
				\multicolumn{1}{c}{50} & \multicolumn{1}{c}{50} & \multicolumn{1}{c}{50} & \multicolumn{1}{c}{50} \\
				\addlinespace[4pt]
				9.32\% & 4.37\% & 1.97\% & 3.90\% \\

			\end{tabular} &
			
			\begin{tabular}[t]
				{
					D{.}{.}{-1}
					D{.}{.}{-1}
					D{.}{.}{-1}
					D{.}{.}{-1}
				}
				
				\multicolumn{4}{c}{\textbf{Panel B. Death Cases}}  \\
				
				\midrule
				\multicolumn{1}{c}{\textbf{Chi-squared}} & \multicolumn{1}{c}{\textbf{Kuiper}} & \multicolumn{1}{c}{\textbf{M}} & \multicolumn{1}{c}{\textbf{D}} \\				
				
				\midrule
				
				152.75 &  0.83 &  0.47 &  0.80\tnote{{\textcolor{white}{*}}} \\
				(  0.15) & ( 0.35) & ( 0.42) & ( 0.36) \\
				\addlinespace[4pt]
				
				-46.18\tnote{*} & -0.32 & -0.23 & -0.31 \\
				(  0.07) & ( 0.24) & ( 0.31) & ( 0.26) \\
				\addlinespace[4pt]
				
				17.92\tnote{***} &  0.19\tnote{***} &  0.15\tnote{***} &  0.18\tnote{***} \\
				(  0.00) & ( 0.00) & ( 0.00) & ( 0.00) \\
				\addlinespace[4pt]
				
				\multicolumn{1}{c}{50} & \multicolumn{1}{c}{50} & \multicolumn{1}{c}{50} & \multicolumn{1}{c}{50} \\
				\addlinespace[4pt]
				63.52\% & 49.82\% & 36.57\% & 43.69\% \\

				\midrule
				-15.83\tnote{**} & -0.17 & -0.15 & -0.17 \\
				(  0.04) & ( 0.10) & ( 0.13) & ( 0.11) \\
				\addlinespace[4pt]
				
				-58.27\tnote{**} & -0.48 & -0.39 & -0.48 \\
				(  0.04) & ( 0.15) & ( 0.21) & ( 0.16) \\
				\addlinespace[4pt]
				
				17.66\tnote{***} &  0.19\tnote{***} &  0.15\tnote{***} &  0.18\tnote{***} \\
				(  0.00) & ( 0.00) & ( 0.00) & ( 0.00) \\
				\addlinespace[4pt]
				
				\multicolumn{1}{c}{50} & \multicolumn{1}{c}{50} & \multicolumn{1}{c}{50} & \multicolumn{1}{c}{50} \\
				\addlinespace[4pt]
				65.01\% & 51.43\% & 38.19\% & 45.38\% \\

				\midrule
				86.46\tnote{*} &  1.91\tnote{***} &  1.72\tnote{**} &  1.85\tnote{**} \\
				(  0.06) & ( 0.01) & ( 0.02) & ( 0.01) \\
				\addlinespace[4pt]
				
				-32.10 & -0.11 & -0.05 & -0.11\tnote{{\textcolor{white}{*}}} \\
				(  0.15) & ( 0.40) & ( 0.45) & ( 0.40) \\
				\addlinespace[4pt]
				
				16.41\tnote{***} &  0.17\tnote{***} &  0.13\tnote{***} &  0.16\tnote{***} \\
				(  0.00) & ( 0.00) & ( 0.00) & ( 0.00) \\
				\addlinespace[4pt]
				
				\multicolumn{1}{c}{50} & \multicolumn{1}{c}{50} & \multicolumn{1}{c}{50} & \multicolumn{1}{c}{50} \\
				\addlinespace[4pt]
				64.68\% & 56.09\% & 42.58\% & 49.68\% \\

				\midrule
				78.50\tnote{*} &  1.85\tnote{***} &  1.60\tnote{**} &  1.72\tnote{**} \\
				(  0.08) & ( 0.01) & ( 0.02) & ( 0.02) \\
				\addlinespace[4pt]
				
				-34.48 & -0.15 & -0.10 & -0.16\tnote{{\textcolor{white}{*}}} \\
				(  0.13) & ( 0.36) & ( 0.42) & ( 0.36) \\
				\addlinespace[4pt]
				
				16.49\tnote{***} &  0.17\tnote{***} &  0.13\tnote{***} &  0.16\tnote{***} \\
				(  0.00) & ( 0.00) & ( 0.00) & ( 0.00) \\
				\addlinespace[4pt]
				
				\multicolumn{1}{c}{50} & \multicolumn{1}{c}{50} & \multicolumn{1}{c}{50} & \multicolumn{1}{c}{50} \\
				\addlinespace[4pt]
				64.32\% & 55.67\% & 41.77\% & 48.86\% \\

				\midrule
				18.22 &  0.61 &  0.92 &  0.81\tnote{{\textcolor{white}{*}}} \\
				(  0.37) & ( 0.22) & ( 0.13) & ( 0.16) \\
				\addlinespace[4pt]
				
				-37.04 & -0.18 & -0.06 & -0.14\tnote{{\textcolor{white}{*}}} \\
				(  0.13) & ( 0.35) & ( 0.45) & ( 0.38) \\
				\addlinespace[4pt]
				
				17.16\tnote{***} &  0.18\tnote{***} &  0.14\tnote{***} &  0.17\tnote{***} \\
				(  0.00) & ( 0.00) & ( 0.00) & ( 0.00) \\
				\addlinespace[4pt]
				
				\multicolumn{1}{c}{50} & \multicolumn{1}{c}{50} & \multicolumn{1}{c}{50} & \multicolumn{1}{c}{50} \\
				\addlinespace[4pt]
				62.75\% & 50.31\% & 38.28\% & 44.74\% \\

			\end{tabular} \\
			
			\midrule[0.3pt]\bottomrule[1.5pt]
		\end{tabular*}

		\begin{tablenotes}
			\singlespacing
			\item[] The table presents the results using equation 6 with U.S. state level data. In addition to state GDP and HE\_GDP values, we also include three dummy variables indicating the political party affiliation in each state, which equals one if Republican is dominant, and 0 otherwise. Panel A shows the results for the cumulative number of confirmed cases, while panel B shows the results for the cumulative number of deaths. To avoid small coefficients, we divide No. of Days values by 100 for all models. All models are estimated using OLS regression. \textit{P}-values for a one-tailed \textit{t}-test are in parentheses. ***, ** and * denote significance at the 1\%, 5\% and 10\% levels, respectively. All variable definitions are in Appendix A2. 
		\end{tablenotes}
		
	\end{threeparttable}
\end{sidewaystable}
\par
}

\subsection{Second Digit Tests}

The NBL can be extended to digits beyond the first (\cite{o2017offsite}
and \cite{hussain2010application}). Beyond the second digit, the
theoretical distribution quickly converges to uniform. \textcite{diekmann2007not}
notes that, when fabricating data, test subjects also naturally lean
toward smaller first digits, resulting in Benford-like distributions
of fabricated data. He suggests that in some cases the second-digit
test may provide a clearer assessment of data manipulation. Therefore,
we repeat our tests but use the second-digit goodness-of-fit measures
instead of the leading digit. Our sample size drops somewhat, especially
for the number of deaths, because the test requires values higher
than ten. The results are presented in Table 10, again, with two panels:
one for the confirmed number of cases, and one for the number of deaths.
In Panel A, all coefficients are negative and nine out of 16 coefficients
are significant. In Panel B, all coefficients are negative, and, except
for two coefficients for $UHC$, none are significant. We conclude
that second-digit test results accord with our main findings.
 
{\centering
\begin{sidewaystable}
	
	\centering
	\begin{threeparttable}[h]		
		\def\sym#1{\ifmmode^{#1}\else\(^{#1}\)\fi}
		
		\caption{Second Digit Tests}\label{tab_res:2digit} 
		\scriptsize
		\renewcommand{\arraystretch}{0.4}
		\begin{tabular*}{0.9\linewidth}{@{\extracolsep{\fill}}l|l|l}
			\toprule[1.5pt]\midrule[0.3pt]
			
			\begin{tabular}[t]{ll}
				
				& \textcolor{white}{1}\\
				\arrayrulecolor{white}\midrule
				\multicolumn{2}{c}{\textbf{Variable}} \\
				
				\arrayrulecolor{black}\midrule

				\textbf{EIU}\tnote{{\textcolor{white}{*}}} \\
				\textcolor{white}{()} \\
				\addlinespace[3pt]
				
				\textbf{ln(Population)}\tnote{{\textcolor{white}{*}}} \\
				\textcolor{white}{()} \\
				\addlinespace[2pt]
				
				\textbf{No. of Days}\tnote{{\textcolor{white}{*}}}  \\
				\textcolor{white}{()} \\
				\addlinespace[1.8pt]
				
				\textbf{Sample Size} \\
				\addlinespace[2.8pt]
				
				\textbf{Adj. R\textsuperscript{2}} \\
				\midrule

				\textbf{ln(GDP)}\tnote{{\textcolor{white}{*}}} \\
				\textcolor{white}{()} \\
				\addlinespace[2pt]
				
				\textbf{ln(Population)}\tnote{{\textcolor{white}{*}}} \\
				\textcolor{white}{()} \\
				\addlinespace[1.8pt]
				
				\textbf{No. of Days}\tnote{{\textcolor{white}{*}}}  \\
				\textcolor{white}{()} \\
				\addlinespace[1.8pt]
				
				\textbf{Sample Size} \\
				\addlinespace[2.8pt]
				
				\textbf{Adj. R\textsuperscript{2}} \\
				\midrule

				\textbf{HE\_GDP}\tnote{{\textcolor{white}{*}}} \\
				\textcolor{white}{()} \\
				\addlinespace[2pt]
				
				\textbf{ln(Population)}\tnote{{\textcolor{white}{*}}} \\
				\textcolor{white}{()} \\
				\addlinespace[1.8pt]
				
				\textbf{No. of Days}\tnote{{\textcolor{white}{*}}}  \\
				\textcolor{white}{()} \\
				\addlinespace[1.8pt]
				
				\textbf{Sample Size} \\
				\addlinespace[2.8pt]
				
				\textbf{Adj. R\textsuperscript{2}} \\
				\midrule

				\textbf{UHC}\tnote{{\textcolor{white}{*}}} \\
				\textcolor{white}{()} \\
				\addlinespace[3pt]
				
				\textbf{ln(Population)}\tnote{{\textcolor{white}{*}}} \\
				\textcolor{white}{()} \\
				\addlinespace[1.8pt]
				
				\textbf{No. of Days}\tnote{{\textcolor{white}{*}}}  \\
				\textcolor{white}{()} \\
				\addlinespace[1.8pt]
				
				\textbf{Sample Size} \\
				\addlinespace[2.8pt]
				
				\textbf{Adj. R\textsuperscript{2}} \\

			\end{tabular} &

			\begin{tabular}[t]
				{
					D{.}{.}{-1}
					D{.}{.}{-1}
					D{.}{.}{-1}
					D{.}{.}{-1}
				}

				\multicolumn{4}{c}{\textbf{Panel A. Confirmed Cases}}  \\
				
				\midrule
				\multicolumn{1}{c}{\textbf{Chi-squared}} & \multicolumn{1}{c}{\textbf{Kuiper}} & \multicolumn{1}{c}{\textbf{M}} & \multicolumn{1}{c}{\textbf{D}} \\

				\midrule
				
				-6.72 & -0.30\tnote{*} & -0.20 & -0.36\tnote{**} \\
				( 0.22) & ( 0.09) & ( 0.17) & ( 0.05) \\
				\addlinespace[3pt]
				
				-2.39\tnote{**} & -0.06\tnote{**} & -0.05\tnote{*} & -0.06\tnote{**} \\
				( 0.04) & ( 0.04) & ( 0.05) & ( 0.03) \\
				\addlinespace[3pt]
				
				-1.66 & -0.27\tnote{*} & -0.18 & -0.21 \\
				( 0.42) & ( 0.09) & ( 0.17) & ( 0.13) \\
				\addlinespace[3pt]
				
				\multicolumn{1}{c}{159} & \multicolumn{1}{c}{159} & \multicolumn{1}{c}{159} & \multicolumn{1}{c}{159} \\
				\addlinespace[3pt]
				1.02\% & 3.79\% & 2.09\% & 4.29\% \\

				\midrule
				-2.41\tnote{**} & -0.08\tnote{***} & -0.07\tnote{**} & -0.09\tnote{***} \\
				( 0.03) & ( 0.01) & ( 0.01) & ( 0.00) \\
				\addlinespace[3pt]
				
				-1.83\tnote{**} & -0.06\tnote{**} & -0.05\tnote{**} & -0.05\tnote{**} \\
				( 0.05) & ( 0.02) & ( 0.03) & ( 0.03) \\
				\addlinespace[3pt]
				
				0.13 & -0.23 & -0.13 & -0.14\tnote{{\textcolor{white}{*}}} \\
				( 0.49) & ( 0.12) & ( 0.22) & ( 0.23) \\
				\addlinespace[3pt]
				
				\multicolumn{1}{c}{170} & \multicolumn{1}{c}{170} & \multicolumn{1}{c}{170} & \multicolumn{1}{c}{170} \\
				\addlinespace[3pt]
				1.74\% & 5.82\% & 4.08\% & 5.51\% \\

				\midrule
				-0.21 & -0.02 & -0.01 & -0.01\tnote{{\textcolor{white}{*}}} \\
				( 0.39) & ( 0.10) & ( 0.22) & ( 0.26) \\
				\addlinespace[3pt]
				
				-1.34 & -0.04\tnote{*} & -0.03\tnote{*} & -0.03 \\
				( 0.11) & ( 0.08) & ( 0.10) & ( 0.11) \\
				\addlinespace[3pt]
				
				-1.46 & -0.29\tnote{*} & -0.19 & -0.21 \\
				( 0.42) & ( 0.07) & ( 0.14) & ( 0.13) \\
				\addlinespace[3pt]
				
				\multicolumn{1}{c}{167} & \multicolumn{1}{c}{167} & \multicolumn{1}{c}{167} & \multicolumn{1}{c}{167} \\
				\addlinespace[3pt]
				-0.30\% & 3.97\% & 1.94\% & 1.81\% \\

				\midrule
				-13.32 & -0.57\tnote{**} & -0.45\tnote{**} & -0.66\tnote{***} \\
				(  0.12) & ( 0.03) & ( 0.05) & ( 0.01) \\
				\addlinespace[3pt]
				
				-1.93\tnote{**} & -0.06\tnote{**} & -0.05\tnote{**} & -0.05\tnote{**} \\
				(  0.05) & ( 0.03) & ( 0.04) & ( 0.03) \\
				\addlinespace[3pt]
				
				-0.24 & -0.22 & -0.15 & -0.15\tnote{{\textcolor{white}{*}}} \\
				(  0.49) & ( 0.13) & ( 0.20) & ( 0.20) \\
				\addlinespace[3pt]
				
				\multicolumn{1}{c}{169} & \multicolumn{1}{c}{169} & \multicolumn{1}{c}{169} & \multicolumn{1}{c}{169} \\
				\addlinespace[3pt]
				1.01\% & 4.90\% & 3.37\% & 5.35\% \\

			\end{tabular} &
			
			\begin{tabular}[t]
				{
					D{.}{.}{-1}
					D{.}{.}{-1}
					D{.}{.}{-1}
					D{.}{.}{-1}
				}
				
				\multicolumn{4}{c}{\textbf{Panel B. Death Cases}}  \\
				
				\midrule
				\multicolumn{1}{c}{\textbf{Chi-squared}} & \multicolumn{1}{c}{\textbf{Kuiper}} & \multicolumn{1}{c}{\textbf{M}} & \multicolumn{1}{c}{\textbf{D}} \\				
				
				\midrule
				
				-17.49\tnote{*} & -0.65\tnote{**} & -0.53\tnote{**} & -0.64\tnote{***} \\
				(  0.06) & ( 0.01) & ( 0.03) & ( 0.01) \\
				\addlinespace[3pt]
				
				-2.15 & -0.08\tnote{**} & -0.06\tnote{*} & -0.07\tnote{**} \\
				(  0.12) & ( 0.04) & ( 0.08) & ( 0.05) \\
				\addlinespace[3pt]
				
				-6.39 & -0.52\tnote{**} & -0.31 & -0.31 \\
				(  0.27) & ( 0.02) & ( 0.11) & ( 0.12) \\
				\addlinespace[3pt]
				
				\multicolumn{1}{c}{113} & \multicolumn{1}{c}{113} & \multicolumn{1}{c}{113} & \multicolumn{1}{c}{113} \\
				\addlinespace[3pt]
				1.59\% & 10.24\% & 4.73\% & 6.88\% \\

				\midrule
				-2.48\tnote{**} & -0.11\tnote{***} & -0.09\tnote{***} & -0.10\tnote{***} \\
				( 0.02) & ( 0.00) & ( 0.00) & ( 0.00) \\
				\addlinespace[3pt]
				
				-2.50\tnote{**} & -0.10\tnote{***} & -0.08\tnote{***} & -0.09\tnote{***} \\
				( 0.02) & ( 0.00) & ( 0.00) & ( 0.00) \\
				\addlinespace[3pt]
				
				-9.65\tnote{*} & -0.60\tnote{***} & -0.40\tnote{**} & -0.37\tnote{**} \\
				( 0.10) & ( 0.00) & ( 0.03) & ( 0.03) \\
				\addlinespace[3pt]
				
				\multicolumn{1}{c}{116} & \multicolumn{1}{c}{116} & \multicolumn{1}{c}{116} & \multicolumn{1}{c}{116} \\
				\addlinespace[3pt]
				7.48\% & 20.22\% & 13.76\% & 15.66\% \\

				\midrule
				-2.14\tnote{***} & -0.07\tnote{***} & -0.06\tnote{***} & -0.07\tnote{***} \\
				( 0.01) & ( 0.00) & ( 0.00) & ( 0.00) \\
				\addlinespace[3pt]
				
				-1.98 & -0.08\tnote{**} & -0.07\tnote{**} & -0.07\tnote{**} \\
				( 0.10) & ( 0.02) & ( 0.04) & ( 0.03) \\
				\addlinespace[3pt]
				
				-8.13 & -0.59\tnote{**} & -0.38\tnote{*} & -0.35\tnote{*} \\
				( 0.21) & ( 0.01) & ( 0.07) & ( 0.08) \\
				\addlinespace[3pt]
				
				\multicolumn{1}{c}{115} & \multicolumn{1}{c}{115} & \multicolumn{1}{c}{115} & \multicolumn{1}{c}{115} \\
				\addlinespace[3pt]
				5.19\% & 15.88\% & 9.16\% & 11.63\% \\

				\midrule
				-5.42 & -0.57\tnote{*} & -0.41 & -0.51\tnote{*} \\
				( 0.35) & ( 0.06) & ( 0.13) & ( 0.08) \\
				\addlinespace[3pt]
				
				-2.47\tnote{*} & -0.10\tnote{**} & -0.09\tnote{**} & -0.09\tnote{**} \\
				( 0.07) & ( 0.01) & ( 0.02) & ( 0.02) \\
				\addlinespace[3pt]
				
				-3.98 & -0.45\tnote{**} & -0.25 & -0.22 \\
				( 0.35) & ( 0.04) & ( 0.16) & ( 0.20) \\
				\addlinespace[3pt]
				
				\multicolumn{1}{c}{117} & \multicolumn{1}{c}{117} & \multicolumn{1}{c}{117} & \multicolumn{1}{c}{117} \\
				\addlinespace[3pt]
				0.38\% & 10.32\% & 5.29\% & 5.89\% \\

			\end{tabular} \\
			
			\midrule[0.3pt]\bottomrule[1.5pt]
		\end{tabular*}

		\begin{tablenotes}
			\singlespacing
			\item[] The table presents the results of equation 6 using OLS for second-digit goodness-of-fit measures. Panel A shows the results for the cumulative number of confirmed cases, while panel B shows the results for the cumulative number of deaths. To avoid small coefficients, we divide EIU, UHC, and No. of Days values by 100 for all models. Sample sizes vary due to missing values. All models are estimated using OLS regression. \textit{P}-values for a one-tailed \textit{t}-test are in parentheses. ***, ** and * denote significance at the 1\%, 5\% and 10\% levels, respectively. All variable definitions are in Appendix A2. 
		\end{tablenotes}
		
	\end{threeparttable}
\end{sidewaystable}

\par
}

In unreported tests, we also combine our robustness checks, i.e.,
we conduct tests using regional data and second-digit tests for global
cutoff dates and during the 45 days since the first case for each
country. Our findings are not affected by choice of the test method
or time period.

\subsection{Swine Flu Pandemic of 2009-2010}

A natural extension to our study is to see if the negative relationship
between goodness-of-fit measures and economic indicators holds for
other pandemics. The unit of observation in our study is a country,
but pandemics that engulf many countries and for which data are available
are rare in modern history. One natural candidate is the recent swine
flu (H1N1) pandemic of 2009–2010. Swine flu (H1N1) 2009–2010 was a
pandemic that lasted over 19 months between January of 2009 and August
2010. The pandemic affected 58 countries, with tens of thousands (in
some estimates, millions and even hundreds of millions) of people
infected and tens of thousands (in some estimates, hundreds of thousands)
of deaths.

Even though the pandemic happened in relatively recent times and after
the advent of the Internet, surveillance data availability and reporting
was much more limited in 2009 than during COVID-19. Many countries
did not collect daily or even weekly data, reporting was limited,
and there was little public availability of data. As a result, only
a very small number of studies directly test the accuracy or manipulation
of data during the swine flu (H1N1) 2009–2010 pandemic (a notable
exception is the study by \cite{idrovo2011performance}). The WHO,
Pan American Health Organization (PAHO), and the Center for Disease
Control and Prevention (CDC) provide many \emph{estimates} for the
total number of cases and deaths, but these cannot be used with the
Newcomb-Benford test because the test gauges human intervention in
\emph{actual} reported data.

To apply the NBL test, we collect data for 35 countries in the Americas
that provided weekly reports of the number of confirmed cases and
deaths to the PAHO. We obtain the data for the weekly number of confirmed
cases and the distribution of first digits from \textcite{idrovo2011performance}.
The data for the weekly number of deaths is downloaded from the PAHO
website.\footnote{https://www.paho.org/hq/images/atlas/en/atlas.html?detectflash=false.}
We then repeat the analyses and re-estimate regression \ref{eq:1}
for swine flu (H1N1) 2009-2010 data. For the economic indicators,
we use 2009 values. The results are reported in Table 11. Panel A
depicts the results for the number of confirmed cases. Out of 16,
12 coefficients in front of macroeconomic indicators are negative.
It should be noted that the sample size for this test is extremely
small, with at most 35 countries. Obtaining significant results with
such small sample sizes is challenging. Yet, we are able to obtain
significant coefficients for five coefficients, and two more just
barely lack significance. Panel B illustrates the results for the
number of deaths. The sample size for Panel B is even smaller: 14
out of 16 coefficients are negative, and seven are significant. We
conclude that the swine flu (H1N1) 2009–2010 results are largely consistent
with our findings for the COVID-19 pandemic.

{\centering
\begin{sidewaystable}
	
	\centering	
	
	\begin{threeparttable}[h]
		
		\def\sym#1{\ifmmode^{#1}\else\(^{#1}\)\fi}
		
		\caption{Swine Flu Pandemic 2009-2010}\label{tab_res:swineflu} 
		\scriptsize
		\renewcommand{\arraystretch}{0.4}
		\begin{tabular*}{0.9\linewidth}{@{\extracolsep{\fill}}l|l|l}
			\toprule[1.5pt]\midrule[0.3pt]
			
			\begin{tabular}[t]{ll}
				
				& \textcolor{white}{1}\\
				\arrayrulecolor{white}\midrule
				\multicolumn{2}{c}{\textbf{Variable}} \\
				
				\arrayrulecolor{black}\midrule

				\textbf{EIU}\tnote{{\textcolor{white}{*}}} \\
				\textcolor{white}{()} \\
				\addlinespace[3pt]
				
				\textbf{ln(Population)}\tnote{{\textcolor{white}{*}}} \\
				\textcolor{white}{()} \\
				\addlinespace[2pt]
				
				\textbf{No. of Days}\tnote{{\textcolor{white}{*}}}  \\
				\textcolor{white}{()} \\
				\addlinespace[1.8pt]
				
				\textbf{Sample Size} \\
				\addlinespace[2.8pt]
				
				\textbf{Adj. R\textsuperscript{2}} \\
				\midrule

				\textbf{ln(GDP)}\tnote{{\textcolor{white}{*}}} \\
				\textcolor{white}{()} \\
				\addlinespace[2pt]
				
				\textbf{ln(Population)}\tnote{{\textcolor{white}{*}}} \\
				\textcolor{white}{()} \\
				\addlinespace[1.8pt]
				
				\textbf{No. of Days}\tnote{{\textcolor{white}{*}}}  \\
				\textcolor{white}{()} \\
				\addlinespace[1.8pt]
				
				\textbf{Sample Size} \\
				\addlinespace[2.8pt]
				
				\textbf{Adj. R\textsuperscript{2}} \\
				\midrule

				\textbf{HE\_GDP}\tnote{{\textcolor{white}{*}}} \\
				\textcolor{white}{()} \\
				\addlinespace[2pt]
				
				\textbf{ln(Population)}\tnote{{\textcolor{white}{*}}} \\
				\textcolor{white}{()} \\
				\addlinespace[1.8pt]
				
				\textbf{No. of Days}\tnote{{\textcolor{white}{*}}}  \\
				\textcolor{white}{()} \\
				\addlinespace[1.8pt]
				
				\textbf{Sample Size} \\
				\addlinespace[2.8pt]
				
				\textbf{Adj. R\textsuperscript{2}} \\
				\midrule

				\textbf{UHC}\tnote{{\textcolor{white}{*}}} \\
				\textcolor{white}{()} \\
				\addlinespace[3pt]
				
				\textbf{ln(Population)}\tnote{{\textcolor{white}{*}}} \\
				\textcolor{white}{()} \\
				\addlinespace[1.8pt]
				
				\textbf{No. of Days}\tnote{{\textcolor{white}{*}}}  \\
				\textcolor{white}{()} \\
				\addlinespace[1.8pt]
				
				\textbf{Sample Size} \\
				\addlinespace[2.8pt]
				
				\textbf{Adj. R\textsuperscript{2}} \\

			\end{tabular} &

			\begin{tabular}[t]
				{
					D{.}{.}{-1}
					D{.}{.}{-1}
					D{.}{.}{-1}
					D{.}{.}{-1}
				}

				\multicolumn{4}{c}{\textbf{Panel A. Confirmed Cases}}  \\
				
				\midrule
				\multicolumn{1}{c}{\textbf{Chi-squared}} & \multicolumn{1}{c}{\textbf{Kuiper}} & \multicolumn{1}{c}{\textbf{M}} & \multicolumn{1}{c}{\textbf{D}} \\

				\midrule
				
				-0.21\tnote{**} &  0.00 & -0.01 &  0.00 \\
				( 0.03) & ( 0.43) & ( 0.11) & ( 0.46) \\
				\addlinespace[3pt]
				
				-2.55\tnote{*} &  0.05 & -0.08 &  0.02 \\
				( 0.07) & ( 0.35) & ( 0.18) & ( 0.37) \\
				\addlinespace[3pt]
				
				3.53\tnote{***} & -0.02 &  0.11\tnote{*} & -0.02 \\
				( 0.01) & ( 0.44) & ( 0.07) & ( 0.32) \\
				\addlinespace[3pt]
				
				\multicolumn{1}{c}{26} & \multicolumn{1}{c}{26} & \multicolumn{1}{c}{26} & \multicolumn{1}{c}{26} \\
				\addlinespace[3pt]
				
				21.51\% & -11.47\% & 1.32\% & -11.01\% \\

				\midrule
				-2.18\tnote{**} & -0.05 & -0.03 & -0.02 \\
				( 0.02) & ( 0.25) & ( 0.32) & ( 0.32) \\
				\addlinespace[3pt]
				
				-1.19 &  0.01 & -0.04 & -0.01\tnote{{\textcolor{white}{*}}} \\
				( 0.13) & ( 0.44) & ( 0.27) & ( 0.39) \\
				\addlinespace[3pt]
				
				2.47\tnote{***} & -0.01 &  0.04 & -0.02 \\
				( 0.01) & ( 0.45) & ( 0.27) & ( 0.34) \\
				\addlinespace[3pt]
				
				\multicolumn{1}{c}{35} & \multicolumn{1}{c}{35} & \multicolumn{1}{c}{35} & \multicolumn{1}{c}{35} \\
				\addlinespace[3pt]
				35.78\% & -7.44\% & -8.02\% & 3.65\% \\

				\midrule
				-0.56 & -0.05\tnote{*} & -0.04\tnote{*} & -0.02 \\
				( 0.11) & ( 0.08) & ( 0.09) & ( 0.20) \\
				\addlinespace[3pt]
				
				-0.25 &  0.04 & -0.02 &  0.00\tnote{{\textcolor{white}{*}}} \\
				( 0.40) & ( 0.28) & ( 0.37) & ( 0.49) \\
				\addlinespace[3pt]
				
				1.90\tnote{**} & -0.01 &  0.04 & -0.02 \\
				( 0.03) & ( 0.42) & ( 0.25) & ( 0.31) \\
				\addlinespace[3pt]
				
				\multicolumn{1}{c}{35} & \multicolumn{1}{c}{35} & \multicolumn{1}{c}{35} & \multicolumn{1}{c}{35} \\
				\addlinespace[3pt]
				29.82\% & -2.05\% & -2.79\% & 5.13\% \\

				\midrule
				-0.09 & -0.01\tnote{*} &  0.00 &  0.00 \\
				( 0.27) & ( 0.10) & ( 0.30) & ( 0.24) \\
				\addlinespace[3pt]
				
				-0.78 & -0.02 & -0.07 & -0.03\tnote{{\textcolor{white}{*}}} \\
				( 0.26) & ( 0.40) & ( 0.18) & ( 0.26) \\
				\addlinespace[3pt]
				
				2.25\tnote{**} &  0.03 &  0.07 &  0.00 \\
				( 0.03) & ( 0.36) & ( 0.17) & ( 0.50) \\
				\addlinespace[3pt]
				
				\multicolumn{1}{c}{33} & \multicolumn{1}{c}{33} & \multicolumn{1}{c}{33} & \multicolumn{1}{c}{33} \\
				\addlinespace[3pt]
				28.35\% & -3.93\% & -6.77\% & 4.94\% \\

			\end{tabular} &
			
			\begin{tabular}[t]
				{
					D{.}{.}{-1}
					D{.}{.}{-1}
					D{.}{.}{-1}
					D{.}{.}{-1}
				}

				\multicolumn{4}{c}{\textbf{Panel B. Death Cases}}  \\
				
				\midrule
				\multicolumn{1}{c}{\textbf{Chi-squared}} & \multicolumn{1}{c}{\textbf{Kuiper}} & \multicolumn{1}{c}{\textbf{M}} & \multicolumn{1}{c}{\textbf{D}} \\

				\midrule

				-0.25\tnote{**} & -0.01 &  0.00 & -0.01 \\
				( 0.04) & ( 0.10) & ( 0.35) & ( 0.17) \\
				\addlinespace[3pt]
				
				-2.89\tnote{**} & -0.23\tnote{***} & -0.13\tnote{**} & -0.13\tnote{**} \\
				( 0.02) & ( 0.00) & ( 0.02) & ( 0.02) \\
				\addlinespace[3pt]
				
				1.01\tnote{***} &  0.05\tnote{***} &  0.03\tnote{***} &  0.03\tnote{***} \\
				( 0.00) & ( 0.00) & ( 0.01) & ( 0.00) \\
				\addlinespace[3pt]
				
				\multicolumn{1}{c}{23} & \multicolumn{1}{c}{23} & \multicolumn{1}{c}{23} & \multicolumn{1}{c}{23} \\
				\addlinespace[3pt]
				47.95\% & 45.96\% & 19.61\% & 33.05\% \\

				\midrule
				-1.97 & -0.16\tnote{**} & -0.06 & -0.12\tnote{*} \\
				( 0.14) & ( 0.03) & ( 0.19) & ( 0.05) \\
				\addlinespace[3pt]
				
				-1.56\tnote{*} & -0.16\tnote{***} & -0.08\tnote{**} & -0.07\tnote{**} \\
				( 0.08) & ( 0.00) & ( 0.03) & ( 0.04) \\
				\addlinespace[3pt]
				
				0.93\tnote{***} &  0.05\tnote{***} &  0.03\tnote{***} &  0.03\tnote{***} \\
				( 0.00) & ( 0.00) & ( 0.00) & ( 0.00) \\
				\addlinespace[3pt]
				
				\multicolumn{1}{c}{26} & \multicolumn{1}{c}{26} & \multicolumn{1}{c}{26} & \multicolumn{1}{c}{26} \\
				\addlinespace[3pt]
				41.60\% & 43.23\% & 16.95\% & 34.09\% \\

				\midrule
				0.04 & -0.05\tnote{*} &  0.01 & -0.01 \\
				( 0.48) & ( 0.09) & ( 0.38) & ( 0.41) \\
				\addlinespace[3pt]
				
				-1.33 & -0.13\tnote{***} & -0.08\tnote{**} & -0.06\tnote{*} \\
				( 0.12) & ( 0.01) & ( 0.04) & ( 0.10) \\
				\addlinespace[3pt]
				
				0.86\tnote{***} &  0.05\tnote{***} &  0.02\tnote{**} &  0.03\tnote{***} \\
				( 0.00) & ( 0.00) & ( 0.01) & ( 0.00) \\
				\addlinespace[3pt]
				
				\multicolumn{1}{c}{26} & \multicolumn{1}{c}{26} & \multicolumn{1}{c}{26} & \multicolumn{1}{c}{26} \\
				\addlinespace[3pt]
				38.29\% & 37.76\% & 14.38\% & 25.50\% \\

				\midrule
				-0.04 & -0.03\tnote{***} & -0.02\tnote{**} & -0.02\tnote{**} \\
				( 0.45) & ( 0.01) & ( 0.03) & ( 0.04) \\
				\addlinespace[3pt]
				
				-1.93\tnote{*} & -0.20\tnote{***} & -0.11\tnote{**} & -0.10\tnote{**} \\
				( 0.07) & ( 0.00) & ( 0.01) & ( 0.02) \\
				\addlinespace[3pt]
				
				0.93\tnote{***} &  0.06\tnote{***} &  0.03\tnote{***} &  0.04\tnote{***} \\
				( 0.00) & ( 0.00) & ( 0.00) & ( 0.00) \\
				\addlinespace[3pt]
				
				\multicolumn{1}{c}{25} & \multicolumn{1}{c}{25} & \multicolumn{1}{c}{25} & \multicolumn{1}{c}{25} \\
				\addlinespace[3pt]
				38.94\% & 52.47\% & 29.49\% & 37.89\% \\

			\end{tabular} \\
			
			\midrule[0.3pt]\bottomrule[1.5pt]
		\end{tabular*}

		\begin{tablenotes}[flushleft]
			\singlespacing
			\item[] The table presents the results of 2009-2010 Swine Flu Pandemic analysis for 35 PAHO countries. We estimate equation 6 using OLS for first-digit goodness-of-fit measures. Panel A shows the results for the cumulative number of confirmed cases, while panel B shows the results for the cumulative number of deaths. To avoid small coefficients, we divide EIU, UHC, and No. of Days values by 100 for all models. Sample sizes vary due to missing values. \textit{P}-values for a one-tailed \textit{t}-test are in parentheses. ***, ** and * denote significance at the 1\%, 5\% and 10\% levels, respectively. All variable definitions are in Appendix A2. 
		\end{tablenotes}
		
	\end{threeparttable}
\end{sidewaystable}
\par
}

\section{Discussion and Conclusion}

In this paper, we investigate the relationship between the accuracy
of reported data and macroeconomic indicators for a set of 185 countries
affected by the COVID-19 pandemic. We use the deviation from the Newcomb-Benford
law of anomalous numbers as a proxy for data manipulation. For approximately
one-third of countries, we document some evidence of data manipulation,
especially for the death toll. We find the negative relationship between
the four NBL goodness-of-fit measures and four economic indicators.
We also find that the relationship is stronger for the number of deaths
than for the number of confirmed cases. Overall, we conclude that
democratic regimes and more economically developed countries, and
countries with stronger healthcare systems, provide more accurate
data during pandemics. Authoritarian regimes and poorer countries,
on the other hand, are more likely to manipulate data, specifically
the death toll. We do not believe that our results are driven by noise
in the data or the specific method used because they are robust to
alternative testing periods, and are not driven by small countries,
countries with a small number of cases, or extreme deviations from
the NBL. We also show that the relationship holds for 50 countries
that report regional data, for second-digit tests, and for the previous
swine flu (H1N1) pandemic.

The interpretations of our findings assume that deviations from the
Newcomb-Benford law are indicative of data manipulation. Indeed, many
studies in macroeconomic, accounting, finance, and forensic analysis
demonstrate that human intervention and data manipulation create data
sets that violate the NBL. Many naturally occurring processes, on
the other hand, generate data that obey the law. This makes the NBL
a useful tool to detect data manipulation. However, several limitations
to our study should be mentioned. Although we use compliance of data
with the NBL as a proxy for non-manipulation of data, alternative
interpretations of our findings are possible. The test works only
if the data are expected to obey the NBL. Several indicators suggest
that the pandemic data are a good candidate to test with the NBL.
In addition, we use careful techniques to separate the “growth” part
from the data. However, if the pandemic data are not supposed to follow
the Newcomb-Benford law, then the observed relationship can be explained
by the expected deviation from the law based on other factors, like
sample size and the span of the data. We control for these effects
in our tests.

The aim of this paper is not to provide evidence whether a particular
country manipulates data. Such claims require precise estimations
of the goodness-of-fit measure and clear evidence that the country’s
expected distribution is indeed the NBL. The number of days since
the beginning of the pandemic and before a country reaches a plateau
(when the NBL is no longer applicable) is small. Therefore, the goodness-of-fit
measures are estimated with error and conclusions for individual countries
are difficult to state with utmost certainty. In contrast, this study
documents a general relationship between macroeconomic indicators
of countries and their tendency to report inaccurate data. The paper
leads to a question about whether falsifying data during pandemics
is a short-lived strategy for governments. Does it have immediate
payback or is it sustainable over the long run? We should also note
that even though the Newcomb-Benford test is less sensitive to noise
in the data, there is still some chance that the divergence from the
expected distribution is not due to the deliberate supply of falsified
of data but the low quality or structural breaks in the data.

Our paper highlights the importance of independent projects to verify
data supplied by the governments. Further research is needed that
would combine different methods that test for data manipulation, including
the Newcomb-Benford law, biostatistics, moments of distributions,
excess mortality rates, and social media data. Even more important
is research related to methods that can prevent data manipulation
and fraud during pandemics.

\singlespacing 
\printbibliography

@Article{adsera2003you,
  author    = {Adsera, Alicia and Boix, Carles and Payne, Mark},
  journal   = {The Journal of Law, Economics, and Organization},
  title     = {Are you being served? {P}olitical accountability and quality of government},
  year      = {2003},
  number    = {2},
  pages     = {445--490},
  volume    = {19},
  publisher = {Oxford University Press},
}

@Misc{alwine2020manipulation,
  author       = {Alwine, James and Goodrum Sterling, Felicia},
  howpublished = {\url{https://thehill.com/opinion/healthcare/499535-manipulation-of-pandemic-numbers-for-politics-risks-lives}},
  note         = {Accessed: 2020-09-16},
  title        = {Manipulation of pandemic numbers for politics risks lives},
  year         = {2020},
}

@article{benford1938law,
	title={The law of anomalous numbers},
	author={Benford, Frank},
	journal={Proceedings of the American philosophical society},
	pages={551--572},
	year={1938},
	publisher={JSTOR}
}

@Article{breunig2011searching,
  author    = {Breunig, Christian and Goerres, Achim},
  journal   = {Electoral studies},
  title     = {Searching for electoral irregularities in an established democracy: {A}pplying {B}enford's law tests to {B}undestag elections in {U}nified {G}ermany},
  year      = {2011},
  number    = {3},
  pages     = {534--545},
  volume    = {30},
  publisher = {Elsevier},
}

@Article{broz2002political,
  author    = {Broz, Lawrence J.},
  journal   = {International Organization},
  title     = {Political system transparency and monetary commitment regimes},
  year      = {2002},
  pages     = {861--887},
  publisher = {JSTOR},
}

@Article{cantu2010supervised,
  author  = {Cantu, Francisco and Saiegh, Sebastian M.},
  journal = {Available at SSRN 1594406},
  title   = {A supervised machine learning procedure to detect electoral fraud using digital analysis},
  year    = {2010},
}

@Article{deckert2011benford,
  author    = {Deckert, Joseph and Myagkov, Mikhail and Ordeshook, Peter C.},
  journal   = {Political Analysis},
  title     = {Benford's law and the detection of election fraud},
  year      = {2011},
  number    = {3},
  pages     = {245--268},
  volume    = {19},
  publisher = {Cambridge University Press},
}

@Book{demir2018forensics,
  author    = {Demir, Banu and Javorcik, Beata K. Smarzynska},
  publisher = {Centre for Economic Policy Research},
  title     = {Forensics, elasticities and benford's law: detecting tax fraud in international trade},
  year      = {2018},
}

@Article{diekmann2007not,
  author    = {Diekmann, Andreas},
  journal   = {Journal of Applied Statistics},
  title     = {Not the first digit! {U}sing benford's law to detect fraudulent scientific data},
  year      = {2007},
  number    = {3},
  pages     = {321--329},
  volume    = {34},
  publisher = {Taylor \& Francis},
}

@article{djankov2003owns,
	title={Who owns the media?},
	author={Djankov, Simeon and McLiesh, Caralee and Nenova, Tatiana and Shleifer, Andrei},
	journal={The Journal of Law and Economics},
	volume={46},
	number={2},
	pages={341--382},
	year={2003},
	publisher={The University of Chicago Press}
}

@Misc{dragan2020,
  author       = {Dragan, Alexander},
  howpublished = {Https://medium.com/},
  note         = {Accessed: 2020-09-16},
  title        = {Kak uvidet jepidemiju, esli ejo staratelno prjachut. {O}pyt pjati rossijskih regionov. (in {R}ussian).},
  year         = {2020},
}

@Article{durtschi2004effective,
  author  = {Durtschi, Cindy and Hillison, William and Pacini, Carl},
  journal = {Journal of forensic accounting},
  title   = {The effective use of {B}enford’s law to assist in detecting fraud in accounting data},
  year    = {2004},
  number  = {1},
  pages   = {17--34},
  volume  = {5},
}

@Article{egorov2009resource,
  author    = {Egorov, Georgy and Guriev, Sergei and Sonin, Konstantin},
  journal   = {American political science Review},
  title     = {Why resource-poor dictators allow freer media: {A} theory and evidence from panel data},
  year      = {2009},
  pages     = {645--668},
  publisher = {JSTOR},
}

@Article{fearon2011self,
  author    = {Fearon, James D.},
  journal   = {The Quarterly Journal of Economics},
  title     = {Self-enforcing democracy},
  year      = {2011},
  number    = {4},
  pages     = {1661--1708},
  volume    = {126},
  publisher = {Oxford University Press},
}

@Article{formann2010newcomb,
  author    = {Formann, Anton K.},
  journal   = {PloS one},
  title     = {The {N}ewcomb-{B}enford law in its relation to some common distributions},
  year      = {2010},
  number    = {5},
  pages     = {e10541},
  volume    = {5},
  publisher = {Public Library of Science},
}

@article{gehlbach2014government,
	title={Government control of the media},
	author={Gehlbach, Scott and Sonin, Konstantin},
	journal={Journal of Public Economics},
	volume={118},
	pages={163--171},
	year={2014},
	publisher={Elsevier}
}

@Article{geyer2004detecting,
  author    = {Geyer, Christina Lynn and Williamson, Patricia Pepple},
  journal   = {Communications in Statistics-Simulation and Computation},
  title     = {Detecting fraud in data sets using {B}enford's law},
  year      = {2004},
  number    = {1},
  pages     = {229--246},
  volume    = {33},
  publisher = {Taylor \& Francis},
}

@Article{giles2007benford,
  author    = {Giles, David E.},
  journal   = {Applied Economics Letters},
  title     = {Benford's law and naturally occurring prices in certain ebay auctions},
  year      = {2007},
  number    = {3},
  pages     = {157--161},
  volume    = {14},
  publisher = {Taylor \& Francis},
}

@Article{gomez2016monitoring,
  author    = {G{\'o}mez-Camponovo, Mariana and Moreno, Jos{\'e} and Idrovo, {\'A}lvaro J. and P{\'a}ez, Malvina and Achkar, Marcel},
  journal   = {Biom{\'e}dica},
  title     = {Monitoring the {P}araguayan epidemiological dengue surveillance system (2009-2011) using {B}enford's law},
  year      = {2016},
  number    = {4},
  pages     = {583--592},
  volume    = {36},
  publisher = {Instituto Nacional de Salud},
}

@book{gonzalez2009benford,
	title={Benford's law and macroeconomic data quality},
	author={Gonzalez-Garcia, Jesus},
	number={2009-2010},
	year={2009},
	publisher={International Monetary Fund}
}

@Article{goutte2020macroeconomic,
  author  = {Goutte, Stephane and Damette, Olivier},
  journal = {Available at SSRN 3610417},
  title   = {The macroeconomic determinants of {COVID19} mortality rate and the role of post subprime crisis decisions},
  year    = {2020},
}

@article{guriev2019informational,
	title={Informational autocrats},
	author={Guriev, Sergei and Treisman, Daniel},
	journal={Journal of Economic Perspectives},
	volume={33},
	number={4},
	pages={100--127},
	year={2019}
}

@Article{hill1995statistical,
  author    = {Hill, Theodore P. and others},
  journal   = {Statistical science},
  title     = {A statistical derivation of the significant-digit law},
  year      = {1995},
  number    = {4},
  pages     = {354--363},
  volume    = {10},
  publisher = {Institute of Mathematical Statistics},
}

@Article{hill1998first,
  author    = {Hill, Theodore P.},
  journal   = {American Scientist},
  title     = {The first digit phenomenon: {A} century-old observation about an unexpected pattern in many numerical tables applies to the stock market, census statistics and accounting data},
  year      = {1998},
  number    = {4},
  pages     = {358--363},
  volume    = {86},
  publisher = {JSTOR},
}

@Article{horton2018detecting,
  author  = {Horton, Joanne and Krishnakumar, Dhanya and Wood, Anthony},
  journal = {Available at SSRN 3164961},
  title   = {Detecting academic fraud in accounting research: the case of {P}rofessor {J}ames {H}unton},
  year    = {2018},
}

@Article{hussain2010application,
  author  = {Hussain, Syed Azfar},
  journal = {Available at SSRN 1626696},
  title   = {The application of {B}enford's law in forensic accounting: an analysis of credit bureau data},
  year    = {2010},
}

@Article{idrovo2011performance,
  author    = {Idrovo, Alvaro J. and Fern{\'a}ndez-Ni{\~n}o, A. and Boj{\'o}rquez-Chapela, I. and Moreno-Montoya, A.},
  journal   = {Epidemiology and Infection},
  title     = {Performance of public health surveillance systems during the influenza {A} {(H1N1)} pandemic in the {A}mericas: testing a new method based on {B}enford's law},
  year      = {2011},
  number    = {12},
  pages     = {1827--1834},
  volume    = {139},
  publisher = {Cambridge University Press},
}

@Article{idrovo2020data,
  author    = {Idrovo, Alvaro J. and Manrique-Hern{\'a}ndez, Edgar Fabi{\'a}n},
  journal   = {Asia-Pacific Journal of Public Health},
  title     = {Data quality of {C}hinese surveillance of {COVID-19}: {O}bjective analysis based on {WHO}’s situation reports},
  year      = {2020},
  publisher = {SAGE Publications},
}

@Article{islam2006does,
  author    = {Islam, Roumeen},
  journal   = {Economics and Politics},
  title     = {Does more transparency go along with better governance?},
  year      = {2006},
  number    = {2},
  pages     = {121--167},
  volume    = {18},
  publisher = {Wiley Online Library},
}

@Article{judge2009detecting,
  author    = {Judge, George and Schechter, Laura},
  journal   = {Journal of Human Resources},
  title     = {Detecting problems in survey data using {B}enford’s law},
  year      = {2009},
  number    = {1},
  pages     = {1--24},
  volume    = {44},
  publisher = {University of Wisconsin Press},
}

@Article{kaiser2019benford,
  author    = {Kaiser, Micha},
  journal   = {Journal of Economic Surveys},
  title     = {Benford's law as an indicator of survey reliability—{C}an we trust our data?},
  year      = {2019},
  number    = {5},
  pages     = {1602--1618},
  volume    = {33},
  publisher = {Wiley Online Library},
}

@Article{kalaichelvan2012critical,
  author  = {Kalaichelvan, Mohandass and Shawn, Lim Kai Jai},
  journal = {International Research Journal of Finance and Economics},
  title   = {A critical evaluation of the significance of round numbers in major {E}uropean stock indices in light of the predictions from {B}enford’s law},
  year    = {2012},
  number  = {95},
  pages   = {196--210},
}

@Article{koch2020benford,
  author  = {Koch, Christoffer and Okamura, Ken},
  journal = {Available at SSRN 3586413},
  title   = {Benford's law and {COVID}-19 reporting},
  year    = {2020},
}

@InProceedings{kuiper1960tests,
  author    = {Kuiper, Nicolaas H.},
  booktitle = {Nederl. Akad. Wetensch. Proc. Ser. A},
  title     = {Tests concerning random points on a circle},
  year      = {1960},
  number    = {1},
  pages     = {38--47},
  volume    = {63},
}

@Article{lebovic2006democracies,
  author    = {Lebovic, James H.},
  journal   = {Journal of Peace Research},
  title     = {Democracies and transparency: country reports to the {UN} {R}egister of {C}onventional {A}rms, 1992-2001},
  year      = {2006},
  number    = {5},
  pages     = {543--562},
  volume    = {43},
  publisher = {Sage Publications Sage CA: Thousand Oaks, CA},
}

@Article{leemis2000survival,
  author    = {Leemis, Lawrence M. and Schmeiser, Bruce W. and Evans, Diane L.},
  journal   = {The American Statistician},
  title     = {Survival distributions satisfying {B}enford's law},
  year      = {2000},
  number    = {4},
  pages     = {236--241},
  volume    = {54},
  publisher = {Taylor \& Francis Group},
}

@Article{magee2015reconsidering,
  author    = {Magee, Christopher S. P. and Doces, John A.},
  journal   = {International Studies Quarterly},
  title     = {Reconsidering regime type and growth: lies, dictatorships, and statistics},
  year      = {2015},
  number    = {2},
  pages     = {223--237},
  volume    = {59},
  publisher = {Blackwell Publishing Ltd Oxford, UK},
}

@InProceedings{mebane2006election,
  author    = {Mebaner, Walter R. Jr.},
  booktitle = {Summer Meeting of the Political Methodology Society, UC-Davis, July},
  title     = {Election forensics: {V}ote counts and {B}enford’s law},
  year      = {2006},
  pages     = {20--22},
}

@Misc{meyer2020experts,
  author       = {Meyer, Henry},
  howpublished = {\url{https://www.bloomberg.com/news/articles/2020-05-13/experts-question-russian-data-on-covid-19-death-toll}},
  note         = {Accessed: 2020-09-16},
  title        = {Experts question {R}ussian data on {C}ovid-19 death toll},
  year         = {2020},
}

@Article{michalski2013countries,
  author    = {Michalski, Tomasz and Stoltz, Gilles},
  journal   = {Review of Economics and Statistics},
  title     = {Do countries falsify economic data strategically? {S}ome evidence that they might},
  year      = {2013},
  number    = {2},
  pages     = {591--616},
  volume    = {95},
  publisher = {MIT Press},
}

@Article{mitchell1998sources,
  author    = {Mitchell, Ronald B.},
  journal   = {International Studies Quarterly},
  title     = {Sources of transparency: {I}nformation systems in international regimes},
  year      = {1998},
  number    = {1},
  pages     = {109--130},
  volume    = {42},
  publisher = {Wiley Online Library},
}

@Article{morrow2010,
  author  = {Morrow, J.},
  journal = {Working Paper},
  title   = {Benford's law, families of distributions and a test basis},
  year    = {2020},
}

@Article{nigrini1996taxpayer,
  author  = {Nigrini, Mark J.},
  journal = {The Journal of the American Taxation Association},
  title   = {A taxpayer compliance application of {B}enford's law},
  year    = {1996},
  number  = {1},
  pages   = {72},
  volume  = {18},
}

@Book{nigrini2012benford,
  author    = {Nigrini, Mark J.},
  publisher = {John Wiley and Sons},
  title     = {Benford's law: {A}pplications for forensic accounting, auditing, and fraud detection},
  year      = {2012},
  volume    = {586},
}

@Article{nye2007political,
  author    = {Nye, John and Moul, Charles},
  journal   = {The BE Journal of Macroeconomics},
  title     = {The political economy of numbers: on the application of {B}enford's law to international macroeconomic statistics},
  year      = {2007},
  number    = {1},
  volume    = {7},
  publisher = {De Gruyter},
}

@Article{newcomb1881note,
  author  = {Newcomb, Simon},
  journal = {American Journal of Mathematics},
  title   = {Note on the frequency of use of the different digits in natural numbers},
  year    = {1881},
  number  = {1},
  pages   = {39--40},
  volume  = {4},
}

@Article{o2017offsite,
  author  = {O'Keefe, John P. and Yom, Chiwon},
  journal = {Available at SSRN 3013174},
  title   = {Offsite detection of insider abuse and bank fraud among {US} failed banks 1989-2015},
  year    = {2017},
}

@Misc{peng2020statistical,
  author       = {Peng, Yaohao and Nagata, Mateus H.},
  howpublished = {\url{https://lamfo-unb.github.io/2020/04/21/COVID-China-EN/}},
  note         = {Accessed: 2020-09-16},
  title        = {Statistical analysis of the {C}hinese {COVID}-19 data with {B}enford's law and clustering},
  year         = {2020},
}

@Article{pinilla2018pinocchio,
  author    = {Pinilla, Jaime and L{\'o}pez-Valc{\'a}rcel, Beatriz G. and Gonz{\'a}lez-Martel, Christian and Peiro, Salvador},
  journal   = {BMJ open},
  title     = {Pinocchio testing in the forensic analysis of waiting lists: using public waiting list data from {F}inland and {S}pain for testing {N}ewcomb-{B}enford’s law},
  year      = {2018},
  number    = {5},
  pages     = {e022079},
  volume    = {8},
  publisher = {British Medical Journal Publishing Group},
}

@Misc{polson2020manipulated,
  author       = {Polson, Don},
  howpublished = {\url{https://www.redbluffdailynews.com/2020/05/04/manipulated-agenda-driven-data/}},
  note         = {Accessed: 2020-09-16},
  title        = {Manipulated, agenda-driven data},
  year         = {2020},
}

@Article{rauch2011fact,
  author    = {Rauch, Bernhard and G{\"o}ttsche, Max and Engel, Stefan and Br{\"a}hler, Gernot},
  journal   = {German Economic Review},
  title     = {Fact and fiction in {EU}-governmental economic data},
  year      = {2011},
  number    = {3},
  pages     = {243--255},
  volume    = {12},
  publisher = {De Gruyter},
}

@Article{rauch2013libor,
  author  = {Rauch, Bernhard and Goettsche, Max and El Mouaaouy, Florian},
  journal = {Available at SSRN 2363895},
  title   = {{LIBOR} manipulation--empirical analysis of financial market benchmarks using {B}enford's law},
  year    = {2013},
}

@Misc{romaniuk2020can,
  author       = {Romaniuk, Scott N. and Burgers, Tobias},
  howpublished = {\url{https://thediplomat.com/2020/03/can-chinas-covid-19-statistics-be-trusted/}},
  note         = {Accessed: 2020-09-16},
  title        = {Can {C}hina’s {COVID}-19 statistics be trusted?},
  year         = {2020},
}

@Misc{sassoon2020florida,
  author       = {Sassoon, Alessandro M.},
  howpublished = {\url{https://www.usatoday.com/story/news/nation/2020/05/19/florida-covid-19-coronavirus-data-researcher-out-state-reopens/5218897002/}},
  note         = {Accessed: 2020-09-16},
  title        = {Florida's scientist was fired for refusing to 'manipulate' {COVID}-19 data},
  year         = {2020},
}

@Misc{speak2020what,
  author       = {Speak, Clare},
  howpublished = {\url{https://www.thelocal.it/20200403/whats-the-problem-with-italys-official-coronavirus-statistics}},
  note         = {Accessed: 2020-09-16},
  title        = {What's the problem with {I}taly's official coronavirus numbers?},
  year         = {2020},
}

@Article{stambaugh2012using,
  author  = {Stambaugh, Clyde and Tipgos, Manuel A. and Carpenter, Floyd and Smith, Murphy},
  journal = {Internal auditing},
  title   = {Using {B}enford analysis to detect fraud},
  year    = {2012},
  number  = {3},
  pages   = {24--29},
  volume  = {27},
}

@Article{stephens1970use,
  author    = {Stephens, Michael A.},
  journal   = {Journal of the Royal Statistical Society: Series B (Methodological)},
  title     = {Use of the {K}olmogorov--{S}mirnov, {C}ramer--{V}on {M}ises and related statistics without extensive tables},
  year      = {1970},
  number    = {1},
  pages     = {115--122},
  volume    = {32},
  publisher = {Wiley Online Library},
}

@Article{suh2011effective,
  author  = {Suh, Ikseon and Headrick, T. Christopher and Minaburo, Sandra},
  journal = {Journal of Forensic and Investigative Accounting},
  title   = {An effective and efficient analytic technique: a bootstrap regression procedure and {B}enford's law},
  year    = {2011},
}

@Article{tam2007breaking,
  author    = {Cho, Wendy K. Tam and Gaines, Brian J.},
  journal   = {The american statistician},
  title     = {Breaking the ({B}enford) law: {S}tatistical fraud detection in campaign finance},
  year      = {2007},
  number    = {3},
  pages     = {218--223},
  volume    = {61},
  publisher = {Taylor \& Francis},
}

@Misc{economist2020tracking,
  author       = {The Economist},
  howpublished = {\url{https://www.economist.com/graphic-detail/2020/07/15/tracking-covid-19-excess-deaths-across-countries}},
  note         = {Accessed: 2020-09-16},
  title        = {Tracking covid-19 excess deaths across countries},
  year         = {2020},
}

@Article{varian1972benfords,
  author    = {Varian, Hal R.},
  journal   = {American Statistician},
  title     = {Benford's law},
  year      = {1972},
  number    = {3},
  pages     = {65},
  volume    = {26},
  publisher = {AMER STATISTICAL ASSOC 1429 DUKE ST, ALEXANDRIA, VA 22314},
}

@Misc{wood2020,
  author       = {Wood, Graeme},
  howpublished = {\url{https://www.theatlantic.com/ideas/archive/2020/03/irans-coronavirus-problem-lot-worse-it-seems/607663/}},
  note         = {Accessed: 2020-09-16},
  title        = {Iran has far more coronavirus cases than it is letting on},
  year         = {2020},
}

@Article{zhang2020testing,
  author  = {Zhang, Junyi},
  journal = {arXiv preprint arXiv:2002.05695},
  title   = {Testing case number of coronavirus disease 2019 in {C}hina with {N}ewcomb-{B}enford law},
  year    = {2020},
  note    = {Available on: \url{https://arxiv.org/pdf/2002.05695.pdf}},
}

@Misc{cambell2020china,
  author       = {Cambell, C. and Gunia, A.},
  howpublished = {Https://time.com/5813628/china-coronavirus-statistics-wuhan/},
  note         = {Accessed: 2020-09-16},
  title        = {China says it's beating coronavirus. {B}ut can we believe its numbers?},
  year         = {2020},
}

@Misc{aron2020pandemic,
  author       = {Aron, J. and Muellbauer, J.},
  howpublished = {Https://ourworldindata.org/covid-excess-mortality.},
  note         = {Accessed: 2020-09-16},
  title        = {A pandemic primer on excess mortality statistics and their comparability across countries.},
  year         = {2020},
}

@Misc{jaskson2020national,
  author       = {Jackson, A. and Sambridge, M.},
  howpublished = {Https://www.nature.com/articles/d41586-020-01565-5},
  note         = {Accessed: 2020-09-16},
  title        = {National {COVID} numbers-{B}enford’s law looks for errors,},
  year         = {2020},
}

@Article{marchi2006assessing,
  author  = {Marchi, S. and Hamilton, J. T.},
  journal = {Journal of Risk and Uncertainty},
  title   = {Assessing the accuracy of self-reported data: an evaluation of the toxics release inventory.},
  year    = {2006},
  pages   = {57--76},
  volume  = {32},
}

@Unpublished{roukema2020anti,
  author = {Roukema, Boudewijn},
  title  = {Anti-clustering in the national {SARS}-{CoV}-2 daily infection counts},
  year   = {2020},
}

@Misc{Smith2020states,
  author       = {Smith, Michelle R. and Long, Coleen and Amy, Jeff},
  howpublished = {Https://kutv.com/news/coronavirus/states-accused-of-manipulating-covid-19-statistics-to-make-situation-look-better},
  note         = {Accessed: 2020-09-28},
  title        = {States accused of manipulating {COVID}-19 statistics to make situation look better},
  year         = {2020},
}

@Misc{King2020Florida,
  author       = {King, Noel},
  howpublished = {Https://www.npr.org/2020/06/29/884551391/florida-scientist-says-she-was-fired-for-not-manipulating-covid-19-data},
  note         = {Accessed: 2020-09-28},
  title        = {Florida scientist says she was fired for not manipulating {COVID}-19 data},
  year         = {2020},
}

@Article{hollyer2011democracy,
  author  = {Hollyer, James R. and Rosendorff, Peter B. and Vreeland, James Raymond},
  journal = {The Journal of Politics},
  title   = {Democracy and transparancy},
  year    = {2011},
  number  = {4},
  pages   = {1191--1205},
  volume  = {73},
}

@Book{Przeworski1999Democracy,
  author    = {Przeworski, Adam and Stokes, Susan C. and Manin, Bernard},
  publisher = {Cambridge: Cambridge University Press.},
  title     = {Democracy, accountability, and representation.},
  year      = {1999},
}

@Book{schumpeter1942capitalism,
  author    = {Schumpeter, Joseph},
  publisher = {New York: Harper and Brothers Publishers.},
  title     = {Capitalism, socialism, and democracy.},
  year      = {1942},
}

@Article{mani2007democracy,
  author  = {Mani, Anandi and Mukand, Sharun},
  journal = {Journal of Development Economics},
  title   = {Democracy, visibility and public good provision.},
  year    = {2007},
  pages   = {506--529},
  volume  = {83},
}

@Article{kono2006optimal,
  author  = {Kono, Daniel},
  journal = {American Political Science Review},
  title   = {Optimal obfuscation: democracy and trade policy transparency},
  year    = {2006},
  pages   = {369--384},
  volume  = {100},
}

@Book{dahl1971polyarchy,
  author    = {Dahl, Robert},
  publisher = {New Haven, CT: Yale University Press},
  title     = {Polyarchy: participation and opposition},
  year      = {1971},
}

@Book{bueno2003logic,
  author    = {Bueno de Mesquita, Bruce and Smith, Alastair and Siverson, Randolph M. and Morrow, James D.},
  publisher = {Cambridge, MA: The MIT Press},
  title     = {The logic of political survival},
  year      = {2003},
}

@Article{rosenas2019how,
  author  = {Rozenas, Arturas and Stukal, Denis},
  journal = {Journal of Politics},
  title   = {How autocrats manipulate economic news: {E}vidence from {R}ussia's state-controlled television},
  year    = {2019},
  number  = {3},
  pages   = {982--996},
  volume  = {81},
}

\bigskip{}

\bigskip{}

\appendix

\newgeometry{left=1cm,bottom=2cm, top=2cm, right=1cm}

\begin{landscape}
	\section*{Appendix A1}\label{app1:origdata}
	\setcounter{table}{0}
	\renewcommand{\thetable}{A1.\arabic{table}}
	\centering
	\begin{center}
		\begin{threeparttable}[!ht]
			\caption{Original Country-Level Data}
			\label{tab:alldata}
			\tiny
			\renewcommand\arraystretch{1}
			\setlength{\tabcolsep}{3pt}

			\begin{tabular*}{\linewidth}{@{\extracolsep{\fill}}p{3.8cm}llllllllllllllllll}
				\toprule
				Country & Pop.(MM) & EIU & GDP & HE\_GDP & UHC & Cutoff & Tot C. & Tot D. & Days C. & Chi C. & K C. & M C. & D C. & Days D. & Chi D. & K D.  & M D. & D D. \\
				\midrule
				
				\textbf{World} & \textbf{7815.20} & \textbf{54.40} & \textbf{18381.00} & \textbf{-} & \textbf{-} & \textbf{2020-06-08} & \textbf{7118471} & \textbf{406522} & \textbf{139} &  \textbf{12.65} & \textbf{1.58} & \textbf{1.09} & \textbf{1.39} & \textbf{139} & \textbf{40.43} & \textbf{2.77} & \textbf{1.58}  &  \textbf{2.24}\\
				
				Afghanistan &   38.87 & 28.50 &    556.30 & 11.78 & 37.00 & 2020-06-05 & 20917 & 369 & 103 &  12.04 & 1.41 & 1.18 & 1.47 & 76 &   8.17 & 1.04 & 0.42 & 0.89 \\
				Albania\tnote{*} &    2.88 & 58.90 &   4532.89 &   - & 59.00 & 2020-04-25 & 1263 & 34 & 48 &  16.07 & 1.71 & 0.93 & 1.38 & 46 &  60.22 & 2.89 & 2.79 & 3.03 \\
				Algeria\tnote{*} &   43.80 & 40.10 &   4044.30 &  6.37 & 78.00 & 2020-05-26 & 10265 & 715 & 92 &   4.24 & 0.73 & 0.49 & 0.67 & 76 &  57.57 & 3.06 & 1.68 & 2.52 \\
				Andorra &    0.08 &   - &  39134.39 & 10.32 &   - & 2020-03-30 & 852 & 51 & 29 &  18.48 & 1.79 & 1.72 & 1.93 & 9 &   8.74 & 0.67 & 0.63 & 1.05 \\
				Angola &   32.79 & 37.20 &   4095.81 &  2.79 & 40.00 & 2020-05-30 & 92 & 4 & 72 &  12.66 & 1.20 & 0.79 & 1.18 & 63 & 176.01 & 5.02 & 4.90 & 5.58 \\
				Antigua and Barbuda &    0.10 &   - &  15383.42 &  4.53 & 73.00 & 2020-04-07 & 26 & 3 & 26 &  38.86 & 2.27 & 1.41 & 2.13 & 1 &   2.32 & 0.98 & 0.70 & 0.75 \\
				Argentina\tnote{*} &   45.17 & 70.20 &  14591.86 &  9.12 & 76.00 & 2020-06-08 & 23620 & 693 & 98 &   1.90 & 0.57 & 0.56 & 0.61 & 93 &   9.89 & 1.42 & 0.62 & 1.06 \\
				Armenia &    2.96 & 55.40 &   3914.50 & 10.36 & 69.00 & 2020-06-07 & 13325 & 211 & 99 &  11.62 & 1.52 & 1.02 & 1.33 & 74 &   6.92 & 0.57 & 0.45 & 0.76 \\
				Australia\tnote{*} &   25.48 & 90.90 &  54066.47 &  9.21 & 87.00 & 2020-03-30 & 7267 & 103 & 65 &  18.15 & 1.83 & 1.79 & 1.96 & 30 &  15.94 & 0.99 & 0.96 & 1.26 \\
				Austria\tnote{*} &    9.00 & 82.90 &  47431.63 & 10.40 & 79.00 & 2020-03-28 & 16968 & 672 & 33 &   2.64 & 0.44 & 0.38 & 0.61 & 17 &   5.69 & 0.70 & 0.45 & 0.65 \\
				Azerbaijan &   10.13 & 27.50 &   4147.09 &  6.65 & 65.00 & 2020-06-08 & 7876 & 93 & 100 &   2.73 & 0.46 & 0.31 & 0.42 & 88 &   5.32 & 0.87 & 0.37 & 0.74 \\
				Bahamas &    0.39 &   - &  31827.24 &  5.76 & 75.00 & 2020-04-26 & 103 & 11 & 42 &  18.08 & 1.65 & 1.07 & 1.43 & 26 &  61.95 & 2.34 & 1.31 & 1.98 \\
				Bahrain &    1.70 & 25.50 &  23715.48 &  4.75 & 77.00 & 2020-06-06 & 15417 & 27 & 104 &   4.13 & 0.76 & 0.52 & 0.72 & 83 &  76.12 & 2.12 & 1.62 & 2.33 \\
				Bangladesh &  164.59 & 58.80 &   1563.99 &  2.27 & 48.00 & 2020-06-08 & 68504 & 930 & 93 &  15.32 & 1.67 & 0.83 & 1.39 & 83 &   8.14 & 0.88 & 0.71 & 0.84 \\
				Barbados &    0.29 &   - &  17431.60 &  6.78 & 77.00 & 2020-04-07 & 92 & 7 & 22 &  12.66 & 1.12 & 0.69 & 1.01 & 3 &   2.67 & 0.81 & 0.36 & 0.54 \\
				Belarus &    9.45 & 24.80 &   5761.75 &  5.93 & 76.00 & 2020-05-17 & 49453 & 276 & 80 &  11.11 & 1.15 & 0.45 & 0.95 & 48 &   7.99 & 1.25 & 0.80 & 1.20 \\
				Belgium\tnote{*} &   11.59 & 76.40 &  44219.56 & 10.34 & 84.00 & 2020-04-15 & 59348 & 9606 & 72 &  30.20 & 2.49 & 2.28 & 2.48 & 36 &   8.23 & 1.03 & 0.75 & 0.94 \\
				Belize &    0.40 &   - &   4887.56 &  5.64 & 64.00 & 2020-04-13 & 19 & 2 & 22 &   3.18 & 0.53 & 0.31 & 0.54 & 8 &  10.00 & 1.60 & 0.92 & 1.23 \\
				Benin &   12.10 & 50.90 &   1136.59 &  3.72 & 40.00 & 2020-05-12 & 339 & 4 & 58 &  48.45 & 1.67 & 1.37 & 2.23 & 37 &  41.22 & 3.28 & 2.28 & 2.76 \\
				Bhutan &    0.77 & 53.00 &   3286.57 &  3.19 & 62.00 & 2020-06-02 & 59 & 0 & 89 &  71.54 & 2.25 & 1.36 & 2.46 & 0 &    - &  - & - & - \\
				Bolivia &   11.66 & 48.40 &   3351.12 &  6.44 & 68.00 & 2020-06-05 & 13949 & 475 & 87 &   1.95 & 0.59 & 0.52 & 0.58 & 69 &   2.74 & 0.71 & 0.29 & 0.51 \\
				Bosnia and Herzegovina &    3.28 & 48.60 &   5394.59 &  8.93 & 61.00 & 2020-04-04 & 2704 & 160 & 31 &   2.95 & 0.68 & 0.38 & 0.60 & 15 &   6.88 & 0.95 & 0.90 & 1.10 \\
				Botswana &    2.35 & 78.10 &   7893.21 &  6.13 & 61.00 & 2020-05-24 & 42 & 1 & 56 &  71.53 & 3.17 & 3.09 & 3.37 & 55 & 127.71 & 5.31 & 5.18 & 5.57 \\
				Brazil\tnote{*} &  212.47 & 68.60 &   9925.39 &  9.47 & 79.00 & 2020-06-08 & 707412 & 37134 & 104 &   5.32 & 0.94 & 0.36 & 0.68 & 84 &   6.48 & 1.05 & 0.68 & 0.87 \\
				Brunei &    0.44 &   - &  28572.15 &  2.37 & 81.00 & 2020-03-19 & 141 & 2 & 11 &   8.76 & 0.89 & 0.64 & 0.94 & 0 &    - &  - & - & - \\
				Bulgaria &    6.95 & 70.30 &   8228.01 &  8.10 & 66.00 & 2020-04-27 & 2810 & 164 & 51 &  11.79 & 1.68 & 0.89 & 1.26 & 48 &  21.91 & 2.09 & 1.08 & 1.75 \\
				Burkina Faso &   20.86 & 40.40 &    642.43 &  6.92 & 40.00 & 2020-04-11 & 890 & 53 & 33 &  14.09 & 1.78 & 0.90 & 1.34 & 25 &  14.89 & 1.47 & 1.29 & 1.59 \\
				Burundi &   11.87 & 21.50 &    293.00 &  7.52 & 42.00 & 2020-05-18 & 83 & 1 & 49 &  39.04 & 1.59 & 1.46 & 2.21 & 36 &  83.59 & 4.33 & 4.19 & 4.50 \\
				Cabo Verde &    0.56 & 77.80 &   3292.65 &  5.17 & 69.00 & 2020-06-05 & 567 & 5 & 78 &  23.74 & 2.17 & 1.07 & 1.82 & 74 &  29.70 & 2.19 & 1.94 & 2.19 \\
				Cambodia &   16.70 & 35.30 &   1385.46 &  5.92 & 60.00 & 2020-03-23 & 126 & 0 & 57 &  51.46 & 3.24 & 3.16 & 3.45 & 0 &    - &  - & - & - \\
				Cameroon &   26.50 & 28.50 &   1421.59 &  4.67 & 46.00 & 2020-06-07 & 8060 & 212 & 94 &   8.70 & 0.78 & 0.77 & 1.18 & 75 &  31.70 & 1.77 & 1.67 & 2.20 \\
				Canada\tnote{*} &   37.72 & 92.20 &  45066.16 & 10.57 & 89.00 & 2020-04-22 & 97779 & 7910 & 88 &  18.05 & 1.13 & 0.84 & 1.23 & 45 &   5.53 & 0.99 & 0.96 & 1.10 \\
				Central African Republic &    4.82 & 13.20 &    449.79 &  5.82 & 33.00 & 2020-06-08 & 1850 & 5 & 86 &  48.37 & 1.96 & 1.63 & 2.46 & 17 &  17.70 & 1.28 & 0.81 & 1.42 \\
				Chad &   16.39 & 16.10 &    664.30 &  4.49 & 28.00 & 2020-05-09 & 839 & 70 & 52 &  11.67 & 0.93 & 0.90 & 1.10 & 12 &   7.37 & 1.01 & 0.54 & 0.91 \\
				Chile\tnote{*} &   19.11 & 80.80 &  15037.35 &  8.98 & 70.00 & 2020-06-07 & 138846 & 2264 & 97 &   3.90 & 0.64 & 0.42 & 0.62 & 78 &   2.09 & 0.52 & 0.23 & 0.39 \\
				China\tnote{*} & 1439.32 & 22.60 &   8759.04 &  5.15 & 79.00 & 2020-02-14 & 84635 & 4645 & 22 &   5.49 & 0.72 & 0.56 & 0.82 & 22 &   1.71 & 0.36 & 0.19 & 0.39 \\
				Colombia\tnote{*} &   50.85 & 71.30 &   6375.93 &  7.23 & 76.00 & 2020-06-08 & 40847 & 1373 & 95 &   5.07 & 0.88 & 0.34 & 0.68 & 79 &   3.55 & 0.61 & 0.33 & 0.55 \\
				Comoros &    0.87 & 31.50 &   1320.54 &  7.38 & 52.00 & 2020-05-24 & 141 & 2 & 25 &  24.25 & 1.58 & 1.09 & 1.78 & 19 &  44.12 & 3.19 & 3.05 & 3.27 \\
				Congo (Brazzaville) &    5.51 & 31.10 &   1702.57 &  2.93 & 39.00 & 2020-05-27 & 683 & 22 & 74 &  40.53 & 2.44 & 1.38 & 1.97 & 56 &  51.47 & 2.64 & 1.01 & 2.31 \\
				Congo (Kinshasa) &   89.37 & 31.10 &   1702.57 &  2.93 & 39.00 & 2020-06-04 & 4106 & 88 & 86 &   9.23 & 1.11 & 0.63 & 0.98 & 76 &  46.33 & 1.90 & 1.37 & 2.25 \\

				Costa Rica &    5.09 & 81.30 &  11752.54 &  7.33 & 77.00 & 2020-06-07 & 1342 & 11 & 94 &  50.60 & 2.96 & 1.06 & 2.15 & 81 &  51.30 & 1.65 & 1.62 & 1.91 \\
				Cote d'Ivoire &     - & 40.50 &   1557.18 &  4.45 & 47.00 & 2020-06-08 & 3881 & 38 & 90 &  17.40 & 1.82 & 1.04 & 1.56 & 72 &  38.40 & 2.25 & 1.65 & 2.02 \\
				Croatia\tnote{*} &    4.11 & 65.70 &  13412.34 &  6.79 & 71.00 & 2020-04-01 & 2247 & 104 & 37 &   5.48 & 0.77 & 0.74 & 0.86 & 14 &  17.89 & 1.04 & 0.82 & 1.38 \\
				Cuba &   11.33 & 28.40 &   8541.21 & 11.71 & 83.00 & 2020-04-13 & 2200 & 83 & 33 &   9.19 & 1.25 & 0.68 & 1.05 & 27 &  15.94 & 1.38 & 1.13 & 1.48 \\
				Cyprus &    1.21 & 75.90 &  26338.69 &  6.68 & 78.00 & 2020-04-04 & 970 & 18 & 27 &   7.84 & 0.66 & 0.42 & 0.81 & 14 &   9.39 & 0.73 & 0.66 & 1.03 \\
				Czechia\tnote{*} &   10.71 & 76.90 &  20379.90 &  7.23 & 76.00 & 2020-04-02 & 9697 & 328 & 33 &  11.43 & 1.07 & 0.85 & 1.04 & 12 &   8.29 & 0.68 & 0.43 & 0.79 \\
				Denmark\tnote{*} &    5.79 & 92.20 &  57141.06 & 10.11 & 81.00 & 2020-04-08 & 12162 & 593 & 42 &   4.62 & 0.73 & 0.38 & 0.57 & 26 &   2.99 & 0.40 & 0.26 & 0.47 \\
				Djibouti &    0.99 & 27.70 &   2930.70 &  3.32 & 47.00 & 2020-06-02 & 4278 & 31 & 77 &  27.68 & 2.36 & 2.03 & 2.27 & 54 &  66.41 & 3.31 & 2.92 & 3.28 \\
				Dominica &    0.07 &   - &   7274.72 &  5.88 &   - & 2020-03-26 & 18 & 0 & 5 &   5.65 & 0.92 & 0.50 & 0.78 & 0 &    - &  - & - & - \\
				Dominican Republic &   10.84 & 65.40 &   7609.35 &  6.14 & 74.00 & 2020-06-08 & 20126 & 539 & 100 &  17.83 & 1.62 & 1.59 & 1.89 & 84 &  37.39 & 2.66 & 1.34 & 2.13 \\
				Ecuador\tnote{*} &   17.63 & 63.30 &   6213.50 &  8.26 & 77.00 & 2020-04-24 & 43378 & 3642 & 55 &  16.66 & 1.51 & 0.87 & 1.40 & 42 &  12.17 & 1.58 & 0.72 & 1.19 \\
				Egypt &  102.21 & 30.60 &   2440.51 &  5.29 & 68.00 & 2020-06-06 & 35444 & 1271 & 114 &   8.39 & 1.30 & 1.19 & 1.30 & 91 &   5.69 & 1.06 & 0.46 & 0.79 \\
				El Salvador &    6.48 & 61.50 &   3902.24 &  7.23 & 76.00 & 2020-06-02 & 3104 & 56 & 76 &   6.49 & 1.14 & 0.59 & 0.90 & 64 &  19.11 & 1.47 & 0.75 & 1.42 \\
				Equatorial Guinea &    1.40 & 19.20 &   9667.91 &  3.11 & 45.00 & 2020-05-19 & 1306 & 12 & 66 &   4.90 & 0.74 & 0.69 & 0.81 & 28 &  30.99 & 1.26 & 0.93 & 1.75 \\
				Eritrea &    3.54 & 23.70 &       - &  2.87 & 38.00 & 2020-04-04 & 39 & 0 & 15 &  12.23 & 1.05 & 0.90 & 1.24 & 0 &    - &  - & - & - \\

				\bottomrule
			\end{tabular*}
			\begin{tablenotes}
				\item[] 
			\end{tablenotes}
		\end{threeparttable}

	\end{center}
\end{landscape}

\newgeometry{left=2cm,bottom=2cm, top=2cm, right=1cm}
\begin{landscape}
	\centering
	\begin{center}
		\begin{threeparttable}[!ht]
			\tiny
			\renewcommand\arraystretch{1}
			\setlength{\tabcolsep}{3pt}

			\begin{tabular*}{\linewidth}{@{\extracolsep{\fill}}p{3.8cm}llllllllllllllllll}
				\toprule
				Country & Pop. & EIU & GDP & HE\_GDP & UHC & Cutoff & Tot C. & Tot D. & Days C. & Chi C. & K C. & M C. & D C. & Days D. & Chi D. & K D.  & M D. & D D. \\
				
				\midrule
				
				Estonia\tnote{*} &    1.33 & 79.00 &  20337.85 &  6.43 & 75.00 & 2020-04-05 & 1940 & 69 & 39 &   6.30 & 0.75 & 0.68 & 0.91 & 12 &  10.30 & 1.35 & 1.27 & 1.47 \\
				Eswatini &    1.16 & 31.40 &   3953.09 &  6.93 & 63.00 & 2020-05-01 & 340 & 3 & 49 &  47.41 & 2.35 & 1.25 & 1.92 & 16 &  37.15 & 2.95 & 2.80 & 3.00 \\
				Ethiopia &  114.77 & 34.40 &    768.43 &  3.50 & 39.00 & 2020-06-08 & 2156 & 27 & 88 &  10.52 & 1.46 & 1.33 & 1.49 & 65 &  80.47 & 3.35 & 2.09 & 3.16 \\
				Fiji &    0.90 & 58.50 &   6101.03 &  3.50 & 64.00 & 2020-04-07 & 18 & 0 & 20 &  33.20 & 1.50 & 1.43 & 1.67 & 0 &    - &  - & - & - \\
				Finland\tnote{*} &    5.54 & 92.50 &  46191.93 &  9.21 & 78.00 & 2020-04-10 & 7001 & 323 & 73 &  29.69 & 2.39 & 2.34 & 2.54 & 21 &   4.55 & 0.77 & 0.58 & 0.76 \\
				France\tnote{*} &   65.27 & 81.20 &  38679.13 & 11.31 & 78.00 & 2020-04-16 & 192330 & 29212 & 84 &  14.46 & 1.38 & 0.84 & 1.36 & 62 &  14.98 & 1.73 & 1.69 & 1.85 \\
				Gabon &    2.22 & 36.10 &   7212.54 &  2.78 & 49.00 & 2020-05-29 & 3101 & 21 & 77 &   7.04 & 0.96 & 0.78 & 0.96 & 71 &  63.04 & 3.49 & 3.40 & 3.81 \\
				Gambia &    2.41 & 43.30 &    679.78 &  3.28 & 44.00 & 2020-05-08 & 28 & 1 & 53 &  49.67 & 2.06 & 1.24 & 2.29 & 47 & 109.13 & 4.92 & 4.79 & 5.15 \\
				Georgia &    3.99 & 54.20 &   4357.01 &  7.60 & 66.00 & 2020-04-18 & 812 & 13 & 53 &  13.80 & 1.16 & 1.01 & 1.19 & 15 &  51.99 & 2.49 & 2.36 & 2.59 \\
				Germany\tnote{*} &   83.77 & 86.80 &  44240.04 & 11.25 & 83.00 & 2020-04-02 & 186109 & 8695 & 67 &  20.67 & 1.85 & 1.81 & 2.10 & 25 &   6.72 & 0.72 & 0.51 & 0.82 \\
				Ghana &   31.03 & 66.30 &   2025.89 &  3.26 & 47.00 & 2020-05-14 & 9910 & 48 & 62 &   6.18 & 0.73 & 0.39 & 0.74 & 55 &  32.70 & 1.51 & 0.93 & 1.60 \\
				Greece\tnote{*} &   10.43 & 74.30 &  18883.46 &  8.04 & 75.00 & 2020-04-02 & 3049 & 182 & 37 &  20.91 & 1.47 & 0.91 & 1.45 & 23 &   7.84 & 0.99 & 0.45 & 0.76 \\
				Grenada &    0.11 &   - &  10163.63 &  4.76 & 72.00 & 2020-03-29 & 23 & 0 & 8 &  20.77 & 1.67 & 0.90 & 1.31 & 0 &    - &  - & - & - \\
				Guatemala &   17.89 & 52.60 &   4470.61 &  5.81 & 55.00 & 2020-06-08 & 7502 & 267 & 87 &   7.50 & 1.01 & 0.56 & 0.85 & 85 &  15.96 & 1.38 & 1.35 & 1.59 \\
				Guinea &   13.11 & 31.40 &    855.57 &  4.12 & 37.00 & 2020-06-03 & 4216 & 23 & 83 &  12.79 & 1.46 & 0.73 & 1.15 & 50 &  27.73 & 1.60 & 0.86 & 1.59 \\
				Guinea-Bissau &    1.96 & 26.30 &    736.73 &  7.24 & 40.00 & 2020-05-10 & 1389 & 12 & 47 &  21.88 & 1.89 & 1.48 & 1.85 & 15 &  12.69 & 1.63 & 1.42 & 1.56 \\
				Guyana &    0.79 & 61.50 &   4586.05 &  4.95 & 72.00 & 2020-04-17 & 154 & 12 & 37 &  14.40 & 1.42 & 0.68 & 1.20 & 37 &  43.71 & 1.64 & 1.29 & 2.19 \\
				Haiti &   11.39 & 45.70 &    765.73 &  8.04 & 49.00 & 2020-06-04 & 3538 & 54 & 77 &  24.63 & 1.35 & 0.80 & 1.35 & 61 &  13.66 & 1.41 & 0.82 & 1.27 \\
				Holy See &     - &   - &       - &   - &   - & 2020-03-28 & 12 & 0 & 23 &  31.62 & 2.41 & 2.31 & 2.62 & 0 &    - &  - & - & - \\
				Honduras &    9.89 & 54.20 &   2453.73 &  7.86 & 65.00 & 2020-05-27 & 6450 & 262 & 78 &  16.17 & 1.96 & 1.07 & 1.65 & 63 &   9.93 & 1.16 & 1.14 & 1.36 \\
				Hungary\tnote{*} &    9.66 & 66.30 &  14457.61 &  6.88 & 74.00 & 2020-04-13 & 4014 & 548 & 41 &   3.38 & 0.59 & 0.35 & 0.55 & 30 &   4.34 & 0.87 & 0.72 & 0.90 \\
				Iceland &    0.34 & 95.80 &  71314.77 &  8.33 & 84.00 & 2020-04-02 & 1807 & 10 & 35 &   6.22 & 0.86 & 0.53 & 0.87 & 17 &  18.44 & 1.76 & 1.46 & 1.64 \\
				India\tnote{*} & 1379.20 & 69.00 &   1981.27 &  3.53 & 55.00 & 2020-06-08 & 265928 & 7473 & 131 &  44.85 & 2.19 & 2.15 & 2.32 & 90 &   7.06 & 1.27 & 0.85 & 1.06 \\
				Indonesia\tnote{*} &  273.35 & 64.80 &   3836.91 &  2.99 & 57.00 & 2020-06-08 & 32033 & 1883 & 99 &   9.29 & 1.31 & 0.96 & 1.16 & 90 &  20.75 & 2.02 & 1.36 & 1.95 \\
				Iran\tnote{*} &   83.93 & 23.80 &   5627.75 &  8.66 & 72.00 & 2020-04-02 & 173832 & 8351 & 44 &   4.16 & 0.62 & 0.49 & 0.66 & 44 &   3.68 & 0.78 & 0.64 & 0.74 \\
				Iraq &   40.16 & 37.40 &   5205.29 &  4.17 & 61.00 & 2020-06-08 & 13481 & 370 & 106 &   6.01 & 0.81 & 0.49 & 0.80 & 97 &  26.40 & 1.99 & 0.82 & 1.47 \\
				Ireland\tnote{*} &    4.93 & 92.40 &  69649.88 &  7.18 & 76.00 & 2020-04-16 & 25207 & 1683 & 48 &   3.85 & 0.50 & 0.37 & 0.62 & 37 &   7.17 & 1.23 & 0.57 & 0.99 \\
				Israel &    9.20 & 78.60 &  40541.86 &  7.41 & 82.00 & 2020-04-03 & 18032 & 298 & 43 &   6.48 & 0.57 & 0.38 & 0.63 & 14 &   4.57 & 0.79 & 0.74 & 0.85 \\
				Italy\tnote{*} &   60.47 & 75.20 &  32326.84 &  8.84 & 82.00 & 2020-03-26 & 235278 & 33964 & 56 &  30.64 & 2.21 & 1.60 & 2.18 & 35 &   2.70 & 0.32 & 0.27 & 0.41 \\
				Jamaica &    2.96 & 69.60 &   5069.18 &  5.99 & 65.00 & 2020-04-26 & 599 & 10 & 47 &   6.08 & 0.49 & 0.42 & 0.65 & 39 &  17.50 & 1.55 & 1.10 & 1.43 \\
				Japan\tnote{*} &  126.50 & 79.90 &  38331.98 & 10.94 & 83.00 & 2020-04-18 & 17060 & 920 & 88 &  15.46 & 1.25 & 1.23 & 1.57 & 66 &  18.13 & 1.24 & 0.77 & 1.33 \\
				Jordan &   10.20 & 39.30 &   4162.82 &  8.12 & 76.00 & 2020-03-28 & 831 & 9 & 26 &  21.03 & 1.97 & 1.80 & 2.02 & 2 &   4.64 & 1.22 & 0.99 & 1.06 \\
				Kazakhstan\tnote{*} &   18.76 & 29.40 &   9247.58 &  3.13 & 76.00 & 2020-06-03 & 12859 & 56 & 83 &  11.99 & 1.63 & 0.88 & 1.33 & 71 &  54.61 & 2.50 & 2.27 & 2.48 \\
				Kenya &   53.69 & 51.80 &   1568.20 &  4.80 & 55.00 & 2020-06-08 & 2872 & 85 & 88 &   9.58 & 0.98 & 0.53 & 0.94 & 75 &  15.72 & 1.40 & 0.82 & 1.22 \\
				Kosovo &     - &   - &   3948.05 &   - &   - & 2020-04-20 & 1263 & 31 & 26 &   6.44 & 0.81 & 0.43 & 0.75 & 26 &  26.88 & 2.27 & 1.80 & 2.23 \\

				Kuwait &    4.27 & 39.30 &  29759.53 &  5.29 & 76.00 & 2020-05-22 & 32510 & 269 & 89 &   6.87 & 0.95 & 0.87 & 1.13 & 49 &   5.57 & 0.92 & 0.89 & 1.05 \\
				Kyrgyzstan &    6.52 & 48.90 &   1242.77 &  6.19 & 70.00 & 2020-06-01 & 2032 & 23 & 76 &  13.80 & 1.63 & 1.05 & 1.59 & 60 &  59.26 & 2.63 & 1.93 & 2.87 \\
				Laos &    7.27 & 21.40 &   2423.85 &  2.53 & 51.00 & 2020-03-28 & 19 & 0 & 5 &  13.60 & 1.06 & 0.74 & 1.12 & 0 &    - &  - & - & - \\
				Latvia\tnote{*} &    1.89 & 74.90 &  15548.08 &  5.96 & 71.00 & 2020-04-01 & 1088 & 26 & 31 &   8.77 & 1.18 & 0.84 & 1.10 & 0 &    - &  - & - & - \\
				Lebanon &    6.83 & 43.60 &   7838.34 &  8.20 & 73.00 & 2020-03-27 & 1350 & 30 & 36 &  10.29 & 1.43 & 1.03 & 1.25 & 18 &  35.69 & 2.48 & 1.36 & 2.18 \\
				Lesotho &    2.14 & 65.40 &   1226.61 &  8.76 & 48.00 & 2020-06-03 & 4 & 0 & 22 &  27.87 & 2.34 & 1.73 & 2.02 & 0 &    - &  - & - & - \\
				Liberia &    5.05 & 54.50 &    698.70 &  8.16 & 39.00 & 2020-06-08 & 370 & 30 & 85 &  32.09 & 2.30 & 1.13 & 1.89 & 66 &  46.40 & 2.44 & 2.39 & 2.75 \\
				Libya &    6.87 & 20.20 &   5756.42 &   - & 64.00 & 2020-06-08 & 332 & 5 & 77 & 113.06 & 2.84 & 2.60 & 2.86 & 68 &  70.34 & 2.42 & 2.24 & 2.71 \\
				Liechtenstein &    0.04 &   - &       - &   - &   - & 2020-03-23 & 82 & 1 & 20 &   6.90 & 0.94 & 0.89 & 1.04 & 0 &    - &  - & - & - \\
				Lithuania &    2.72 & 75.00 &  16840.94 &  6.46 & 73.00 & 2020-04-04 & 1720 & 71 & 37 &   5.32 & 0.82 & 0.80 & 0.93 & 15 &  13.90 & 1.28 & 0.55 & 1.02 \\
				Luxembourg &    0.63 & 88.10 & 107361.31 &  5.48 & 83.00 & 2020-03-28 & 4040 & 110 & 29 &   7.35 & 0.82 & 0.79 & 1.04 & 15 &  30.14 & 1.64 & 1.09 & 1.41 \\
				Madagascar &   27.64 & 56.40 &    515.29 &  5.50 & 28.00 & 2020-06-04 & 1094 & 9 & 77 &  12.67 & 1.62 & 1.35 & 1.63 & 19 &  32.81 & 1.60 & 1.53 & 2.09 \\
				Malawi &   19.10 & 55.00 &    356.72 &  9.65 & 46.00 & 2020-06-03 & 443 & 4 & 63 &  17.30 & 1.67 & 1.02 & 1.50 & 58 & 117.60 & 4.16 & 3.25 & 3.91 \\
				Malaysia\tnote{*} &   32.34 & 71.60 &  10254.23 &  3.86 & 73.00 & 2020-04-08 & 8329 & 117 & 75 &  18.67 & 1.53 & 1.36 & 1.56 & 23 &  12.90 & 1.63 & 0.82 & 1.26 \\
				Maldives &    0.54 &   - &   9540.63 &  9.03 & 62.00 & 2020-06-01 & 1916 & 8 & 86 &  38.69 & 2.83 & 2.60 & 2.93 & 34 &  43.76 & 2.62 & 1.66 & 2.12 \\
				Mali &   20.21 & 49.20 &    828.51 &  3.79 & 38.00 & 2020-06-02 & 1547 & 92 & 70 &   4.96 & 0.87 & 0.52 & 0.71 & 66 &  28.53 & 1.80 & 1.09 & 1.65 \\
				Malta &    0.44 & 79.50 &  27283.54 &  9.34 & 82.00 & 2020-04-11 & 630 & 9 & 36 &  11.56 & 1.26 & 0.75 & 1.10 & 4 &   4.51 & 1.02 & 0.65 & 0.78 \\
				Mauritania &    4.64 & 39.20 &   1145.55 &  4.40 & 41.00 & 2020-06-07 & 1104 & 59 & 86 & 108.53 & 3.61 & 1.83 & 3.01 & 70 &  49.76 & 3.17 & 3.10 & 3.35 \\
				Mauritius &    1.27 & 82.20 &  10484.91 &  5.72 & 63.00 & 2020-04-09 & 337 & 10 & 23 &   6.78 & 1.12 & 0.64 & 0.91 & 20 &  52.24 & 1.63 & 1.53 & 2.13 \\
				Mexico\tnote{*} &  128.85 & 60.90 &   9278.42 &  5.52 & 76.00 & 2020-06-08 & 120102 & 14053 & 102 &  10.09 & 1.35 & 0.59 & 1.06 & 82 &   2.46 & 0.49 & 0.25 & 0.42 \\
				Moldova &    4.03 & 57.50 &   3509.69 &  7.01 & 69.00 & 2020-06-07 & 9807 & 353 & 92 &  10.57 & 1.48 & 1.22 & 1.39 & 82 &  16.93 & 1.79 & 1.17 & 1.52 \\
				Monaco &    0.04 &   - & 167101.76 &  1.77 &   - & 2020-03-27 & 99 & 4 & 28 &  16.56 & 1.50 & 1.05 & 1.44 & 0 &    - &  - & - & - \\
				Mongolia &    3.27 & 65.00 &   3669.42 &  4.00 & 62.00 & 2020-05-18 & 194 & 0 & 70 &  48.69 & 1.84 & 1.55 & 2.70 & 0 &    - &  - & - & - \\
				Montenegro &    0.63 & 56.50 &   7784.07 &   - & 68.00 & 2020-04-06 & 324 & 9 & 21 &   9.55 & 1.13 & 0.72 & 0.99 & 15 &  17.72 & 2.14 & 1.13 & 1.65 \\
				Morocco &   36.88 & 51.00 &   3036.33 &  5.25 & 70.00 & 2020-04-22 & 8302 & 208 & 52 &   7.75 & 0.88 & 0.67 & 0.91 & 44 &   9.04 & 1.36 & 1.17 & 1.30 \\
				Mozambique &   31.19 & 36.50 &    461.42 &  4.94 & 46.00 & 2020-06-08 & 433 & 2 & 79 &  29.29 & 1.42 & 0.78 & 1.33 & 15 &  41.51 & 2.55 & 2.42 & 2.58 \\
				Myanmar &   54.41 & 35.50 &   1249.83 &  4.66 & 61.00 & 2020-04-19 & 244 & 6 & 24 &  16.54 & 1.11 & 0.97 & 1.37 & 20 &  18.13 & 1.40 & 0.91 & 1.42 \\
				Namibia &    2.54 & 64.30 &   5646.46 &  8.55 & 62.00 & 2020-03-29 & 31 & 0 & 16 &  17.04 & 1.20 & 0.95 & 1.43 & 0 &    - &  - & - & - \\
				Nepal &   29.10 & 52.80 &    911.44 &  5.55 & 48.00 & 2020-06-08 & 3762 & 14 & 136 &  46.90 & 3.03 & 2.75 & 3.13 & 24 &   5.23 & 0.40 & 0.36 & 0.61 \\
				Netherlands\tnote{*} &   17.13 & 90.10 &  48554.99 & 10.10 & 86.00 & 2020-04-14 & 47945 & 6035 & 48 &   5.79 & 1.10 & 0.66 & 0.92 & 40 &   8.48 & 1.13 & 0.78 & 1.02 \\

				\bottomrule
			\end{tabular*}
			\begin{tablenotes}
				\item[] 
			\end{tablenotes}
		\end{threeparttable}

	\end{center}
\end{landscape}

\begin{landscape}
	\centering
	\begin{center}
		\begin{threeparttable}[!ht]
			\tiny
			\renewcommand\arraystretch{1}
			\setlength{\tabcolsep}{3pt}

			\begin{tabular*}{\linewidth}{@{\extracolsep{\fill}}p{3.8cm}llllllllllllllllll}
				\toprule
				Country & Pop. & EIU & GDP & HE\_GDP & UHC & Cutoff & Tot C. & Tot D. & Days C. & Chi C. & K C. & M C. & D C. & Days D. & Chi D. & K D.  & M D. & D D. \\
				
				\midrule
				New Zealand &    5.00 & 92.60 &  42260.13 &  9.17 & 87.00 & 2020-04-05 & 1504 & 22 & 38 &  19.42 & 1.22 & 1.13 & 1.29 & 8 &  18.58 & 2.14 & 1.98 & 2.12 \\
				Nicaragua &    6.62 & 35.50 &   2159.16 &  8.65 & 73.00 & 2020-05-26 & 1118 & 46 & 69 &  24.82 & 2.08 & 1.07 & 1.73 & 61 &  35.48 & 1.54 & 1.36 & 2.07 \\
				Niger &   24.14 & 32.90 &    375.87 &  7.74 & 37.00 & 2020-04-11 & 973 & 65 & 23 &   5.99 & 0.85 & 0.43 & 0.76 & 18 &  17.75 & 1.63 & 1.55 & 1.86 \\
				Nigeria &  205.79 & 41.20 &   1968.56 &  3.76 & 42.00 & 2020-06-04 & 12801 & 361 & 98 &   4.11 & 0.39 & 0.38 & 0.61 & 74 &  12.69 & 1.74 & 0.93 & 1.38 \\
				North Macedonia &    2.08 & 59.70 &   5417.64 &  6.06 & 72.00 & 2020-06-08 & 3152 & 156 & 104 &  35.12 & 2.56 & 2.52 & 2.73 & 79 &  22.16 & 1.67 & 1.00 & 1.46 \\
				Norway\tnote{*} &    5.42 & 98.70 &  75496.75 & 10.45 & 87.00 & 2020-03-29 & 8561 & 239 & 33 &   3.38 & 0.77 & 0.71 & 0.78 & 16 &  16.79 & 0.81 & 0.77 & 1.10 \\
				Oman &    5.10 & 30.60 &  15130.50 &  3.85 & 69.00 & 2020-06-06 & 17486 & 81 & 104 &   4.57 & 0.84 & 0.56 & 0.83 & 68 &   7.20 & 0.93 & 0.55 & 0.90 \\
				Pakistan\tnote{*} &  220.62 & 42.50 &   1464.99 &  2.90 & 45.00 & 2020-06-08 & 108317 & 2172 & 104 &  11.40 & 1.33 & 0.91 & 1.24 & 82 &   5.15 & 0.75 & 0.47 & 0.65 \\
				Palestine &    5.05 & 38.90 &   3254.49 &   - &   - & 2020-04-07 & 473 & 3 & 34 &   6.81 & 1.02 & 0.47 & 0.80 & 13 &  30.19 & 2.68 & 2.52 & 2.71 \\
				Panama &    4.31 & 70.50 &  15166.12 &  7.32 & 79.00 & 2020-06-08 & 16854 & 398 & 91 &  14.82 & 1.43 & 0.95 & 1.22 & 90 &  21.80 & 2.02 & 1.03 & 1.55 \\
				Papua New Guinea &    8.94 & 60.30 &   2695.25 &  2.47 & 40.00 & 2020-04-22 & 8 & 0 & 34 &  29.77 & 2.22 & 1.16 & 1.90 & 0 &    - &  - & - & - \\
				Paraguay &    7.13 & 62.40 &   5680.58 &  6.65 & 69.00 & 2020-05-09 & 1145 & 11 & 63 &   9.84 & 0.78 & 0.62 & 0.99 & 50 &  40.69 & 1.47 & 1.10 & 1.89 \\
				Peru\tnote{*} &   32.94 & 66.00 &   6710.51 &  5.00 & 77.00 & 2020-06-01 & 199696 & 5571 & 88 &   6.34 & 1.09 & 0.80 & 1.07 & 74 &   4.53 & 0.56 & 0.44 & 0.63 \\
				Philippines\tnote{*} &  109.49 & 66.40 &   2981.93 &  4.45 & 61.00 & 2020-06-04 & 22474 & 1011 & 127 &  37.35 & 1.90 & 1.88 & 2.15 & 124 &  35.11 & 2.60 & 1.42 & 2.26 \\
				Poland\tnote{*} &   37.85 & 66.20 &  13856.98 &  6.54 & 75.00 & 2020-06-08 & 27160 & 1166 & 97 &  19.92 & 2.14 & 1.20 & 1.82 & 89 &  21.07 & 1.53 & 0.92 & 1.32 \\
				Portugal\tnote{*} &   10.20 & 80.30 &  21437.35 &  8.97 & 82.00 & 2020-04-03 & 34885 & 1485 & 33 &   5.07 & 0.77 & 0.68 & 0.87 & 18 &   5.49 & 0.84 & 0.61 & 0.82 \\
				Qatar &    2.81 & 31.90 &  61264.40 &  2.61 & 68.00 & 2020-06-03 & 70158 & 57 & 96 &  14.14 & 1.37 & 1.01 & 1.38 & 68 &  19.86 & 1.68 & 1.64 & 1.95 \\
				Romania\tnote{*} &   19.24 & 64.90 &  10807.68 &  5.16 & 74.00 & 2020-04-17 & 20604 & 1339 & 52 &   4.42 & 0.83 & 0.51 & 0.82 & 27 &   7.23 & 0.97 & 0.70 & 0.88 \\
				Russia\tnote{*} &  145.93 & 31.10 &  10750.59 &  5.34 & 74.00 & 2020-05-12 & 476043 & 5963 & 103 &  43.33 & 2.49 & 2.45 & 2.63 & 55 &   6.29 & 0.96 & 0.87 & 1.13 \\
				Rwanda &   12.93 & 31.60 &    762.91 &  6.57 & 57.00 & 2020-04-30 & 451 & 2 & 48 &  27.79 & 2.24 & 1.96 & 2.30 & 0 &    - &  - & - & - \\

				Saint Kitts and Nevis &    0.05 &   - &  19155.43 &  5.04 &   - & 2020-03-31 & 15 & 0 & 7 &  18.54 & 1.56 & 1.42 & 1.74 & 0 &    - &  - & - & - \\
				Saint Lucia &    0.18 &   - &  10039.67 &  4.55 & 68.00 & 2020-04-04 & 19 & 0 & 22 &  17.03 & 1.72 & 0.88 & 1.38 & 0 &    - &  - & - & - \\
				Saint Vincent and the Grenadines &    0.11 &   - &   7212.96 &  4.49 & 71.00 & 2020-04-09 & 27 & 0 & 27 &  31.36 & 2.54 & 2.28 & 2.58 & 0 &    - &  - & - & - \\
				San Marino &    0.03 &   - &  48494.55 &  7.36 &   - & 2020-04-22 & 687 & 42 & 56 &  17.08 & 1.60 & 1.09 & 1.37 & 51 &  44.50 & 2.68 & 1.91 & 2.32 \\
				Sao Tome and Principe &    0.22 &   - &   1811.01 &  6.23 & 55.00 & 2020-05-31 & 513 & 12 & 56 & 129.38 & 3.39 & 3.02 & 3.76 & 31 &  29.65 & 1.69 & 0.98 & 1.55 \\
				Saudi Arabia\tnote{*} &   34.78 & 19.30 &  20803.74 &  5.23 & 74.00 & 2020-05-22 & 105283 & 746 & 82 &   7.48 & 0.92 & 0.50 & 0.73 & 60 &   8.59 & 1.07 & 0.57 & 0.94 \\
				Senegal &   16.71 & 58.10 &   1367.22 &  4.13 & 45.00 & 2020-05-16 & 4427 & 49 & 76 &  14.80 & 1.39 & 0.76 & 1.23 & 46 &  39.40 & 2.11 & 1.46 & 1.98 \\
				Serbia &    8.74 & 64.10 &   6284.19 &  8.43 & 65.00 & 2020-04-19 & 11896 & 250 & 45 &   3.44 & 0.68 & 0.29 & 0.55 & 31 &   4.16 & 0.89 & 0.62 & 0.84 \\
				Seychelles &    0.10 &   - &  15683.66 &  5.01 & 71.00 & 2020-03-20 & 11 & 0 & 7 &   7.88 & 1.15 & 0.80 & 1.07 & 0 &    - &  - & - & - \\
				Sierra Leone &    7.97 & 48.60 &    499.38 & 13.42 & 39.00 & 2020-05-30 & 1001 & 49 & 61 &  11.35 & 1.37 & 0.56 & 1.01 & 38 &  22.09 & 1.60 & 1.19 & 1.46 \\
				Singapore &    5.85 & 60.20 &  60297.79 &  4.44 & 86.00 & 2020-04-26 & 38296 & 25 & 95 &  12.58 & 1.33 & 0.63 & 1.03 & 37 &   9.78 & 0.87 & 0.58 & 0.91 \\
				Slovakia &    5.46 & 71.70 &  17510.09 &  6.74 & 77.00 & 2020-04-19 & 1531 & 28 & 44 &   6.81 & 0.70 & 0.52 & 0.75 & 14 &  23.83 & 1.85 & 1.75 & 1.97 \\
				Slovenia &    2.08 & 75.00 &  23442.70 &  8.19 & 79.00 & 2020-04-02 & 1485 & 109 & 29 &  11.85 & 1.17 & 0.88 & 1.14 & 20 &  15.02 & 1.89 & 1.56 & 1.76 \\
				Somalia &   15.86 &   - &    309.06 &   - & 25.00 & 2020-06-03 & 2368 & 84 & 80 &  14.31 & 1.66 & 1.44 & 1.70 & 57 &  38.97 & 2.50 & 1.21 & 1.97 \\
				South Africa\tnote{*} &   59.26 & 72.40 &   6132.48 &  8.11 & 69.00 & 2020-06-08 & 50879 & 1080 & 96 &   5.88 & 1.03 & 0.42 & 0.75 & 74 &  11.68 & 0.92 & 0.60 & 0.96 \\
				South Korea\tnote{*} &   51.27 & 80.00 &  29803.23 &  7.60 & 86.00 & 2020-03-04 & 11852 & 274 & 43 &  13.89 & 1.55 & 1.13 & 1.34 & 14 &   7.15 & 1.22 & 0.74 & 1.03 \\
				South Sudan &   11.19 &   - &       - &  9.76 & 31.00 & 2020-05-27 & 1604 & 19 & 53 &  41.58 & 2.13 & 1.77 & 2.23 & 13 &  51.36 & 2.18 & 1.59 & 2.19 \\
				Spain\tnote{*} &   46.75 & 82.90 &  28100.85 &  8.87 & 83.00 & 2020-03-31 & 241717 & 28752 & 60 &  16.91 & 1.51 & 1.48 & 1.62 & 29 &   7.61 & 0.69 & 0.50 & 0.74 \\
				Sri Lanka &   21.41 & 62.70 &   4104.63 &  3.81 & 66.00 & 2020-05-30 & 1857 & 11 & 125 &  68.03 & 3.75 & 3.52 & 3.94 & 64 & 261.49 & 4.74 & 2.79 & 4.12 \\
				Sudan &   43.78 & 27.00 &   3015.02 &  6.34 & 44.00 & 2020-05-27 & 6242 & 372 & 76 &  13.52 & 1.55 & 0.75 & 1.23 & 76 &  20.36 & 2.14 & 1.39 & 1.83 \\
				Suriname &    0.59 & 69.80 &   5379.12 &  6.23 & 71.00 & 2020-06-07 & 128 & 2 & 86 & 108.04 & 4.70 & 4.54 & 5.04 & 66 & 153.25 & 5.81 & 5.68 & 6.10 \\
				Sweden\tnote{*} &   10.10 & 93.90 &  53744.43 & 11.02 & 86.00 & 2020-06-08 & 45133 & 4694 & 130 &  26.10 & 2.38 & 1.57 & 1.96 & 90 &  33.28 & 2.49 & 1.24 & 1.92 \\
				Switzerland\tnote{*} &    8.65 & 90.30 &  80450.05 & 12.35 & 83.00 & 2020-03-25 & 30972 & 1923 & 30 &   3.26 & 0.44 & 0.37 & 0.56 & 21 &   6.41 & 0.92 & 0.80 & 1.00 \\
				Syria &   17.47 & 14.30 &       - &   - & 60.00 & 2020-05-28 & 144 & 6 & 68 &  70.22 & 2.98 & 2.11 & 2.60 & 61 & 127.47 & 4.68 & 3.25 & 4.25 \\
				Tajikistan &    9.52 & 19.30 &    806.04 &  7.23 & 68.00 & 2020-05-21 & 4609 & 48 & 22 &   4.33 & 0.81 & 0.37 & 0.66 & 20 &  15.10 & 1.88 & 1.00 & 1.53 \\
				Tanzania &   59.62 & 51.60 &   1004.84 &  3.65 & 43.00 & 2020-04-20 & 509 & 21 & 36 &   9.62 & 1.35 & 0.69 & 1.14 & 21 &  12.86 & 1.07 & 1.02 & 1.47 \\
				Thailand\tnote{*} &   69.79 & 63.20 &   6578.19 &  3.75 & 80.00 & 2020-04-03 & 3119 & 58 & 73 &  25.04 & 1.92 & 1.27 & 1.69 & 34 &  41.92 & 2.97 & 2.88 & 3.18 \\
				Timor-Leste &    1.32 & 71.90 &   1294.72 &  3.88 & 52.00 & 2020-04-20 & 24 & 0 & 30 &  33.10 & 2.64 & 2.55 & 2.75 & 0 &    - &  - & - & - \\
				Togo &    8.27 & 33.00 &    626.09 &  6.20 & 43.00 & 2020-05-18 & 497 & 13 & 74 &  25.78 & 2.29 & 1.01 & 1.62 & 53 &  42.35 & 1.03 & 1.01 & 1.91 \\
				Trinidad and Tobago &    1.40 & 71.60 &  16238.19 &  6.98 & 74.00 & 2020-03-27 & 117 & 8 & 14 &  22.08 & 1.82 & 1.13 & 1.61 & 3 &   3.32 & 1.06 & 0.63 & 0.78 \\
				Tunisia &   11.81 & 67.20 &   3482.19 &  7.23 & 70.00 & 2020-04-06 & 1087 & 49 & 34 &   5.32 & 0.49 & 0.40 & 0.60 & 19 &   8.68 & 0.83 & 0.52 & 0.88 \\
				Turkey &   84.28 & 40.90 &  10513.65 &  4.22 & 74.00 & 2020-04-16 & 171121 & 4711 & 37 &   7.09 & 0.78 & 0.41 & 0.76 & 31 &   5.64 & 0.79 & 0.62 & 0.87 \\
				Uganda &   45.64 & 50.20 &    631.52 &  6.19 & 45.00 & 2020-06-02 & 646 & 0 & 74 &  24.86 & 1.49 & 1.18 & 1.54 & 0 &    - &  - & - & - \\
				Ukraine\tnote{*} &   43.75 & 59.00 &   2640.68 &  7.00 & 68.00 & 2020-06-07 & 28077 & 805 & 97 &  17.39 & 1.94 & 1.50 & 1.79 & 87 &  16.01 & 1.60 & 1.09 & 1.43 \\
				United Arab Emirates &    9.88 & 27.60 &  39811.63 &  3.33 & 76.00 & 2020-05-24 & 39376 & 281 & 117 &  21.48 & 1.30 & 0.98 & 1.32 & 66 &  27.21 & 1.94 & 1.89 & 2.05 \\
				United Kingdom\tnote{*} &   67.87 & 85.20 &  40361.42 &  9.63 & 87.00 & 2020-04-14 & 288834 & 40680 & 75 &  41.99 & 1.62 & 1.22 & 1.71 & 40 &   2.92 & 0.50 & 0.45 & 0.52 \\
				United States\tnote{*} of America &  330.89 & 79.60 &  59957.73 & 17.06 & 84.00 & 2020-04-10 & 1960897 & 110990 & 80 &  14.30 & 1.13 & 1.11 & 1.48 & 42 &   6.43 & 0.79 & 0.67 & 0.89 \\
				Uruguay &    3.47 & 83.80 &  16437.24 &  9.30 & 80.00 & 2020-03-28 & 845 & 23 & 16 &   1.40 & 0.50 & 0.30 & 0.43 & 1 &   2.32 & 0.98 & 0.70 & 0.75 \\
				Uzbekistan &   33.44 & 20.10 &   1826.57 &  6.41 & 73.00 & 2020-04-17 & 4440 & 18 & 34 &   3.41 & 0.75 & 0.39 & 0.67 & 22 &  35.64 & 2.73 & 1.52 & 2.29 \\
				Venezuela &   28.44 & 28.80 &       - &  1.18 & 74.00 & 2020-06-07 & 2473 & 22 & 86 &  17.73 & 1.53 & 0.98 & 1.52 & 73 &  76.57 & 3.79 & 2.58 & 3.23 \\
				Vietnam &   97.29 & 30.80 &   2365.62 &  5.53 & 75.00 & 2020-03-28 & 332 & 0 & 66 &  27.55 & 2.28 & 2.23 & 2.45 & 0 &    - &  - & - & - \\
				Western Sahara &    0.60 &   - &       - &   - &   - & 2020-04-05 & 9 & 1 & 1 &   9.32 & 1.26 & 0.90 & 0.99 & 0 &    - &  - & - & - \\
				Yemen &   29.78 & 19.50 &    963.49 &   - & 42.00 & 2020-06-05 & 496 & 112 & 57 &  15.55 & 1.74 & 1.57 & 1.72 & 37 &   6.02 & 0.70 & 0.68 & 0.88 \\
				Zambia &   18.35 & 50.90 &   1534.87 &  4.47 & 53.00 & 2020-05-14 & 1200 & 10 & 58 &  12.94 & 1.36 & 0.89 & 1.36 & 43 &  43.03 & 2.05 & 1.77 & 2.18 \\
				Zimbabwe &   14.85 & 31.60 &   1602.40 &  6.64 & 54.00 & 2020-06-02 & 287 & 4 & 75 &  32.17 & 1.94 & 1.69 & 2.02 & 72 & 213.06 & 4.82 & 4.13 & 4.61 \\

				\bottomrule
			\end{tabular*}
			\begin{tablenotes}
				\item[] This table shows original dataset from 185 countries. * denotes countries with regional data. 
			\end{tablenotes}
		\end{threeparttable}

	\end{center}
\end{landscape}

\restoregeometry

\newgeometry{left=1cm,bottom=2cm, top=2cm, right=1cm}
\begin{landscape}
	\centering
	\renewcommand{\thetable}{A1.\arabic{table}}
	\begin{center}
		\begin{threeparttable}[!ht]
			\caption{Original State-Level Data in the U.S.}
			\label{tab:statedata}
			\tiny
			\renewcommand\arraystretch{1}
			\setlength{\tabcolsep}{3pt}

			\begin{tabular*}{\linewidth}{@{\extracolsep{\fill}}llllllllllllllllllll}
				\toprule
				State & Pop. & GDP & HE\_GDP & Won\_Rep. & Senate\_Rep. & Governor\_Rep. & Cutoff & Tot C. & Tot D. & Days C. & Chi C. & K C. & M C. & D C. & Days D. & Chi D. & K D.  & M D. & D D. \\
				\midrule
				Alabama &  4.90 & 37261.00 & 0.20 & 1 & 1 & 1 & 2020-05-29 & 21626 & 739 & 4476 &  12.55 & 1.12 & 0.59 & 1.09 & 2263 &  212.20 &  6.72 &  3.88 &  5.42 \\
				Alaska &  0.73 & 63971.00 & 0.17 & 1 & 1 & 1 & 2020-06-06 & 593 & 11 & 1040 & 362.00 & 8.75 & 6.63 & 7.86 & 330 &  235.74 &  6.09 &  3.94 &  5.74 \\
				Arizona &  7.28 & 38590.00 & 0.17 & 1 & 1 & 1 & 2020-06-10 & 29852 & 1094 & 1300 &  28.70 & 1.93 & 1.58 & 1.87 & 679 &   47.17 &  2.44 &  1.35 &  1.98 \\
				Arkansas &  3.02 & 36368.00 & 0.20 & 1 & 1 & 1 & 2020-06-10 & 10366 & 164 & 5710 &  76.74 & 3.01 & 3.01 & 3.47 & 1557 &  645.54 & 11.58 & 10.42 & 11.49 \\
				California & 39.51 & 58619.00 & 0.13 & 0 & 0 & 0 & 2020-06-10 & 139736 & 4854 & 5059 & 133.08 & 4.36 & 2.81 & 3.82 & 3026 &  299.67 &  7.79 &  4.07 &  6.47 \\
				Colorado &  5.76 & 52795.00 & 0.13 & 0 & 0 & 0 & 2020-04-30 & 28495 & 1571 & 2326 &  47.19 & 2.01 & 1.36 & 2.32 & 945 &  138.80 &  5.76 &  3.69 &  4.80 \\
				Connecticut &  3.57 & 64511.00 & 0.15 & 0 & 0 & 0 & 2020-04-22 & 44347 & 4120 & 333 &  13.93 & 1.48 & 0.67 & 1.16 & 246 &   27.93 &  1.74 &  1.00 &  1.65 \\
				Delaware &  0.97 & 63664.00 & 0.16 & 0 & 0 & 0 & 2020-04-28 & 10056 & 413 & 153 &   1.03 & 0.18 & 0.18 & 0.24 & 98 &   18.60 &  1.54 &  0.99 &  1.39 \\
				Florida & 21.48 & 39543.00 & 0.20 & 1 & 1 & 1 & 2020-06-10 & 67370 & 2800 & 5627 &  75.30 & 4.05 & 2.16 & 3.34 & 3653 &   32.56 &  1.73 &  1.50 &  1.90 \\
				Georgia & 10.62 & 44723.00 & 0.15 & 1 & 1 & 1 & 2020-05-27 & 54115 & 2344 & 10328 & 184.93 & 6.31 & 2.93 & 4.62 & 6274 & 1433.65 & 17.01 & 15.77 & 17.11 \\
				Hawaii &  1.42 & 51277.00 & 0.14 & 0 & 0 & 0 & 2020-04-03 & 685 & 17 & 98 &  29.75 & 2.49 & 1.87 & 2.32 & 4 &    4.51 &  1.02 &  0.65 &  0.78 \\
				Idaho &  1.79 & 35466.00 & 0.20 & 1 & 1 & 1 & 2020-04-04 & 3259 & 85 & 335 &  74.65 & 3.77 & 2.85 & 3.39 & 44 &   32.13 &  2.72 &  1.32 &  2.13 \\
				Illinois & 12.67 & 54091.00 & 0.15 & 0 & 0 & 0 & 2020-05-07 & 129829 & 6094 & 3893 & 141.21 & 5.22 & 2.44 & 3.90 & 1392 &  351.55 &  8.26 &  8.23 &  8.76 \\
				Indiana &  6.73 & 45317.00 & 0.18 & 1 & 1 & 1 & 2020-04-30 & 38332 & 2172 & 3506 & 104.28 & 4.19 & 4.18 & 4.67 & 1819 &  313.60 &  7.54 &  7.51 &  8.09 \\
				Iowa &  3.16 & 50315.00 & 0.16 & 1 & 1 & 1 & 2020-05-07 & 22502 & 629 & 3538 & 173.25 & 5.68 & 5.06 & 5.64 & 726 &  238.04 &  6.63 &  6.59 &  7.05 \\
				Kansas &  2.91 & 46982.00 & 0.16 & 1 & 1 & 0 & 2020-05-08 & 10846 & 240 & 2848 & 175.76 & 5.30 & 2.77 & 4.11 & 567 &  174.02 &  5.55 &  5.51 &  5.94 \\
				Kentucky &  4.47 & 38985.00 & 0.21 & 1 & 1 & 0 & 2020-06-06 & 11872 & 483 & 7829 & 330.49 & 7.17 & 7.16 & 8.00 & 2479 &  498.85 & 10.69 &  6.12 &  8.34 \\
				Louisiana &  4.65 & 43917.00 & 0.18 & 1 & 1 & 0 & 2020-04-07 & 44013 & 2854 & 1110 &  41.26 & 3.06 & 2.04 & 2.58 & 486 &  126.66 &  4.83 &  4.79 &  5.12 \\
				Maine &  1.34 & 38921.00 & 0.24 & 0 & 0 & 0 & 2020-05-26 & 2636 & 100 & 1087 & 117.91 & 4.68 & 4.21 & 4.72 & 333 &   80.16 &  3.42 &  2.40 &  3.21 \\
				Maryland &  6.05 & 55404.00 & 0.16 & 0 & 0 & 1 & 2020-05-24 & 59463 & 2867 & 1626 &  19.89 & 1.36 & 1.35 & 1.95 & 1110 &   89.65 &  4.01 &  3.99 &  4.33 \\
				Massachusetts &  6.89 & 65545.00 & 0.16 & 0 & 0 & 1 & 2020-04-29 & 104156 & 7454 & 744 &  10.38 & 1.03 & 0.48 & 0.89 & 498 &   25.41 &  1.85 &  1.40 &  1.72 \\
				Michigan &  9.99 & 43372.00 & 0.19 & 1 & 1 & 0 & 2020-04-07 & 65180 & 5952 & 1152 &  82.69 & 3.80 & 3.78 & 4.05 & 311 &  116.59 &  5.13 &  3.76 &  4.47 \\
				Minnesota &  5.64 & 53704.00 & 0.17 & 0 & 1 & 0 & 2020-05-25 & 28865 & 1234 & 4744 & 154.47 & 5.63 & 4.28 & 5.03 & 1173 &  361.37 &  8.66 &  7.73 &  8.49 \\
				Mississippi &  2.98 & 31881.00 & 0.24 & 1 & 1 & 1 & 2020-06-02 & 18480 & 868 & 5874 &  50.59 & 1.82 & 1.60 & 2.37 & 3655 &  320.81 &  7.40 &  6.02 &  7.01 \\
				Missouri &  6.14 & 43317.00 & 0.19 & 1 & 1 & 1 & 2020-04-12 & 15183 & 848 & 1491 & 160.41 & 5.57 & 4.04 & 4.95 & 265 &  257.86 &  7.21 &  7.14 &  7.71 \\
				Montana &  1.07 & 39356.00 & 0.21 & 1 & 1 & 0 & 2020-04-01 & 568 & 19 & 234 &  33.42 & 2.28 & 2.26 & 2.55 & 8 &    7.09 &  1.22 &  0.92 &  1.10 \\
				Nebraska &  1.93 & 53114.00 & 0.16 & 1 & 0 & 1 & 2020-05-04 & 15839 & 191 & 1840 & 167.23 & 6.11 & 4.85 & 5.53 & 287 &  340.20 &  8.02 &  7.95 &  8.66 \\
				Nevada &  3.08 & 43820.00 & 0.15 & 0 & 0 & 0 & 2020-04-22 & 10155 & 460 & 362 &  68.61 & 3.67 & 3.15 & 3.62 & 109 &   63.93 &  3.43 &  3.37 &  3.71 \\
				New Hampshire &  1.36 & 51794.00 & 0.19 & 0 & 0 & 1 & 2020-05-07 & 5177 & 301 & 551 &  26.89 & 2.39 & 0.82 & 1.75 & 172 &   55.11 &  2.95 &  2.19 &  2.67 \\
				New Jersey &  8.88 & 57084.00 & 0.16 & 0 & 0 & 0 & 2020-04-09 & 165280 & 12377 & 605 &   8.18 & 1.06 & 0.75 & 0.98 & 365 &   17.82 &  1.70 &  1.37 &  1.66 \\
				New Mexico &  2.10 & 41348.00 & 0.17 & 0 & 0 & 0 & 2020-05-24 & 9248 & 410 & 1719 & 116.54 & 4.23 & 4.21 & 4.59 & 586 &  164.67 &  5.92 &  5.11 &  5.64 \\
				New York & 19.45 & 64579.00 & 0.15 & 0 & 0 & 0 & 2020-04-09 & 380225 & 30342 & 1532 &  38.67 & 2.77 & 1.81 & 2.43 & 603 &   82.35 &  4.00 &  3.97 &  4.22 \\
				North Carolina & 10.49 & 44325.00 & 0.16 & 1 & 1 & 0 & 2020-06-07 & 38168 & 1041 & 7372 &  80.75 & 4.09 & 1.59 & 2.97 & 4001 &  518.50 &  9.95 &  6.10 &  8.39 \\
				North Dakota &  0.76 & 62837.00 & 0.16 & 1 & 1 & 1 & 2020-05-22 & 2941 & 73 & 1924 & 482.35 & 9.58 & 8.34 & 9.35 & 318 &  255.91 &  8.00 &  5.12 &  6.61 \\
				Ohio & 11.69 & 47567.00 & 0.18 & 1 & 1 & 1 & 2020-04-21 & 39572 & 2457 & 2485 &  45.47 & 2.65 & 1.70 & 2.45 & 955 &  254.59 &  7.33 &  6.49 &  7.13 \\
				Oklahoma &  3.96 & 44623.00 & 0.17 & 1 & 1 & 1 & 2020-04-27 & 7479 & 355 & 2083 & 111.39 & 4.80 & 3.42 & 4.22 & 716 &  198.16 &  6.24 &  5.10 &  5.81 \\
				Oregon &  4.22 & 50582.00 & 0.16 & 0 & 0 & 0 & 2020-06-10 & 5059 & 169 & 2655 & 146.08 & 3.19 & 3.18 & 4.25 & 866 &  126.94 &  3.78 &  2.86 &  4.07 \\
				Pennsylvania & 12.80 & 50997.00 & 0.18 & 1 & 1 & 0 & 2020-04-11 & 76843 & 6060 & 1387 &  36.55 & 2.71 & 2.70 & 2.89 & 396 &   93.98 &  4.67 &  3.21 &  3.98 \\
				Rhode Island &  1.06 & 47639.00 & 0.20 & 0 & 0 & 0 & 2020-05-05 & 15756 & 812 & 303 &  16.33 & 0.84 & 0.76 & 1.22 & 95 &   38.15 &  2.26 &  2.09 &  2.54 \\
				South Carolina &  5.15 & 37063.00 & 0.20 & 1 & 1 & 1 & 2020-06-10 & 15758 & 575 & 3791 &  58.72 & 3.04 & 2.17 & 2.69 & 2359 &  295.40 &  6.46 &  6.44 &  7.46 \\

				South Dakota &  0.88 & 48076.00 & 0.19 & 1 & 1 & 1 & 2020-05-13 & 5596 & 69 & 1981 & 411.55 & 8.54 & 8.51 & 9.15 & 202 &  194.77 &  6.38 &  5.92 &  6.43 \\
				Tennessee &  6.83 & 43267.00 & 0.17 & 1 & 1 & 1 & 2020-05-04 & 27830 & 431 & 3869 &  86.43 & 3.45 & 1.41 & 2.59 & 1041 &  567.83 & 10.36 & 10.31 & 11.00 \\
				Texas & 29.00 & 53795.00 & 0.13 & 1 & 1 & 1 & 2020-06-10 & 79928 & 1860 & 15838 & 299.09 & 7.30 & 6.22 & 7.09 & 6252 & 1271.36 & 16.53 & 13.20 & 15.01 \\
				Utah &  3.21 & 44636.00 & 0.13 & 1 & 1 & 1 & 2020-06-07 & 12893 & 128 & 1884 &  17.78 & 1.48 & 0.84 & 1.51 & 622 &  281.90 &  7.56 &  7.01 &  7.62 \\
				Vermont &  0.62 & 43946.00 & 0.23 & 0 & 0 & 1 & 2020-04-09 & 1094 & 55 & 298 &  12.61 & 1.44 & 0.89 & 1.21 & 73 &   24.10 &  2.00 &  1.88 &  2.23 \\
				Virginia &  8.54 & 51736.00 & 0.15 & 0 & 0 & 0 & 2020-05-31 & 52160 & 1502 & 8474 &  50.69 & 2.87 & 1.65 & 2.44 & 3828 &  705.46 & 10.93 & 10.74 & 11.60 \\
				Washington &  7.61 & 56831.00 & 0.14 & 0 & 0 & 0 & 2020-04-01 & 24628 & 1189 & 672 &  94.70 & 3.95 & 3.92 & 4.40 & 198 &   49.48 &  2.98 &  2.94 &  3.21 \\
				West Virginia &  1.79 & 36315.00 & 0.26 & 1 & 1 & 1 & 2020-04-10 & 2183 & 85 & 555 &  67.79 & 3.46 & 1.39 & 2.52 & 34 &   66.72 &  3.86 &  3.73 &  3.99 \\
				Wisconsin &  5.82 & 47266.00 & 0.18 & 1 & 1 & 0 & 2020-05-30 & 21593 & 670 & 4485 & 120.43 & 4.25 & 2.12 & 3.38 & 1897 &  881.27 & 12.70 & 12.65 & 13.57 \\
				Wyoming &  0.58 & 58821.00 & 0.14 & 1 & 1 & 1 & 2020-05-21 & 979 & 18 & 1194 & 231.41 & 5.31 & 5.28 & 6.76 & 141 &  176.35 &  5.52 &  5.44 &  6.14 \\

				\bottomrule
			\end{tabular*}
			\begin{tablenotes}
				\item[] 
			\end{tablenotes}
		\end{threeparttable}

	\end{center}
\end{landscape}

\restoregeometry

\section*{Appendix A2}\label{app2:varDefinition}
\setcounter{table}{0}
\renewcommand{\thetable}{A2.\arabic{table}}

\begin{center}
	\begin{table}[H]
		\begin{threeparttable}
			\scriptsize
			\renewcommand{\arraystretch}{1.25}
			
			\caption{Variable Definitions}\label{tab0:VarDefine}
			\begin{tabular*}{\textwidth}{l@{\extracolsep{\fill}}p{13cm}}
				\toprule
				\textbf{Variable Names}& \textbf{Definitions} \\
				
				\midrule
				
				Chi-sq. Conf.  & Chi-squared statistic based on the cumulative number of confirmed
				cases, calculated as in formula 1.\\
				\addlinespace[4pt]
				Kuipier Conf.  & Kuiper statistic based on the cumulative number of confirmed cases,
				calculated as in formulae 2 and 3.\\
				\addlinespace[4pt]
				M Conf.  & M-statistic based on the cumulative number of confirmed cases, calculated
				as in formula 4.\\
				\addlinespace[4pt]
				D Conf.  & D-statistic on the cumulative number of confirmed cases, calculated
				as in formula 5.\\
				\addlinespace[4pt]
				Chi-sq. Death  & Chi-squared statistic based on the cumulative number of deaths, calculated
				as in formula 1. \\
				\addlinespace[4pt]
				Kuiper Death  & Kuiper statistics based on the cumulative number of deaths, calculated
				as in formulae 2 and 3. \\
				\addlinespace[4pt]
				M Death  & L-statistics based on the cumulative number of deaths, calculated
				as in formula 4. \\
				\addlinespace[4pt]
				D Death  & D-statistics based on the cumulative number of deaths, calculated
				as in formula 5. \\
				\addlinespace[4pt]
				EIU  & \emph{The Economist Intelligence Unit }Democracy Index. It is presented on a scale of 0 to 100. The index consists of five components. \\
				\addlinespace[4pt]
				ELECT & Electoral pluralism \\
				\addlinespace[4pt]
				GVMT & Functioning of the government \\
				\addlinespace[4pt]
				PART & Political participation \\
				\addlinespace[4pt]
				CULT & Political culture \\
				\addlinespace[4pt]
				LIBERT & Civil liberties \\
				\addlinespace[4pt]

				GDP per capita  & Gross domestic product divided by midyear population. GDP is the sum
				of gross value added by all resident producers in the economy plus
				any product taxes and minus any subsidies not included in the value
				of the products. It is calculated without making deductions for depreciation
				of fabricated assets or for depletion and degradation of natural resources.
				Data are in current U.S. dollars.\\
				\addlinespace[4pt]
				HE\_GDP  & Level of current health expenditure (\% of GDP). Level of current
				health expenditure expressed as a percentage of GDP. Estimates of
				current health expenditures include healthcare goods and services
				consumed during each year. This indicator does not include capital
				health expenditures such as buildings, machinery, IT and stocks of
				vaccines for emergency or outbreaks.\\
				\addlinespace[4pt]
				UHC  & Coverage index for essential health services. UHC is the coverage
				index for essential health services (based on tracer interventions
				that include reproductive, maternal, newborn and child health, infectious
				diseases, noncommunicable diseases and service capacity and access).
				It is presented on a scale of 0 to 100.\\
				\addlinespace[4pt]
				FH\_DEM & A dummy variable, which equals one if the country is on the list of electoral democracies by the \textit{Freedom House}, and zero otherwise. \\
				\addlinespace[4pt]
				FH\_AV & The sum of two measures of democracy, political freedom and civil liberties, provided by the \textit{Freedom House} \\
				\addlinespace[4pt]

				No of Days Conf.  & Number of non-zero days of daily new confirmed cases.\\
				\addlinespace[4pt]
				No of Days Death.  & Number of non-zero days for daily new deaths.\\
				\addlinespace[4pt]
				Population  & Population of a country.\\
				\addlinespace[4pt]
				Cutoff value & The earliest date with the maximum 7-day moving average number of
				new confirmed cases for the country. \\

				\bottomrule
			\end{tabular*}
			\begin{tablenotes}
				\item[\textdagger] The first 8 variables are goodness-of-fit measures: Chi-sq.
				Conf., Kuipier Conf., M Conf., D Conf., Chi-sq. Death, Kuiper Death,
				M Death and D Death. They are calculated with 3 cutoff points: using
				the cutoff for the growth part, using 80 days since January 22, 2020,
				and using 45 calendar days since the first nonzero case for individual
				countries. In addition, we apply a so-called ``window'' approach.
			\end{tablenotes}
			
		\end{threeparttable}
	\end{table}
\end{center}

\end{document}